Krzysztof Wołk

**Building an Internet Radio System**

# Building an Internet Radio System

with Interdisciplinary factored system for automatic content recommendation.

with Interdisciplinary factored system for automatic content recommendation.

***Krzysztof Wołk was supported by the European Community from the European Social Fund within the Interkadra project UDA-POKL-04.01.01-00-014/10-00.***

## Table of Contents

with Interdisciplinary factored system for automatic content recommendation.

# **GLOSSARY**

**4G** – fourth generation mobile cell based on the transmission of data with IP protocol, which uses LTE and WiMAX technology.

**AAC (Advanced Audio Coding)** – a sound compression algorithm designed as the successor to MP3.

**AM (Amplitude Modulation)** –a type of modulation that encodes instantaneous signal amplitude changes, where the output signal is a narrowband signal suitable for transmission, for example, via radio.

**Broadcasting** – a type of transmission that sends data via a single port packets, which can be received by any other ports connected to network.

**DAB (Digital Audio Broadcasting)** – a land digital radio technology that enables the transmission of multiple signals in stereo high quality, enabling transmission and reception of radio programs in digital form.

**DNAS (Distributed Network Audio Server)** – a server that enables scattered transmission of audio data using SHOUTcast technology.

**DRM (Digital Rights Management)** – a system allowing electronic access control of digital data rights.

**FM (Frequency Modulation)** –the modulation of the carrier with temporary changes in the frequency that enables the receiver to filter out more noise than in the AM.

with Interdisciplinary factored system for automatic content recommendation.

**GNU General Public License** – a license used for open and free software.

**HSPA (High Speed Packet Access)** – a wireless standard known as 3G broadband, widely used in mobile phones.

**Hyper-V** –software created Microsoft that is used to create virtual machines and is available in a free version and built-in Windows Server systems in a series of 2008 and newer.

**IWA (International Webcasting Association)** – a non-profit organization that is bringing together Internet broadcasters.

**Camouflage** – a phenomenon occurring in the auditory system that involves an increase **in** the detection signal masked by the presence of another signal called a Masker. You can distinguish simultaneous signal broadcast where masking occurs. When a Masker is masking the original signal, the Masker occurs immediately after or before the masked signal. These phenomena are very closely related to the adaptation of the auditory system. It is mainly used in perceptual encoding.

**Podcasting – a** form of broadcasting accomplished by the publication of an RSS channel and the opportunity to subscribe to the broadcast.

**RemoteFX** – a technology developed by Microsoft for Windows 7 and 2008 **(**introduced with Service Pack 1), that provides the functionality of virtual machines running on Hyper-V to support multimedia tasks, including 3D graphics processing.

**RIAA (Recording Industry Association of America)** – a U.S. organization that protects copyrights.

with Interdisciplinary factored system for automatic content recommendation.

**RDS (Radio Data System)** – a standard designed to allow transfer of digital information using regular FM broadcast.

**RSS (Really Simple System)** – a simple form of information distribution used primarily to inform users of websites, etc. when new content is available.

**SAWP -** Association Artists Contractors Songs Music - an organization dedicated to the regulation of rights similar to the copyrights holders.

**SDK** – a set of tools used by a developer for application implementation.

**SHOUTcast Broadcaster Tool** – a set of tools designed for online broadcasters.

**Spave** – the producers of Audio-Video material Association video and phonograms for the fight against piracy.

**SSB (Single Side Band)** - a type of amplitude modulation, characterized by saving bandwidth and power, which is sent on only one sideband without a carrier.

**STOART** – the Association of Performing Artists, an organization dealing with collective copyrights and related rights.

**Streaming** - a technology which allows to play media in real-time.

**Web Service** – a network service, which allows the creation of distributed applications and independent platforms, using protocols such as HTTP and SOAP, and XML. It allows you to integrate applications written in different languages.

with Interdisciplinary factored system for automatic content recommendation.

**WiMAX (Worldwide Interoperability for Microwave Access)** – a technology for wireless data transmission faster than standard 3G.

**WIPO (World Intellectual Property Organization)** – an organization that coordinates and creates regulations related to intellectual property protection.

**WMS (Windows Media Services)** - a set of tools developed by Microsoft to broadcast multimedia contents online.

*X11 (X Window System)* – a GUI windows-based software for computers.

# PURPOSE OF STUDY

Automatic systems for music content recommendation and recognition have assumed a new role in recent years. These systems have easily transformed themselves from being just a convenient, standalone tool into an inseparable element of modern living. In addition, not only do these systems strongly influence human moods and feelings with the selection of proper music content, but they also provide significant commercial and advertising opportunities. This research aims to examine and implement two such systems available for the automatic recognition and recommendation of music and advertisement content for Internet radio. Through analysis of the practical issues of application fields and spheres of influence, conclusions will be drawn about the possible perspectives on and future role of such systems. Other content adaptation that is based on music genres will be discussed, as well.

There are several factors that play a huge role in characterizing this phenomenon. First, unprecedented technological developments lead to the introduction of new spheres of musical experiences, such as the Internet and mobile devices. Second, the invaluable and indisputable influence that music has on human behavior provides an unparalleled opportunity to access music contents and genres. As expected, these processes were noticed by the music industry, as well as the advertising industry. Thus, in their persistent effort to reach new and sustainable numbers of listeners, automatic music content tools for recommendation and recognition proved to be invaluable allies. This research will introduce and explain the two different types of automatic content recommendation systems, along with an implementation of an Internet radio system, which such systems include. In addition, there shall be an immense emphasis on commercial applications, advertising opportunities, and the orientation of business spheres. The effects of music on human psyche will not be neglected. However, these effects shall be examined primarily in the context of the purposes and objectives of automatic content recognition and recommendation. Furthermore, this research will also attempt to provide an appropriate justification for the creation and development of such systems. The issues that are faced in practical applications (including the future perspectives) of these systems will also be examined.

Another aim of this study is to provide an innovative Internet radio implementation as compared to traditional radio and other Internet broadcast solutions. This will include automatic content recommendation systems for listeners and marketing companies, as well as the usage of a voice synthesizer in automatic program scheduling, and discussion of how it relates to the following issues:

- available formats and software required to listen to Internet radio broadcasts;
- technologies used in the operation of Internet radio;
- legal and financial aspects related to the construction of Internet radio;
- review and analysis of Internet radio markets in selected countries ;
- the future of Internet radio in view of the development of other electronic media;
- the connection between the listener and radio advertising;
- one radio project implementation;
- the results of performance tests and reliability;
- an approximation of hardware requirements.

with Interdisciplinary factored system for automatic content recommendation.

The purpose of this study is not only to examine the functioning and evolution of radio communication technology in the world, but also to discuss the need for its development and design. This study will prove that it is worth the investment in innovation and provide further implementation ideas related to Internet radio. This will help not only to promote a new kind of radio communications and the general progress of civilization, but also will reflect market interest from the public, focused on comfort and demanding newer and newer solutions to meet their needs.

with Interdisciplinary factored system for automatic content recommendation.

# INTRODUCTION

The need for communication between people is as old as human history, and is associated with satisfying the needs of not only information but also recreation, which combines the search for ever newer and better forms in order to improve the quality of life.

In the past, due to the huge distance between human settlements, it was hard to communicate except through the use of visual communication. With signals at high amplitudes, such as fire, whose task was to convey specific messages or to call for a community event, this need was satisfied. To some extent, it was a good method, but it did not solve the problem of communication beyond the range of sight and hearing, which led to further research in methods of achieving it.

In the mid-nineteenth century the first prototype and radio model was introduced, the telegraph (Figure 1). This led to the birth of media as we know it today, methods that were still waiting to be invented so that we can communicate better and more comfortably.

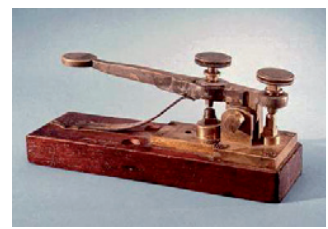

**Figure 1. Figure 1. Telegraph from the nineteenth century**

Today's media allow us to receive information from all over the world, wherever we are. They have a global reach. Radio, which is the main focus of this work, is not only a carrier of both political and social information, but it also fulfills the mission of entertainment and adaptation to market needs, not only general needs, but also for the specific needs of individuals.

with Interdisciplinary factored system for automatic content recommendation.

Almost every element of radio's techniques evolved in design and capabilities over the years as a result of man's endless ingenuity, including the changing reality and technical environment. Its development continues to introduce pioneering solutions in terms of components, adaptation in ecology and environmental protection, durability, and equipment uptime.

Internet radio is a new way of communication. In practice, Internet radio is often amateurish, and traditional broadcasters use the new means of communication as an additional broadcast channel.

Internet radio, which is the main focus of this work, will be described in terms of its history, construction, technical capabilities, and issues related to communications technology and implementation.

This work consists of a theoretical part, which describes the theoretical issues of Internet radio, as well as a practical part, which will present an actual model in this area.

In the first chapter the history and construction of traditional radio, and Internet radio will be presented. These two products will be compared in terms of techniques, advantages, and disadvantages to show the idea of innovation, as well as to demonstrate the opportunities and possible threats.

In the second chapter of this document will characterize it in terms of technical formats and software availability, which will be followed by the detailed description of the technical requirements. In addition, we will discuss issues such as the economic and legal aspects.

The third section will include information about Internet radio in the world market. Some of the international markets, such as Germany, Sweden,

United Kingdom, United States and Poland, the world's largest broadcasting markets, as well as those operating in Poland will be examined. This will also be examined in the context of Internet radio and other electronic media, their growth, and competitiveness, including analysis of commercial products that are currently attracting attention in the marketplace.

In the fourth chapter the radio will be examined with respect to the listener and advertising. It will be discussed in the context of music, the main genres and trends emerging in the field. This chapter will present the relationship between music and human psychology, as well as the relationship between music and marketing, including the results of public opinion research and conclusions that emerge on the topic of music links to marketing and advanced content personalization.

In the fifth chapter technical aspects of the operation of Internet radio will be discussed, such as technology transmission, methods of streaming audio compression, dependence on required quality programs, ways to improve the quality of recordings, network requirements, and useful network protocols.

The sixth chapter will present an actual Internet radio implementation in practice. This chapter will address technology issues such as the BASS library, .NET, C#, Silverlight, Mono, and MoonLight. Next, the structure and scheme of Internet radio application installation, references to the users interface application, and then the structure of code and implementation will be discussed. In the last section, detailed information on the need and the ability to test the performance and reliability of Internet radio designed will be examined, as well as the results of this analysis and research, including the approximation of hardware requirements.

with Interdisciplinary factored system for automatic content recommendation.

This paper will help us understand Internet radio and the issues associated with it. It will also address the legal, financial, and --- above all --- technical issues, such as the description of technologies used by Internet Radio and implementation. In this document references to other electronic media that present opportunities and pose threats will also be addressed.

with Interdisciplinary factored system for automatic content recommendation.

# Chapter 1. History and operation of radio

Radio history can be dated back to the mid-nineteenth century, when the phenomenon of radio waves was discovered. They are an essential medium to transmitting a radio signal. From the point of view of physics, they are a kind of electromagnetic wave. Their existence initially was predicted only in theory by James Clerk Maxwell (Figure 2) in 1864. The presence of the effect of waves was explained and also substantiated by Maxwell's equations. He reported phenomena such as electricity and magnetism, and also defined the relationship between them, which are now known under the name of their creator - the laws of Maxwell.

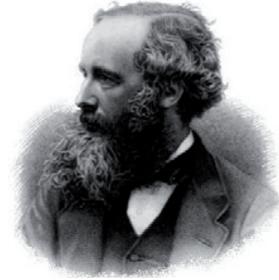

**Figure 3. James Maxwell**

Maxwell's theory was turned into practice in 1887, when Heinrich Hertz (Figure 3) experimentally confirmed the existence of these waves. A study conducted by the experiment has shown that they are light. However, it is interesting that the Hertz said that *"(...) These waves will never have a practical application, the discovery has only the importance of pure theory"*[1]. Nevertheless, history proved him wrong.

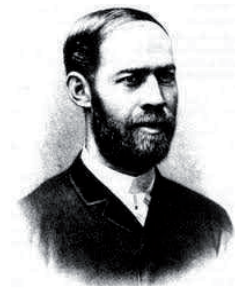

**Figure 2. Heinrich Hertz**

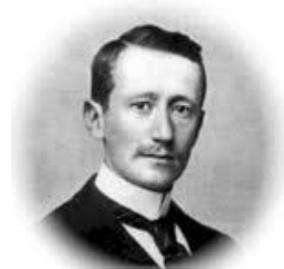

**Figure 4. Guglielmo Marconi**

Twenty years later, Guglielmo Marconi (Figure 4) developed and patented the first transceiver system. This allowed the sending of coded

---

[1] Quote taken from a speech by Heinrich Hertz

with Interdisciplinary factored system for automatic content recommendation.

signals at a distance - Morse code. During this time, it was possible to achieve a "Wireless telegraph". For this purpose, the signals were used as electromagnetic waves, previously discovered in Maxwell's theory and experimentally proven by Hertz.

with Interdisciplinary factored system for automatic content recommendation.

## 1.1. A brief history of radio around the world

At the beginning of the twentieth century the first radio broadcasts began to appear. This revolution began in the United States. However, they were given to very few users at the beginning. Furthermore, the first customers were mainly radio operators on ships, because only they had the appropriate receivers. Nevertheless, at the beginning of the 20th century, the first regularly broadcasting station, which was supposed to be targeted to any owner of a home receiver, was introduced.

The date of birth of the radio on Polish territory is considered to be the 1 February 1925. On that day at 18:00 on the 385m wave, one station started broadcasting radio on behalf of the Polish Engineering Society, which dealt with the production of radio engineering equipment. The first broadcast lasted only an hour, and the radio itself was created two days earlier, with its signal emitted from Warsaw. Further programs were broadcast on a 480m wave, and the announcer began with the words: *"Hello, hello, Polish Radio Warsaw, wave 480"*[2]. On the air soon after, the first messages from the Polish Telegraph Agency started to appear, which included weather forecasts. It is noteworthy that the radio program was created and broadcasted live. Initially, the station broadcasted to a fairly limited area, a circle with a radius of about 200km. However, after a short period of time and as a result of improvements in the radio transmitter, it was possible to listen to it even in Vilnius and Krakow.

In August 1925 the first license for radio broadcasting in Poland was granted. It is surprising that it was not received by the Polish Society of Radio

---

[2]http://pl.wikipedia.org/wiki/Polskie_Radio

with Interdisciplinary factored system for automatic content recommendation.

Engineering, which already had a lot of facilities and --- above all --- experience in radio broadcasting, but rather by the first Polish Radio company associated with the "Power and Light" company. Perhaps these facts determined the decisions relating to the granting of the license.

As of April 18, 1926, Polish Radio started broadcasting regular programs in Warsaw, on the 1111m wave. In the subsequent years, radio stations were launched in other cities, including Katowice, Krakow, Vilnius, Lviv and Toruń. Before the war, a transmitter was also built in Raszyn (see Figure 5) with the respectable power of 120kW and 200m. At that time it was the highest emitter in Europe and allowed coverage of the whole country.

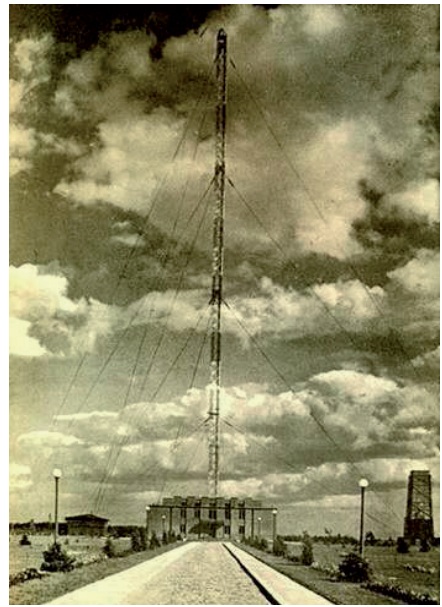

**Figure 5. The transmitter in Raszyn**

The outbreak of war halted the development of broadcasting in Poland, as consumers had been banned by the Nazis. During the Warsaw uprising the insurgent radio station "Lightning" began broadcasting after a few years of silence. In the period between 8 August 1944 and 4 October 1944, the station broadcast 4 hours of daily information on the latest news of the uprising, the world situation, and literary programs and music. The first emission that occurred was "the Bee". It was installed inside a station wagon, which was in a part of Lublin. Its "Bee" activities began on with Interdisciplinary factored system for automatic content recommendation.

Aug. 11, 1944, and the last broadcast from Lublin was on March 9, 1945. After one week the station started broadcasting from Warsaw, unfortunately ending its operations shortly after that in 15 November 1945. [3]

A period of rapid expansion of the Polish radio network came after the war. A Polish company was founded under the already recognized name of Polish Radio. In 1950, each of the 17 provincial cities had their own radio station owners. The previously mentioned Raszyn transmitter, which during the war was blown up by a group of soldiers wanting to prevent the spread of Nazi propaganda, was also rebuilt. Due to the prevailing political system, the broadcasts of Polish Radio were subject to strong censorship. This was the reason for the uprising in May 1952. The radio "Free Europe" had its headquarters in Munich. Its signal was transmitted from the area of Lisbon. Although the country's government tried to jam the radio, because it gave the politically incorrect content, they could not prevent it from being registered on the pages of history.

In the period 1952-2000 the country developed a working radio broadcasting on short-wave (FM), with sixty transmitters operating in the band 87.5-100MHz (according to international agreements) that replaced transmitters operating in the lower band 66-73MHz. For the first time the stereo radio broadcast was introduced.

Political changes at the end of the eighties and nineties in the twentieth century also contributed to the development of a radiophone market in Poland. On 15 January 1990, the first private radio station in Krakow called "Radio

---

[3]Information on the history of the radio is available on the website:
http://radiopolska.pl/portal/staticpages/index.php?page=historia

Little Poland" began broadcasting. Poland in this way became the leading country in Eastern Europe. By adjusting the receivers to 70.6 MHz, the only program initially heard came from the French radio "FUN". Over time, the program introduced leading Polish songs. Shortly after this, the radio station changed its name to "RMF FM", under which it is known today. In 1990, a number of other commercial radio stations were introduced:

- in Opatów "Radio Opatów";

- in Zakopane "Alex";

- in Warsaw "Radio S" including "Solidarność", which is now split into two radio stations known as "Eska" and "Radio Zet".

By 1994 a number of radio stations had appeared in Poland, including commercial stations. In addition, in order to regulate the market, the National Council of Radio Broadcasting and Television was introduced. This Council introduced the country's first licensing process for broadcasters such as "RMF FM", "Radio Zet" and "Radio Maryja" which received national licenses.

From the beginning of 2000 to the present, only the upper band VHF, or 87.5-108 MHz, has been used in Poland. The only exceptions are radio stations that, due to the lack of upper frequencies, could not emit in this band. Currently in Poland there are about 200 radio stations, of which just over 10% are part of "Polish Radio". The other stations are private, government, or church-owned. On the national level, the most popular are "RMF FM", "Polish Radio" and "Radio Zet." The radio frequency distribution is shown in the following table:

with Interdisciplinary factored system for automatic content recommendation.

Table 1.

| Frequency | Purpose | Issue |
|---|---|---|
| 27MHz | CB | mainly AM |
| 28MHz | Shortband Radio 10m | mainly SSB |
| 50MHz | Shortband Radio 6m | mainly SSB |
| 65-74MHz | UKF | FM Radio |
| 87.5-108MHz | UKF | FM Radio |
| 144-146MHz | Shortband Radio 2m | mainly FM |
| 175-230MHz | VHF3 | digital |
| 430MHz | Shortband Radio 70cm | mainly FM |
| 450MHz | NMT | digital-analog |
| 470-860MHz | UHF | digital |
| 900MHz | GSM | digital |

**Table 1. Overall distribution of radio frequencies**

[4]

It is worth mentioning that in modern times, one of the most important innovations in this area is digital radio. It allows transfer of sound using modulation. This type of transmission can be received by DAB[5] (Digital Audio

---

[4]Table 1 Outline of radio frequency allocation.
http://grzesiek21.republika.pl/podzial_czestotliwosci.htm
[5]http://pl.wikipedia.org/wiki/Radio_cyfrowe

Broadcasting)[6], Internet, and satellite TV. In Poland, unfortunately --- although being tested at the moment --- there is no station officially broadcasting in this way. The main advantage of the new technology is a much better sound, as well as the possibility of providing a much larger amount of information than current RDS systems, while not requiring the tuning of radios after the first sync. The beginning of Internet radio can be assumed to be 1995, when the Conference of European digital radio reserved the 87.5-108MHz frequency band. The first digital receivers went on sale in 1998. 1995 is also the date of the first digital broadcast made by the BBC. However, it turned out that the new technology at transmission speeds of 128kb per second has a lower quality than regular FM stereo broadcasting. The MPEG-2 codec was required to achieve at least the same quality at 160kb per second, and CD quality at 260kB per second. This became the reason why in 2007 the update from DAB[7] to DAB+ was performed. It uses the AAC+ codec, which needs only 64kb per second to transmit sound of satisfactory quality. Unfortunately, the use of DAB+ requires the purchase of a new type of receiver. Despite the advantages of this new technology, it is still far less popular than regular FM radio. Figure 6 shows the countries where there is a digital broadcasting.

---

[6] http://en.wikipedia.org/wiki/Digital_Audio_Broadcasting
[7] http://pl.wikipedia.org/wiki/Digital_Audio_Broadcasting

with Interdisciplinary factored system for automatic content recommendation.

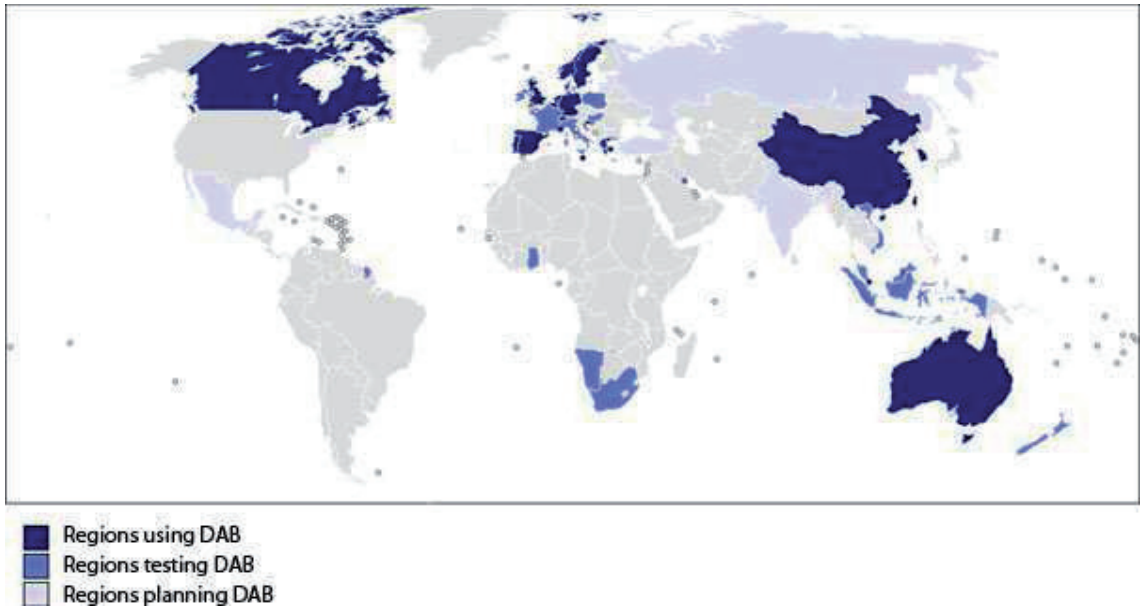

**Figure 6. The range of digital radio[8]**

In our country at the moment, digital channels are only being tested. It is run by several broadcasters, which are listed in Table 2.

| State | Localization | Channel | Content | Compression |
|---|---|---|---|---|
| MZ | Warsaw | 7B (190,640 MHz) | Polskie Radio Program I<br>Polskie Radio Program II<br>Polskie Radio Program III<br>Polskie Radio Program IV<br>RMF FM<br>RMF Classic<br>Radio ZET<br>Chilli ZET<br>TOK FM<br>Radio dla Ciebie<br>Radio Eska<br>Radio Maryja | DAB+ (AAC+) |

**Table 2. Testing of digital transmitters inPoland[9]**

---

[8] http://en.wikipedia.org/wiki/Digital_Audio_Broadcasting

with Interdisciplinary factored system for automatic content recommendation.

## 1.2. History of Internet Radio

In the 90's the Internet significantly changed the conditions and opportunities for the development of new radio stations. The phenomenon of the global network, the Internet, is considered to be one of the wonders of the modern world. Howeverproviding a clear definition of the Internet is extremely difficult. One possible definition is given by Bryan Pfaffenberger, which was formulated as follows:

*"Worldwide, a rapidly growing system of related networks. It offers many services such as remote users to login, file transfer, electronic mail, internet and newsgroups."*[10]

It is also important that the Internet uses TCP / IP protocol. Every computer connected to the Internet has its own, clearly assigned number called an Internet address, also known as an IP address. Thanks to this technology, a computer connected to the network can easy find and exchange data with other machines. Undoubtedly, the Internet has become a new and --- more importantly --- commonly available means of telecommunication that has the transfer speed of your phone, and even television.

The following processes began in connection with the development of network services on the Internet,:

---

[9] http://pl.wikipedia.org/wiki/Digital_Audio_Broadcasting

[10] B. Pfaffenbergera, Słownik terminów komputerowych opracowany na podstawie 6 wyd. Webster's New Works Dictionary of Computer Terms, 1999, s. 111 [w: ] J. Grzenia, Komunikacja językowa w Internecie, Wydawnictwo Naukowe PWN, Warszawa, 2006, s. 19-20.

with Interdisciplinary factored system for automatic content recommendation.

*"(…)technological convergence of media understood as a process of technology transfer characteristic of the traditional electronic media, telecommunications and information technology"*[11]

Media convergence was made possible by the access of more and more people to the Internet, which allows us to receive and transfer multimedia content. Thanks to Wi-Fi hotspots, WiMAX, HSPA and 4G network usage, the Internet is rapidly becoming ubiquitous and a public means of communication.

Internet radio is referred to in the literature as a Web-Net-radio. This form of the message is defined:

*"(…) the transmission and reception of audio files by streaming over the Internet in real time. In this sense, Internet radio is both an online business run by traditional radio stations that broadcast in ether program also available live on the web, as well as Internet radio stations operating activities only on the Internet. This should be distinguished from the possibility of downloading files on demand and the ability to play audio files on demand."*[12]

The term streaming refers to technical aspects affecting the webcasting and defines the means of data transmission. Streaming technology allows direct transmission of files, making it possible to receive audio and live video. To achieve this goal it is necessary to have special software, so-called streaming player or media player, which allows you to read data in the process of downloading. This technique is based on combined data, transmission, compression, and subsequent decompression, which are now an essential condition for the operation of Internet radio. It is important to note the simulcasting technique, which involves simultaneous broadcasting by traditional radio stations on the air and on the Internet. An example is the RMF MAXXX.

---

[11] W. Kolodziejski, P. Keel, Internet Radio, the report of the National Council of Radio and Television, Department of European Policy and International Relations, No. 14/2005, March 2005, p.1

[12] W. Kolodziejski, P. Keel, Internet Radio, the report of the National Council of Radio and Television, Department of European Policy and International Relations, No. 14/2005, March 2005, p.3

With the development of webcasting, new terms such as broadcasting appeared. It was created by "Television without Frontiers" Directive 13, by which it means the transmission of data using wired communications, wireless, satellite, including code, as well as encoded television programs and radio broadcasts intended for the public. It can be said that this is the original form that includes:

*"(…) but does not include communication services providing items of information or other messages on individual demand (…)"*[13]

It is possible to distinguish among them, especially financial services, electronic databases, etc.

---

http://www.cie.gov.pl/www/serce.nsf/0/2CF3BCFA064CB2DBC1256E7F004CC92F?Open&RestrictToCategory=14 Council Directive of 3 October 1989 on the coordination of certain provisions laid down by law, regulation or administrative action in Member States concerning the pursuit of television transmission (89/552/EEC), as amended by Directive 97/36/EC of the European Parliament and of the Council of 30 June 1997 with Interdisciplinary factored system for automatic content recommendation.

The origins of Internet radio and the Internet itself date back to the first half of the 90s. The first online audio broadcast (live) took place in March 1992. This event is known as the MBONE Audio Multicast[14]. Not long after this, only a year later the first Internet radio was implemented and launched. The event took place in the United States because its founder Carl Malamud [15] (Figure 7) was located there. Initially, it took the name of "Internet Talk Radio." The project was very innovative, but we cannot say that was a full-fledged product. The first Internet radio operators around the clock had become established in the same country in 1995 - "Radio HK". It was focused on providing content to independent music listeners. Its creators were Hajjar / Kaufman New Media Lab, who collaborated with an advertising agency operating in California. In the same year there was also the first live news broadcast over the Internet. The initiative was taken by the American network "ABC Radio".

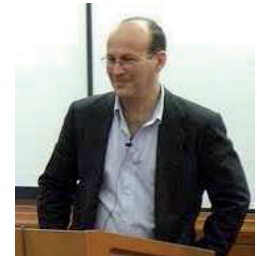

**Figure 7. Carl Malamud**

Since then, the number of online radio stations has grown rapidly. In addition, most of the traditional radio stations began broadcasting their programs in parallel on a network. This process resulted from the fact that there were about 1,500 radio stations that broadcasted over the Internet in 1998. They represented 100 countries of the world, while in 1996 the number of Internet radio stations was 178 and came from 32 countries. Currently, it is estimated that there are several thousands of radio stations worldwide that broadcast their program throughout the web.[16]

---

[14] http://en.wikipedia.org/wiki/Mbone

[15] http://en.wikipedia.org/wiki/Carl_Malamud

A critical moment in the history of Internet radio was 1997, when MP3 audio compression technology was created. Although it had poor compression, it was officially adopted as a standard by the Moving Picture Experts Group[17]. Without a doubt, MP3 is the most common format. In the remainder of this document, it will be specifically discussed.

At the turn of the twenty-first century, Internet radio has become an indispensable component of life for many Internet users. Prime Time (the time with the greatest audience) Internet broadcasting includes the two hours after the morning peak for classical radio stations. It turns out that a lot of people use Internet radio at work. In some companies, it has become such a serious problem that they block access to servers streaming real audio and MP3 formats. Even more so, the introduction of locks allowed for relief in corporate Internet connections, in many cases up to ten percent. This clearly shows how popular this type of service can be.[18]

It is no wonder that Internet radio has been recognized by the business world. In the 90's the company AudioNet.com was founded. Its founders were Mark Coban and Todd Wagner. Currently, it is better known by the name Broadcast.com. In the beginning it was merely an amateur initiative, but due to the huge interest that surprised even the authors, it was quickly transformed into a professional venture. The company very quickly was on the stock exchange. At the moment, it has hundreds of exclusive radio stations and dozens of TV stations, as well as the rights to broadcast sport events over the Internet. In 1990 the company was absorbed by the greatest contemporary Broadcast.com giant Internet service, Yahoo, for the sum of 5.7 billion dollars.

---

[16] W. Kolodziejski, P. Keel, Internet Radio, the report of the National Council of Radio and Television, Department of European Policy and International Relations, No. 14/2005, March 2005, p. 5-7.

[17] http://www.mpeg.org

[18] M. Reaper, AM Zarychta, Internet radio: history, function, evolution and trends. A few notes for the uninitiated, [in] New media and traditional media, newspapers, advertising, Internet, edited by M. Jelińskiego, Publisher "Adam Marshall", Torun 2009, p. 58

With such a rapid tempo of development, the new medium of communication was quickly recognized not only by government agencies but also by various market participants. This is the reason why in 1990, the International Webcasting Association (IWA) was established to operate worldwide. The primary interest of this organization is the development of the industry itself, as well as mediation in the relationship between the institutions.[19]

---

[19] W. Kolodziejski, P. Keel, Internet Radio, the report of the National Council of Radio and Television, Department of European Policy and International Relations, No. 14/2005, March 2005, p. 6

with Interdisciplinary factored system for automatic content recommendation.

## 1.3. Technical Prerequisites to Internet Radio

The development of Internet radio would not have been feasible without the development of the Internet. Access to the Internet was important, because it requires that a radio have the ability to communicate with a global network. Currently, the most commonly used devices are mainly computers, both desktop and mobile (laptops and notebooks). We cannot forget about mobile phones, in particular the so-called smart phones, and PDAs, as well as some of the more advanced MP3 and MP4 playersMP3MP4. The most famous example of this type of device is the iPod series made by Apple. In 2010 and 2011 tablets became the new trend in the development of Internet connectivity. The main purpose of using tablets is to provide the user with access to a variety of informative contents, especially through global media networks. The same Internet connection can be made both in the traditional way --- for example, via a cable connected to the device --- an through the use of wireless technology. Examples of such technologies are radio waves, such as Wi-Fi, microwaves, such as WiMAX, and satellite signal or mobile phone infrastructure, which allows transfer of several megabits per second via HSPA.

The development of the Internet was necessary but not sufficient for Internet Radio. The emergence of Internet radio would not have been possible without significant development of technology to transfer sound in a stream in real-time over long distances. For this purpose it is necessary to have an Internet connection with high bandwidth, the so-called broadband. Unambiguous definition of technical specifications that enable use of a fast Internet broadband connection is extremely difficult because of the very rapid changes in this area.

with Interdisciplinary factored system for automatic content recommendation.

***"Own, different from each other and the time-varying definitions, based on the bandwidth requirements for applying the International Telecommunications Union (ITU) or the Organisation for Economic Cooperation and Development (OECD). In view of the fact that the definition of broadband and evaluates technical progress, following the example of the European Commission, you can always use the current, dynamic definition: the term broadband is high-speed, always-on Internet connection that enables the delivery of innovative content and services. Compared to traditional calls through broadband access is immediate and large amounts of data can be transmitted almost instantaneously, reducing time to market and increasing efficiency for the user."*** [20]

The importance of access to specific specifications of this medium cannot be overstated. This is due primarily to the cost of sending text. Especially in the case of multimedia, which requires high-speed Internet connections, costs are rising proportionally to the number of users interested in certain media. Fortunately, the growing competition in the ISP market is resulting in progressively lower prices in combination with lower operating costs for every Internet broadcaster.

Technological engineering progress in the efficient transmission of audio was also an important factor. Among other things, there are newer and more efficient methods of data compression. With effective compression, smaller data

---

[20] "Broadband" refers to high-speed 'always-on' connections to the Internet that support the delivery of innovative content and services. Compared to traditional narrowband connections, broadband access is immediate and large volumes of data can be almost instantly transmitted, reducing waiting time and improving efficiency for users, the European Commission Communication "Connecting Europe at High Speed: National Broadband Strategies, May 26, 2004, available on the website: http://www.csi.map.es/csi/pdf/com_broadband_en.pdf.

packets need to be transmitted. We therefore need less bandwidth to send the same amount of data.

The effect of the introduction of data compression was the development of streaming, the technology that allows you to play audio and video during streaming. However, we did not have the Internet data packets, which can play information that is transmitted to your computer in real time. This was not possible until the introduction of technical advances that allowed streaming playback live over the Internet. Without the use of this technology, Internet radio would not be possible.

Dissemination of streaming technology was made possible by proprietary software that transmits and receives radio broadcast using streaming technology. The first of these programs was presented in April 1996 when the Real Networks Company, the author of the well known Real Player software and codecs characterized by a good compression ratio. The resulting competitive streaming media formats were introduced by the major IT giants. Microsoft has developed the WMA (Windows Media Audio) file format and Windows Media Player. Apple Computer demonstrated its media player and codec in a series of Quick Time. Another technology. SHOUTcast was developed as a method for streaming music files in MP3 format.

The technology used to broadcast Internet radio is called podcasting. It uses a combination of elements specific to Internet radio with different methods of data collection. Podcasting enables automatic selection and downloading of those programs, which may be played in the future, for example, on a portable multimedia device. An important feature is that this method allows you to omit the personal computer distribution program as a receiver, which allows users to enjoy their mobility. Development of Internet radio forced the transformation of the audiovisual market. It resulted in special devices that make it possible to listen to Internet radio stations without using computers. They usually use Wi-Fi.[21]

## 1.4. Classical Radio vs. Internet Radio

The difference between the classic radio and Internet radio is not only in technology. Broadcasting on the Internet in most countries of the world, as well as in Poland, is not regulated by law. First of all, there is no need to apply for a license as in the case of traditional radio. This is because there is no need for allocation of frequencies. The Internet is a medium common to all, and everyone around the world can use it. A similar issue is the regulation of program content. For example, there are no restrictions on the length of the band advertising. In the case of Internet radio, these items are simply removed from legal jurisdiction. In the majority of cases, these regulations are solely the responsibility of the sender.

Internet has made the radio medium e suitable for the global structure. Waldemar Dubaniowski at the conference "Audiovisual services on the Internet and copyright protection in the digital environment" noted that the modern development of media such as broadcasting and television has chosen a different direction. Moreover, in the case of Internet radio there was a sudden increase in development cost. [22] Internet radio is very small in comparison with classical

---

[21] W. Kolodziejski, P. Keel, Internet Radio, the report of the National Council of Radio Broadcasting and Television Department European Policy and International Relations, No. 14/2005, March 2005, p. 7-14.

[22] Dubaniowski W., Development of new online media (TV and Internet radio), webcasting - analysis of current international practice conference "Audiovisual services on the Internet and copyright protection in the digital environment, Dec. 17, 2002, Warsaw, [a:] M. Reaper, AM

radio. However, please note the copyright notice and ZAiKS, which is dedicated to one of the following chapters. In fact amateur solutions need a computer with broadband Internet access, a microphone, and special software that is available at affordable prices. In addition, there is a full range of free and open source applications running in the cloud that allow people without special hardware or technical expertise to start broadcasting. Problems may arise when the radio gains popularity and large numbers of users listen to it. Then the cheap, amateur measures become insufficient. This technology involves the transmission of data to a local server where network packets are transferred to individual users. The great thing about Internet radio is its independence. Increasingly, a large number of independent and even amateur radio providers are broadcasting. They are also a source of independent information, opinions and comments.

Internet radio amateurs can inexpensively start broadcasting activities. There are many radio stations that were originally supposed to be a hobby and eventually became prosperous stations (for example Last.fm). Emerging radio stations on the Internet became the symbol of a truly independent radio sector.

Małgorzata Kosiarz and Anna M. Zarychta state that:

*"(…)A sign of our times has, inter alia, the creation of the emerging sector is truly independent radio - that created by amateurs passionate about new technological capabilities, but also the created by people with a sense of passion and mission, feeling the need expressed their views, which elsewhere*

---

Zarychta, Internet radio: history, function, evolution and trends. A few notes for the uninitiated, [in] New media and traditional media, newspapers, advertising, Internet, edited by M. Jelińskiego, Publisher "Adam Marshall", Torun 2009, p. 59.

with Interdisciplinary factored system for automatic content recommendation.

***can not express . It is tempting to say that it's just a modern trend (...), but a kind of a cultural phenomenon and commercial"*** [23]

Because of its independence and special nature, Internet broadcasting may be very narrowly targeted to very specific groups , for example, students of film music, religious songs, shanties, etc. In this way, a growing number of Internet radio stations are filling gaps that existed years in traditional broadcasting. They correspond to specific requirements and needs of the audience that traditional radio stations, for various reasons, will never be able to provide.

In summary, traditional radio stations have a large range, good audio quality, but lack a convenient return path for tested and proven technology and the marginal costs associated with receivers. In addition, they are regulated. They are required to have licenses that can only be issued by the National Council of Radio and Television. On the other hand, Internet broadcasting can be characterized as a form of communication with unlimited range (as it can be picked up from anywhere in the world), and a return channel that is interactive with listeners. However, it is still in a stage of development and growth and requires more expensive receivers. In addition, Internet broadcasting is not subject regulation and does not require a license like a radio station.[24]

---

[23] M. Kosiarz, Anna M. Zarychta, Internet radio: history, function, evolution and trends. A few notes for the uninitiated, [in] New media and traditional media, newspapers, advertising, Internet, edited by M. Jelińskiego, Publisher "Adam Marshall", Torun 2009, p. 63.

[24] K. Goldhammer, A. Zerdick, Rundfunk Online - Entwicklung und des Internets Perspektiven Hörfunk-und für Fernsehanbieter, Berlin, 1999, p.21, [a:] W. Kolodziejski, P. Keel, Internet Radio, reports National Broadcasting Council Television, Department of European Policy and International Relations, No. 14/2005, March 2005, p.5.

with Interdisciplinary factored system for automatic content recommendation.

## Chapter 2. Internet Radio Today

Internet radio is a relatively new form of communication. Just like any other form of communication, it has evolved gradually to arrive at its current shape. Its development proceeded with the progress of civilization and technology. After all, it is just dependent on the changing reality and the technical environment, which, thanks to dynamic changes, began to displace solutions used in traditional radios.

### 2.1. Characteristics of formats and software needed to listen to Internet radio

Today, in the era of pervasive globalization and universal access to the Internet, virtually every traditional radio station has its counterpart on the network. It is estimated that the Internet now has about 300,000 radio stations. On average, there is a new station on the Internet every 15 minutes. In recent years, a number of large radio stations broadcast their programs only on the network and are competitive with traditional radio. An example is the station "Live 365", which offers 222 channels of music, 451 classical and jazz stations. and 792 stations that offer pop music.[25]

---

[25] S. Jedrzejewski, Radio in the digital world, [in] New media and traditional media, newspapers, advertising, Internet, edited by M. Jelinski, Publisher "Adam Marshall", Torun 2009, p. 54.

with Interdisciplinary factored system for automatic content recommendation.

### 2.1.1. Characteristics of Internet radio formats

According to S. Jędrzejewski, modern Internet radio can offer the following services:

- Radio Stream – simulcast, the parallel transmission of FM or AM;
- Web Stream – streaming radio programs via the Internet;
- Radio on Demand – individual programs or software packages that allow listening to the radio at a specific time;
- e-Radio – Radio enriched - stream radio, Web radio streaming or on-demand with value added, for example, links to Web sites or information, with the option of downloading or copying music or other program-related products;
- i-Radio – interactive radio, where listeners decide about the content and order programs via simple-to-use interfaces. The listeners can "build" their own radio programs of individual elements (music style, number and frequency and duration of services, weather, local or general, radio drivers, SCN, etc.). Such a "personal radio station" can be controlled manually by each individual listener's profile, including offer of a service that reflects the individual tastes of the recipient;
- Blog–Radio (also known as a wiki-radio), where the user creates the program itself, creating a new community;
- Podcasting (Pod Radio), which will be described in detail later in this paper. [26]

---

[26] S. Jedrzejewski, Radio in the digital world, [in] New media and traditional media, newspapers, advertising, Internet, edited by M. Jelinski, Publisher "Adam Marshall", Torun 2009, p. 54-55.

with Interdisciplinary factored system for automatic content recommendation.

Listening to Internet radio is playing back streaming, for example, by using players that are capable of playing audio already being downloaded from the Internet. When using this technology there is no need to send the whole file to be able to play it. However, it should be taken into consideration that not every audio file format should be sent via Internet streaming.

The first playable file format was a real audio stream (a file with the extension *. rm). It has been designed to take into account the time used by modems to allow people to listen to audio data with modems that have a capacity of only 14.4 kbps (kilobits per second). Subsequently, there were other file formats, such as *.acc, *.wma, *.ogg, *.mp3 and others.

Preparing the appropriate type of files does not guarantee success in the sound transmission through the Internet. The recipient's stream file must comply with Internet technical parameters. The point is that its capacity must be sufficient to receive the bit stream at a certain rate.

Most commonly used streaming audio files can be easily encoded for the most common bandwidths. Naturally, the lower the bandwidth, the worse the quality of the transmitted sound is. In most commercial applications it is assumed, therefore, that the best compromise between Internet usage and audio compression is broadcasting on the order of 128kbps bandwidth.[27] This will ensure that the value offered by transmission is right for Internet resources. This has been confirmed in practice by the largest online broadcasters such as Radio RMF FM, and Last.fm.

---

[27] http://www.lastfm.pl/forum/85180/_/622911, http://forum.dobreprogramy.pl/bitrate-radia-rmf-maxxx-t325463.html

If we have speed fast enough for high-speed continuous playback, file transfer should not pose any problems. However, normally the data flowing through the network is not always at the same speed. Therefore, all programs available on the market use so-called write caching. A buffer is memory space that is obtained from the stored network information to be used in the near future. So, if the media player buffering function is implemented correctly, the listener is not exposed to a sudden interruption of transmission, even in the case of a momentary break of communication with the server, because as long as the buffers have data and are not completely empty, playback remains smooth. [28]

### 2.1.2. Features of available software

The first player that allowed streaming files, as already mentioned, was Real Player. It allowed receiving Real Audio files.[29] Later, there were more players on the market, such as Apple Quick Time[30] (files *.mov) and Microsoft Windows Media Player[31] (files *.wma).

Now, by far the most popular music player is Winamp by NullSoft [32]. It allows you to listen to both files and streaming media. Most of the available radios transmit sound files formats supported by this program. In the case of

---

[28] NPlus.pl, streaming and transmission of Internet, information is available on the website: http://nplus.pl/przesylanie-strumieniowe-i-przesylanie-internetem.

[29] NPlus.pl, streaming and transmission of Internet, information is available on the website: http://nplus.pl/przesylanie-strumieniowe-i-przesylanie-internetem

[30] Information available on the website: http://www.apple.com/quicktime/download/

[31] Information available on the website: http://www.microsoft.com/windows/windowsmedia/pl/

[32] Information available on the website: http://www.winamp.com

Linux-based operating system users, the recommended players are: XMMS[33], Amarok[34] or EXAILE[35], while the Audion[36] player is recommended for Mac OS X users.

At the beginning of each streaming solution, creation of your own file format was required to use a self-made player. Their standards were closed and reserved only for their own products, as exemplified by the Real Player from Real Audio format or Windows Media Player format *.wma. Currently, players usually allow freedom to play all types of files, which reduces the need to install multiple programs to perform the same tasks.

In addition, most Internet broadcasters have built-in music players on their websites that support streaming through a web browser. With just a click on the "Listen" button, without having to install any software, you can listen to the Internet radio. In most cases, these capabilities provide us with products such as Adobe Flash and Microsoft Silverlight, on which, to a certain extent, their work was based.

In addition to programs for streaming playback, there is also a wide range of applications that allow a computer user to become an Internet broadcaster. The simplest form of broadcasting for your own radio is Shoutcast technology, which is also largely explained in this work. It uses sound to transmit MPEG Layer3, commonly known as MP3.

---

[33] Information available on the website: http://www.xmms.org/
[34] Information available on the website: http://amarok.kde.org/
[35] Information available on the website: http://www.exaile.org/
[36] Information available on the website: http://www.panic.com/audion/

with Interdisciplinary factored system for automatic content recommendation.

As it is a generally accepted standard for people to be able to listen to radio in the system, they need only to have a Shoutcast in any media player compatible with streaming files encoded in MP3 format. Practically every popular application has this feature. So you can use the built-in Windows Media Player, available in the popular XMMS Linux on Mac OS X Quick Time Media Player. During the writing of this document, an authoring system has been developed that allows the client to not only listen to broadcasts, but also to receive advertising and provide valuable feedback to the sender of radio program by voting for their favorite programs.

In order to transmit a signal, it is necessary to create a server. For example, in the case of Winamp users, you only need to install an additional SHOUTcast Source for the Winamp Plug-In[37]. After configuring the add-in, which is quite simple and described in detail on the manufacturer's website, it is possible to start broadcasting any playlists stored on your computer. The recipients only need the program or can enter the address http://127.0.0.1:8000/audition.pls directly in the web browser, where 127.0.0.1 is the IP address of any suitable transmission of Shoutcast services of the server, and 8000 is the default port, which is used for client-server communication.

**2.1.3. Podcasting**

Podcasting is a form of Internet radio broadcasting that is published on the Internet in the form of regularly posted sections using an RSS feed (Really Simple Syndication). Podcasting is also sometimes called offline radio, because it is possible to download the entire program onto your computer in the form of any audio file like MP3, which allows you to restore it later without having to connect to the network[38].

---

[37] For information on add-on SHOUTcast Source for Winamp Plug-In and instructions for installation in the wines are available on the website: http://www.winamp.com/online-service/Shoutcast-radio/10100

The term podcast is a combination of words derived from the words "under," which constitutes an abbreviation for iPod (a popular media player, created in the U.S. by Apple), and "cast" (derived from the word meaning broadcast transmission, or transfer). This term was first formulated by Ben Hammersley[39], who worked as a journalist in "The Guardian" [40].

In recent years Podcasting has become a very popular way of getting information without having to tediously search multiple websites. The system works through RSS tags, to which surfers subscribe. The user introduces himself to a program that supports RSS feeds that interest him, and the application synchronizes itself with the server, so that the latest information on interesting issues is downloaded to the device.

The idea of podcasting was presented for the first time In October 2003. This took place at BloggerCon, the first global conference of bloggers. On the second day of this conference, Kevin Marks (Figure 8) (the author of the blog "Epeus follower" and an engineer who has worked with companies such as Google and Apple) presented the first application (implemented in AppleScript) that allowed downloading audio files from the

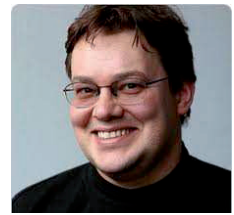

**Figure 8. Kevin Marks**

---

[38] S. Jedrzejewski, Radio in the digital world, [in] New media and traditional media, newspapers, advertising, Internet, edited by M. Jelinski, Publisher "Adam Marshall", Torun 2009, p. 55.

[39] http://en.wikipedia.org/wiki/Ben_Hammersley

[40] http://www.guardian.co.uk/media/2004/feb/12/broadcasting.digitalmedia

Web using RSS feeds. This idea was adopted by the popular iTunes music player where they copied audio to portable iPods with synchronization. Initially, the program written by Marks was called iPodder, and later was converted into a currently very popular podcast called unit Juice[41].

Currently, the network offers a range of different podcasts. These include not only professional productions, streamed versions of radio and television programs, but also amateur audio blogs.

Over time podcasting has also become very popular in Poland. Among other things, the podcast has its own platform for Polish Radio. It provides downloadable versions of some radio programs, including "Salon Polityczny Trójki" or "Sygnały Dnia"[42]. Its podcast channels are also included by commercial radio stations such as RMF FM, Radio Zet, TOK FM. TVN 24 has pieces of their programs available in the form of video-podcast[43].

Podcasting is a relatively new form of streaming, but it appears that this technology has a future. The dynamics of the development of podcasting can be visualized with the following example. A renowned American journalist and blogger on 28 September 2004 demonstrated that the word "podcast" appeared only 24 times as a keyword in the world's most popular search engines, such as Google.com. Less than a month later, on 18 October 2004, the same search returned more than 100,000 links [44].

---

[41] Information available on the website: http://juicereceiver.sourceforge.net/

[42] Information available on the website: http://www.polskieradio.pl/podcasting/

[43] Information available on the website: http://www.tvn24.pl/podcast.html

[44] Podcast - what is it?, The information available on the website:
http://podcastsportowy.wordpress.com/podcast-a-co-to-jest/

with Interdisciplinary factored system for automatic content recommendation.

## 2.2. Economic conditions and financial model for Internet radio

Internet radio has become the domain of small, amateur radios in the international media. Thanks to the Internet, the barriers of political boundaries are being eliminated. Especially active is the Internet radio as a voice of the people in the countries where there is a strong political censorship. We are talking mainly about China, Iran, and North Korea[45]. Furthermore, the lack of clear regulatory law, which will be described a bit later in this paper, creates additional barriers to entry of new operators. In such a situation, it is difficult to consider these in the context of broadcasting business ventures. Typically, this type of radio station is aimed at a small group of customers, who make up a close-knit community of, and the radio station itself is maintained by voluntary contributions from listeners and so-called webcasters.

In addition to small broadcasters, who are the majority, there are also "big webcasters", for which broadcasting is not a hobby, but who focus on achieving measurable gains. In terms of economic factors, such activities already have had a significant impact on the conduct of business. This will be given particular attention in the remainder of this chapter.

An important economic factor is conditions for the development of this medium, which does not require a license. This offsets barriers for new broadcasting media services to enter the market on the network. In addition, this significantly reduces the cost of starting a business. Also, traditional

---

[45] http://pl.wikipedia.org/wiki/Cenzura_Internetu_w_Chi%C5%84skiej_Republice_Ludowej

broadcasters who intend to start broadcasting on the Internet do not have to apply for a license.

Another factor that greatly facilitates entry into the market is the lack of a need for expertise such as programming or having your own server. In this way, the cost of starting up this type of business is small, and the barriers to entry are virtually nonexistent. This is due to the fact that on the market there are a large numbers of companies that provide webcasting services, thereby people who have no expertise, or do not have their own equipment, to broadcast. Service providers also offer other services, and often are open to negotiations, for example, in regulating the fees associated with copyrights. They also offer professional advice on issues that are closely related to the funding model. Examples of such companies are AbRadio Czech, British and Swedish Astra Nordic Web Radio. With these companies, it is easier to enter the market with a large number of online radio operators who receive well-planned and professional commercial offers.

It also comes to our attention that getting a large mass audience is technologically determined. If the broadcaster broadcasts from a single computer (often the small broadcaster is the founder), then the private computer radio signal can be received by a maximum of a dozen or less listeners at the same time. In this case, the chance of obtaining an income, even with the advertising, is diminished.

Each online radio station is trying to increase its number of listeners. The difference between traditional and online radio in this aspect is a very significant one. In a traditional radio station, listener expansion is usually a derivative of programming and relevant marketing campaigns. When it comes to Internet radio, there is also the problem of investment in infrastructure.

On the other hand, the advertiser a selected group of sophisticated recipients during working hours. When the hours of work come to an end, however, Internet radio has a significant lower number of listeners compared to the traditional radio.

An interesting issue is no doubt the financial model used in today's Internet radio. In a traditional radio station, funding is primarily derived from advertising. Due to the small Internet radio audience, this model will not be as effective. It is worth also noting that advertising on the Internet, despite its dynamic development, still has a very limited percentage of profit compared to expenditures on other media of communication. In addition, because of the potential of radio provided for an international audience, it is extremely important for advertisers who do not have time to deal with a specially chosen group of customers. For the same reason, advertising of certain products on the Internet is pointless because of the very nature of the Internet as a global network. Only in the case of web radio with a very narrow profile will advertisers be willing to offer their products, because it will be easier to hit their targeted clients.

Other sources of funding for Internet radio include: e-commerce, subscriptions, membership fees, sales of producing content, or Pay-per-use services. Service Pay-Per-Use is one of the innovative technologies of automated measurement. As part of this technology:

with Interdisciplinary factored system for automatic content recommendation.

> ***"(...) The fees collected from the users depending on the used computing power, Internet bandwidth and disk space."*** [46]

Another form of obtaining additional revenue for Internet radio is to work with existing music stores such as SmoothJazz.com[47] or Amazon.com[48]. In the process of issuing a piece of music on the radio, you have to place a link to a music store where you can purchase a CD containing the band to which you are listening at the moment. An alternative way of financing Internet radio is through subscriptions[49]. If you are paying a subscription, you can gain access to all broadcast programs, as well as in many cases additional services provided in the package. Nevertheless, this way of financing Internet radio is not prevalent. In addition, it is not likely that it can become widespread due to the very nature of the Internet as a medium for free access, and due to increasing competition from other radio stations that offer their programs for free. Another interesting alternative is the compulsory license fee paid by voluntary contributions from members. Thanks to them, in many cases, students receive additional discounts at Internet music stores. Of course, in contrast to the radio license fee payment, this is not a condition for access to the broadcasted content[50]. The use of this type of funding strategy also helps to build relationships in the community gathered around the radio. A relationship between radio and its listeners based on a sense of community is being created.

---

[46] "Cloud computing, or business in the clouds", the information available on the website: http://nowetechnologie.comarch.com/trendy-i-strategie/cloud-computing-software-as-a-service/

[47] SmoothJazz.com, the information available on the website: http://smoothjazz.com/

[48] Amazon.com, dostępne na stronie internetowej: http://www.amazon.com/

[49] Examples of Internet radio stations, which is funded by the license fee is to be paid by students RadioIO, the information available on the website: http://www.radioio.com/

[50] An example of radio in which one of the sources of income are voluntary contributions paid by students is SmoothJazz.com, the information available on the website: http://www.smoothjazz.com/donations/

As previously mentioned, Internet radio stations can also create and sell their copyright contents to others, often to stronger economic players in the network. In this way, it is possible to partially, or even completely, cover the costs of their activities. The risk depends on the quality of the particular products and recordings, and audience interest in the program curriculum, which can be, for example, educational programs or all sorts of live events, etc. Quite rarely used services include Pay-Per-Use. The main reason for this is the fact that members of the community are not willing to pay for content on the Internet when they can get it for free. Moreover, if we consider the idea of radio, this type of payment method would at most be used when making archival programs or relationships with major, live events. Otherwise, we would have to deal with the ordinary form of subscription.

Despite the fact that the least financial resources are spent on advertising campaigns on the Internet, it turns out that the source of funding for Internet radio is actually dominated by income from advertising. According to a study of L. Ha and R. Ganahl, it turns out that more than 95% of broadcasters from South Korea, as well as more than 80% of the leading U.S. webcasters are financed from this source. The next most significant way to obtain e-cash from a broadcast is derived from e-commerce. Other sources of funding are statistically negligible[51].

---

[51] L. Ha, R. Ganahl, "Determinants of Webcasting Business Models: A comparative study of South Korean and United States Leading Webctasters", [in] Proceedings of the 6th World Media Economics Conference, Montréal, Canada, 12-15 May 2004, p. 8.

with Interdisciplinary factored system for automatic content recommendation.

## 2.3. Legal aspects of Internet Radio

Internet radio is a new type of medium, and therefore most countries have not yet introduced relevant legislation. Currently, the existing interpretation of the law was formulated in relation to traditional forms of media. This must be transferred to new media. Guidelines for regulation relate mainly to problems such as:

- Protection of minors;
- Consumer protection;
- Protection of journalistic independence and objectivity;
- Copyright protection.

Therefore, questions arise whether regulations should be developed for electronic media, and especially for Internet radio. Another question is to what extent should it be regulated. Some countries have already taken action in this area. However, there is concern about whether they should rely on skilled transnational bodies to supplement shortcomings in European regulations.

It seems that the current state of Internet technology regulation is necessary to limit potential chaos. Future regulation should take into account changes that may occur due to dynamic technological progress.

Some steps for the introduction of legal standards are already being carried out in the "Directive for Television without Frontiers", as well as the "European Convention on television without borders." Given that the subject of this paper is primarily Internet radio, we need a clear statement that both these initiatives intend to update the law relating only to the television. Many of the demands associated with Internet TV seem to be also motivating issues related to Internet radio. At the moment, unfortunately, webcasting is only partially covered by a typical adjustment of standard media regulations. The likely result of the development of laws may be, and perhaps should be, definition and validation of an entirely new regulatory model for the phenomenon of media access. Currently webcasting service shall be treated as belonging to the

*"(...) Information Society, fills the definition laid down in Directive 98/34/WE of the European Parliament and of the Council of 22 June 1998 establishing a procedure for the provision of information in the field of technical standards and regulations and of rules on Information Society services, as amended by Directive 98/48 /WE ("information Society service, that is to say, any service normally provided for remuneration, at a distance, by electronic means and at the individual request of a recipient of services") (...)"*

Here is a more often quoted definition, with its direct reference to another document:

*"Directive 2000/31/WE of the European Parliament and of the Council of 8 June 2000 on certain legal aspects of information society services, in particular electronic commerce, in the external market (Directive on electronic commerce), which contains provisions relating to the provision of such services."*

It should also be borne in mind that the content transmitted using media like radio will be similar to the countries' criminal code sanctions and provisions. This is due to the fact that there are no other legal solutions that exist for traditional media.

with Interdisciplinary factored system for automatic content recommendation.

The most difficult task for lawyers is a new kind of media placement in an appropriately selected framework. There are different proposals for implementing this task. One solution is based on the idea of creating different standards for different categories of media. This approach can be justified by the fact that some of the objectives pursued by the directive reflect certain standards which are required simply to protect minors, human dignity and the like, which should be applied to all forms of audiovisual content sharing.

In order to assess the legal aspects of radio operation, we have to take into account the following criteria:

- The size of the audience or the public;
- Type of content transmitted;
- Impact of the shape of the viewer;
- Edited content;
- Relationship shaping public opinion of their content;
- The type of media (for example, to a particular user or group of recipients);
- Technical measures used in the transfer (taking into account the distinction between "point to multipoint" and "point to point").

Referring to the assumptions presented above, it can be concluded that Internet radio meets the majority of them.

In the case of electronic media, especially Internet radio, do not forget the purely practical problems associated with the introduction of any regulation of content transmitted over the Internet. In particular, pay attention to these issues, which have cross-border implications. According to the opinion of D. Goldberg:

with Interdisciplinary factored system for automatic content recommendation.

"The international nature of the Internet and other forms of new media means that future monitoring will need to have an international character or rely on self-regulation" [52]

It is still important to pay attention to the effect of Internet radio regulation on copyrights. It is essential to be able to create and use radio in professional applications and to achieve one hundred percent legality, and hence legal certainty. through the implementation of appropriate practices. The rapid development of this communication medium in recent years has resulted in numerous types of licenses that allow us to use music legally on radio websites. In addition, it is worth paying attention to the discussions that take place in the framework of the WIPO (WIPO - World Intellectual Property Organization). The work on the revision of the Rome Convention of 1961 covers:

- Protection of artists;
- Protection of Performers;
- Protection of Producers of Phonograms;
- Protection of radio and television stations.

In particular, changes to the proposed treaty address the legal protection of webcasting coverage.

---

[52] D. Goldberg, T. Prospero, S. Verhulst, Regulating the Changing Media, quoted by T. McGonagle, "Does the Existing Regulatory Framework for Television Apply to the New Media", information is available at: http://www .ivir.nl / publications / McGonagle / report.html

with Interdisciplinary factored system for automatic content recommendation.

# Chapter 3. Internet radio in Poland and around the world

In the case of Internet radio, it is difficult to talk about the selected market, because the range of each sender is global, as is the Internet itself. Multiple services related to Internet Radio, for example, hosting, are of great importance here. Analyzing Internet radio, it can be assumed that the sender usually directs their programs to people of the same country. This is due to the fact that there were many barriers such as language and cultural barriers that limit broadcasters from targeting listeners outside the country of transmission.

## 3.1. Characteristics of selected global markets

In some countries, the following elements contribute to the development of Internet Radio:

- Regulation;
- Development of infrastructure;
- Demographic factors;
- Cultural factors.

Looking at these factors can lead to determining the properties of Internet radio markets in the country.

It was previously mentioned that the development of the market is constrained by the increasing number of individuals who are using the Internet. Therefore, when characterizing the markets in the countries we cannot ignore the driving force of broadband Internet.

with Interdisciplinary factored system for automatic content recommendation.

The evolution of Internet radio is also inextricably linked with the language in which programs are broadcasted. This factor determines the maximum distance from the source, from which the station can be received. The sender who emits very popular programs in languages that are not international are limited to certain countries. In this way, despite having many opportunities, these radios do not use the opportunity of Internet access. The situation is much better with radio stations that provide the use one of the popular languages such as English and German, which are used by many people in many countries. These radios intend to offer their programs to students from around the world, which is why many of them give their broadcasts 24 hours a day, seven days a week. The language factor is not as important in the case where radios broadcast music, but sometimes jingles, commercials and some voice messages cease to be understandable for some listeners. However, if a suitable radio station broadcasts mainly spoken content, then understanding becomes crucial.

Another very important element that affects the development of the market is the issue of settlement. This applies to the amount of royalties that are paid by webcasters, as well as the regulations that apply to the broadcast of their content. Inhibitory factor may be largely due to higher royalties and fees, which can destroy a small radio owner[53].

### 3.1.1. USA

The United States of America is a country that boasts the largest Internet market in the world. In 2002, the country has recorded 165.7 million Internet users [54]. In the United States the Internet radio market is monitored by two companies. One of them is owned by the arbitrators, the other by Edison Media Research. They publish their results in the form of reports, which include research, audience, and listeners profile.

---

[53] W. Kolodziejski, P. Keel, Internet Radio, the report of the National Council of Radio and Television, Department of European Policy and International Relations, No. 14/2005, March 2005, p. 29-31.

In January 2008, these companies conducted a survey on a group of 1,857 people over the age of 12, which shows that 76% of them have home access to broadband Internet. Analysis of studies for certain years shows a huge increase in the popularity of this medium (in 2007 - 69%, in 2006 - 58%, in 2002 - 21%). These data, as you might guess, largely reflects the number of Internet radio listeners. 19% of Americans say that they are listening to at least once the Internet radio, which since 1998 shows a more than threefold increase (in 1998 only 6% of Americans said that they were listening to the radio on-line at any time). Among those listening to the radio, 46% stated that they listened to it on-line as much as 21% over the past 30 days (which equals to about 54 million listeners). These data clearly show that the market for Internet radio is vast and cannot be ignored, especially in terms of marketing and advertising. It is also worth mentioning that the largest number of regular Internet radio listeners are aged 35-44 years[55].

In the U.S. market, there are several thousand Internet radio stations. Unfortunately, the number of radio stations can only be estimated, because the sender does not have to register with the FCC (Federal Communications

---

[54] L. Ha, R. Ganahl, "Determinants of Webcasting Business Models: A comparative study of South Korean and United States Leading Webctasters", [in] Proceedings of the 6th World Media Economics Conference, Montréal, Canada, 12-15 May 2004

[55] The Infinite Dial 2008. Radio's Digital Platforms AM/FM, Online, Satellite, HD Radio and Podcasting, arbiters & Edison Media Research, the information available on the website: http://www.edisonresearch.com/Infinite% 20Dial% 202008_Presentation.pdf

Commission). However, the U.S. market can be considered as the largest for Internet Radio[56]. This market is very large. Further evidence of this is the fact that the largest U.S. companies (e.g. AOL, RealNetwork, YAHOO) have a role in the creation of web radio. Yet another important element that has a big impact on the U.S. market is the popularity of the English language, and thus also of American culture.

The strength of the online market produced several professional organizations that include webcasters (e.g. Webcaster Alliance, IWA and DIMA). They are supposed to represent the interests of the industry's collective management organizations and public institutions, as well as their integration[57].

### 3.1.2. Great Britain

In 2009 the British Isles have reported that more than 79.8% of the population are Internet users, representing 48.7 million people, and as many as 40% of the population have broadband[58].

The evolution of the Internet market will increase public interest in Internet radio issues. In December 2004, a study was conducted. Among the groups of adults examined, up to 15% have listened to the radio via the Internet at least once[59].

---

[56] BM Comaine, E. Smith, Internet Radio: A New Engine for Content Diversity?, Paper 131, 2001, information is available on the website: http://ebusiness.mit.edu/research/papers/131% 20Compaine,% 20Internet% 20Radio.pdf

[57] W. Kolodziejski, P. Keel, Internet Radio, the report of the National Council of Radio and Television, Department of European Policy and International Relations, No. 14/2005, March 2005, pp. 32-33

[58] Information available on the website: http://www.internetworldstats.com/eu/uk.htm

Estimates show that about 250 traditional broadcasters in the UK provide on-line programs in addition to traditional media[60]. The largest of such broadcasters are the BBC and Virgin. The BBC allows us to both listen to radio broadcasts in the past, and to listen to the Internet radio broadcast live. However, as reported by some electronic media monitoring company, Virgin is the radio network that occupies one of the leading places in the world in terms of online audience.

In the UK, there is no organization for webcasters, so they are not affiliated. Most of the commercial broadcasters and traditional radio stations are the members of the CRCA (Commercial Radio Companies Association). Web Broadcasters can only apply for the status associated with the CRCA[61].

### 3.1.3. Sweden

Sweden as well as all of Scandinavia is a highly-developed, teleinformatic country. In Sweden, up to 92.5% (according to Eurostat statistics) population has access to broadband Internet, which equals more than 8 million people. In 2000 only 50% of the population had Internet service. After comparing these data, it can be concluded that Sweden has made great strides in development.

---

[59] W. Kolodziejski, P. Keel, Internet Radio, the report of the National Council of Radio and Television, Department of European Policy and International Relations, No. 14/2005, March 2005, p. 32-33.

[60] Information available on the website: http://www.radiofeeds.co.uk/

[61] W. Kolodziejski, P. Keel, Internet Radio, the report of the National Council of Radio and Television, Department of European Policy and International Relations, No. 14/2005, March 2005, p. 33-34.

with Interdisciplinary factored system for automatic content recommendation.

This level of infrastructure means that Internet radio is quite a popular medium. Estimates show that about 8.8% of the population (1/3 of all who listened to Internet radio at least once, about 2,100,000 people) listen to Internet radio at least once a week[62].

In Sweden, about 60 radio stations broadcast their programs via the Internet, most of whom are traditional broadcasters. The Swedish market has unfortunately developed a local market, mainly due to the language barrier, since Swedish is not very popular. The largest Internet radio stations are suitable for RIX FM and NRJ.

The majority of traditional broadcasters who broadcast their programs on the Internet obtain about 10% of their listeners in this manner.

One of the most popular Internet radio stations in Sweden is Spraydio, which has a high level of interactivity and was founded by Nordic Web Radio. Nordic Web Radio is a company that has contributed to the development of Internet radio and its popularization in Scandinavia.

Some Internet radio stations are members of Radio Associations. The plan is to create an association that will bring webcasters together [63].

---

[62] Information available on the website: http://www.internetworldstats.com/eu/se.htm

[63] W. Kolodziejski, P. Keel, Internet radio, the report of the National Council of Radio Broadcasting and Television Department European Policy and International Relations, No. 14/2005, March 2005, p. 34-35.

with Interdisciplinary factored system for automatic content recommendation.

### 3.1.4. Germany

In Germany there are over 65 million Internet users, 16% of them listen to radio on the Internet (6% once a month, 10% less). Advanced infrastructure and potential population growth make Germany an attractive market for the development of Internet Radio[64]. The same is true for the broadcasters of traditional radio. Radiophony Website has a chance to influence Austria and the German-speaking part of Switzerland. Radio stations in Germany alone have the chance to reach many millions of Internet users. Currently, there are about 2,000 Internet radio stations. Most of them are small. The large share of the market is still with traditional radio broadcasters.

An example of a large radio that transmits only on the Internet is RadioMelodie.net. It was founded in 2002 and is broadcast from 20 locations including Spain, the United States and Germany. The recipients of this radio are mainly Germans (80%), as well as citizens of Austria, Switzerland and the United States. It is quite an interesting radio, because in addition to music listeners, it also offers interviews with celebrities and original programs --- like traditional radio.

In Germany more than 200 Internet radio stations are affiliated in the Radio-ring. It is an organization that primarily provides a forum for exchanging experiences among broadcasters. In addition, it represents the interests of its members and other public institutions[65].

---

[64] Informacje dostępne na stronie internetowej: http://www.internetworldstats.com/eu/de.htm

## 3.2. Internet Radio in Poland

In Poland, about 50% of households have broadband Internet access, while in 2009 30% had Internet, and the EU average is 56%. The rapid pace of growth may surpass this average in 2011. This perspective seems quite realistic because of the number of projects and funds from the EU that flow into Poland. In 2008 Poland was ranked 20th among European Union countries in number of users with access to broadband Internet. A year later it has already taken 15th place, which indicates a huge growth rate[66].

In 2009 a law was passed to "promote the development of telecommunications networks and services," which is to contribute to overcoming legal barriers to building a network to ensure free access to the Internet. It is assumed that:

- "governments can act as an investor in the telecommunications market in both the networking and telecommunications services;
- simplified and shortened procedures for the creation of regional broadband networks, allowing full use of European funds;

---

[65] W. Kolodziejski, P. Keel, Internet radio, the report of the National Council of Radio Broadcasting and Television Department European Policy and International Relations, No. 14/2005, March 2005, p. 35-36.

[66] Internet access in Poland - European Commission the most recent data, information available on the website: http://www.mswia.gov.pl/portal/SZS/497/8048/Internet_szerokopasmowy_w_Polsce__najnowsze_dane_KE.html

with Interdisciplinary factored system for automatic content recommendation.

- manager of that road at every road investment and modernization will have to take care to carry out work on the development of ICT infrastructure"[67]

The fact that, according to a study conducted by the Office of Electronic Communications (UKE), only a fifth of Internet users use mobile access to the Internet also deserves attention. It should also be noted that the trend of this phenomenon is an increase[68].

Although the data are not very favorable, many experts say Poland has great potential in terms of market access. One factor that is the benefit of a large number of citizens of the country, or with Polish roots, living and working abroad (12-15 million). As a result, despite the Polish infrastructure suitable for the Internet radio, it has a great chance to reach the Polish community. It is also important that the Polish population resides in mainly developed countries, such as Canada, United States and Australia.

Most traditional broadcasters in Poland broadcast their programs via the Internet. Examples of such broadcasters are RMF FM, Radio Zet and Polish Radio [69]. The radio station RMF FM, which was not more than two years ago, has the largest audience and set up a web portal at miastomuzyki.pl address.

---

[67] Internet access in Poland - the most recent data the European Commission, the website of the Ministry of Interior and Administration, the information available on the website: http://www.mswia.gov.pl/portal/SZS/497/8048/Internet_szerokopasmowy_w_Polsce__najnowsze_dane_KE.html

[68] UKE: Analiza dostępu do szerokopasmowego Internetu, Lanpolis.pl, informacje dostępne na stronie internetowej: http://www.lanpolis.pl/2009/04/17/uke-analiza-dostepu-do-szerokopasmowego-internetu/

[69] W. Kołodziejski, P. Stępka, Radio internetowe, raport Krajowej Rady Radiofonii i Telewizji, Departament Polityki Europejskiej i Współpracy z Zagranicą, nr 14/2005, marzec 2005, s. 35.

with Interdisciplinary factored system for automatic content recommendation.

Some of the Internet radio stations are also available via this portal, in addition to traditional broadcasting stations . RMF FM has broadcast over the network since 1996[70]. This is because media such as the Internet broadcasters have the potential to reach out to students outside the country and beyond the reach of radio waves.

Gazeta.pl Portal provides quite an interesting opportunity to listen to Radio TOK FM without leaving the portal. The fact that you can use automatic text information and recorded radio programs with audio makes this service stand out among the others. It gives its audience the opportunity to comment on the program, as well as the use of the wealth of the entire site[71].

On the website www.radiopol.com you can find a list of as many as 300 Internet radio stations (both traditional radio broadcast over the Internet, as well as those suitable only for the Internet) that operate in the Polish market. This page is provided by the Canadian company Radiopol.com. You can also find a list of radio stations outside Polish borders (44 stations), for example, Polish Radio Net, which broadcasts from the United States, broadcasts from Radio Seal of Canada, and Radio Polonia Chicago, as well as radio stations from Latvia, Ukraine, Australia, New Zealand and Germany[72].

Research conducted by Millward Brown SMG / KRC Research on behalf of the Radio Committee presents an image of popularity for Internet radio stations and radio frequencies created by classic broadcasting on the Internet.

---

[70] Information available on the website: http://www.miastomuzyki.pl/

[71] Information available on the website: http://www.gazeta.pl/0,0.html

[72] W. Kolodziejski, P. Keel, Internet radio, the report of the National Council of Radio Broadcasting and Television Department European Policy and International Relations, No. 14/2005, March 2005, p. 36-38.

with Interdisciplinary factored system for automatic content recommendation.

This research was conducted between 7 and 18 May 2010 using an online survey. The invitation was sent to 3,100 people, of which 618 responded.

Statistical studies show that more than two thirds of respondents (nearly 70 percent) are between 15 and 34 years of age, most of whom have secondary or higher education. The structure of the sample reflects the population of Internet users. 89 percent of respondents said they listened to the radio on the Internet, of which 29 percent of radio listening on-line occurred very often, and 60% was occasional. Interest in listening to podcasts or previously issued programs is minimal (respectively 11 and 16 percent). Half of the respondents indicated that they are listening to the radio a few times a week for 2 or 3 hours a day, which is a relatively high percentage in relation to other European countries. Like all over the world, the largest audience period in Poland does not include the most socially active days and hours. More than 85 percent of respondents say they listen to the radio only on weekdays. Almost two thirds of respondents declare that the times at which they listen to Internet radio are in the afternoon and evening, while only a third of respondents listen to the radio at work. Statistics show that online radio is complementary to the FM radio, which dominates the first part of the day. A small number (10 percent) use mobile phone capabilities to listen to Internet radio. It can be assumed that this observation is associated with relatively high fees for packet transmission in mobile networks. In addition, it is worth noting that almost two thirds of respondents listen to radio stations that are also available in the ether, and only a third of the study population listens to radio stations only available on-line. It also turns out that the majority of respondents (74 percent) did not use any special applications such as Media Player to listen to the radio, because the website of the sender web application was sufficient. Only 37% of respondents use media players, the most popular being Winamp (14 percent). Other than with Interdisciplinary factored system for automatic content recommendation.

WinAmp, they are using Windows Media Player (8 percent), OpenFM player (4 percent), Foobar (3 percent), Real Player (2 percent), or other players (6 percent). The vast majority of Internet users agree to listen to the ads on the broadcasts, and only 14 percent would be willing to pay a subscription to be able to get rid of the ads. These statistics clearly show that Internet users feel that the Internet is all they need[73].

### 3.2.1. Major broadcasters in Poland

The Internet radio market in Poland is developing very dynamically. It is estimated that currently there are around 400 radio stations available on the Internet. This number does not include the radio stations that are available in the ether. [74]

The oldest and primary players in the market are PolskaStacja.pl and Megastacja.net. PolskaStacja.pl today has 68 radio channels. We also have the ability to create our own radio channels after logging into their website. Users can enter the name of their favorite artist, and then indicate the channels of music will play that artist's songs. The station has a large number of fans, which can be deduced by from an active community forum. [75]

The second largest player in the market is Megastacja.net, which started its operations seven years ago, running a broadcast on the Internet for 20 people

---

[73] Study Auditorium Internet Radio MillwordBrown SMG / KRC on behalf of the Committee Research

Radio, 2010, available on the website: http://badaniaradiowe.pl/aktualnosci/BINAR_raport.pdf

[74] Information available on the website: http://one.xthost.info/emsoft/shc1.htm

[75] Information available on the website: http://www.polskastacja.pl/

on a music channel. Currently it is playing for more than 25,000 members, including twelve channels (Rock, Mystica, Mix Music, CafeClub, Disco-Polo, Hot, Hip-Hop, Polish Music, Mega Old Dance, Dance, Trance and Romantica). Megastacja.net also has an offer, in addition to music, for authors to conduct live broadcasts. The most popular are those that are carried by Channel Rock. Particularly noteworthy is the fact that the radio station in question was the patron and co-organizer of many concerts, and was one of the few among Polish Internet broadcasters of concerts and live channels.[76]

It is worth paying attention to such radio stations as eMuzyka.pl[77] or RadioFTB.net[78], which do not have such a long history, as discussed above, but their dynamic development indicates that Megastacji.net PolskiejStacji.pl may be threatened by them in the near future.

Internet radio is becoming an increasingly important medium worldwide. This trend has also reached the Polish market. There are many regional stations which are important from the point of view of local communities. With the development of these stations, the number of listeners is growing.

In addition, an extremely fast-growing competition forces the search for new ideas to reach new listeners. One such innovation is the cooperation of Internet radio clubs. This initiative can be extremely cost effective for small, local radio stations, as it offers mutual benefits for both the clubs and the radio itself. Internet radio is an interesting material for broadcast, while the club is free for advertising.

---

[76] Information available on the website: http://megastacja.net/, http://www.e-biznes.pl/inf/2008/23127.php

[77] Information available on the website: http://www.emuzyka.pl/

[78] Information available on the website: http://radioftb.net/

with Interdisciplinary factored system for automatic content recommendation.

Examples of such cooperation are: antyRadio cooperating with the club Liverpool from Wrocław, and Ultrastacja transmitting live from the club HARLEM. This cooperation takes place at various levels. For example, the Liverpool club sponsored antyRadio, while in their broadcasts and website promoting the most important musical events was club Liverpool. However, transmission of live events from the club HARLEM allowed Ultrastacja to attract new students who are may not be in the club. On the other hand, HARLEM club gained new customers from the audience through radio.[79]

### 3.2.2. Radio and ZAIKS

ZAIKS is a Polish association that provides a set of services for collective management of copyright and related rights. ZAIKS is derived from the union of writers, composers and performers, and was founded in 1918. STOART is another organization operating in this area. It was formed in 1995, and its functions include the management of related rights to performances of music, word and music, and collection and distribution of royalties enforced in fields such as: recording, reproduction and re-broadcasting. On the initiative of journalists, music labels and musicians themselves, the association was established in 1991 to provide legal representation for producers and fight piracy. It also is responsible for granting gold records for best-selling publications in Poland. There is also the Association of Performing Artists Music Songs and Music and verbal (SAWP), which is a sister organization to STOART. However, the best-known organization for legal issues in Poland is ZAIKS.

---

[79] Information available on the website:
http://www.radiopol.com/index.php?&pid=334&a_gid=1&a_id=21

with Interdisciplinary factored system for automatic content recommendation.

Under Polish law, the use of Internet radio seems easy, because in official documents it does not require a license, regardless of whether it provides commercial or non-commercial functions. It must be remembered, however, that the franchise is something totally different than the license, because the license is permission to broadcast in a general sense. However, broadcasters need a license to distribute specific tracks. While there are no financial sanctions, imprisonment for two years is possible. Whether the subject of licensing is a Polish or foreign contractor, the authority to which they report is just a ZAIKS. This is why they must submit a written request, after a consideration of the agreement is signed. You cannot do this without incurring any costs. They are not only dependent on the popularity of songs, but also several other factors.

It is also necessary to pay income tax on the radio, depending on the amount of emitted music. The rates range from 1-7% of their income. For example, if the season schedule consists of 10% music, radio station has to pay 1% of income. If the music is 70% and more, they will be charged 7% of revenue. Private, non-commercial broadcasters, who are involved in this as a hobby even need to pay certain fees. In this case, however, the amount depends on the number of listeners. The greater the number of simultaneous connections to the radio, the greater charge is., For example, 50 students cost on the order of 100 PLN per month, while 10,000 students cost 4,000 PLN.

Internet radio can be free only when programs are only provided to immediate family and friends. But it seems problematic to classify, for example, university radio. On the one hand, it can be non-commercial. On the other hand, it brings intangible income; it might be advertising the institution, if it will be made available to third parties.

with Interdisciplinary factored system for automatic content recommendation.

The activities of organizations such as the American RIAA and ZAIKS are still controversial. Many individuals as well as businesses try to fight these organizations and submit a series of complaints. These include interference with freedom of speech and adverse impact on the exchange of information and confidentiality of correspondence. It is also believed that they operate primarily in the interests of narrow groups, while neglecting the majority of the population, as well as ignoring the interests of the artists themselves, for their own benefit. In addition, ZAIKS is accused of incompetence and anti-technological development, contributing to the decline in the quality of commercial music production by negative selection and administration of penalties and damages disproportionate to blame.

It is hard to say which side is right, but the rules that ZAIKS imposes in our country are often ignored or circumvented. We can cite an example of a pizzeria in Lodz, who missed ZAIKS sending out the message "Please do not listen to the radio, it is intended only for employees." Although such a move seems absurd, the matter has not been decided in court for four years. In the case of Internet radio, a vast and growing number of radio stations, thanks to the anonymity afforded by the Internet, explicitly and openly encourages broadcasters to ignore the law and promotes illegal radio broadcasting. In addition, many people are not aware of the legal requirements and do this unconsciously. Certainly, the rigid structure and internal problems of ZAIKS have impacted this practice. ZAIKS, according to a report by Gazeta Wyborcza, has PLN 300 million tax arrears. The legal way to bypass ZAIKS is issuing tracks to non-members. or private contractors providing their works. One of these sites is jamendo.com, where a song with a commercial license for unlimited application costs about 70 euros. Of course, it is necessary to maintain appropriate documentation and communication, for its annual ZAIKS assessment. You should also pay attention to the definition of private radio, intended only for family and friends. The ambiguity of the word "friend" makes it very convenient legally. In a sense, each student is a college's friend, for example, as well as the dean's office staff. It is also possible to issue songs whose copyrights have expired. Under Polish law, the copyright to the song contractor expires 70 years after his death. Therefore, classical music radio stations can operate virtually for free and legally. An interesting solution is the use of works covered by the CC (Creative Commons). CC is a non-profit organization whose mission is to reach a reasonable compromise between full copyright and the unfettered use of works of others. In the face of increasingly restrictive, default rules of copyright law, CC is gaining more and more popularity. A CC license, in theory, should give sufficient flexibility to Internet radio stations, but in practice ZAIKS could often charge the users of the works under this license. All of these legal nuances and limitations meant that it became a popular to establish servers in countries that are not subject to the above-mentioned organizations. This method, although debatable, at the moment is the most appropriate from a legal point of view.

## 3.3. Overview of popular commercial products

You can broadcast your own radio programs from a stream server. First, however, you need to upload your own audio files via the Internet from your computer to the appropriate server. Then, post a link or embed player code on your own website. Among other features, the Stream24.pl portal provides opportunities for previously prepared broadcast to users around the world. You could say that stream24 is the biggest online "frequency" radio. The company works with the world's largest corporations, providing their services with their copyright standards. Among other things in its portfolio, it has Nullsoft SHOUTcase servers, or servers Icecast Xiph Windows Media and Flash Media. Internet world is not enough.

Stream24 customers are primarily both small and big companies, big publishers, and web content from around the world, including the Polish market. Also, many hobbyists decide to use the services of stream24 due to the relatively favorable cost and ease of use. Prices for streaming servers start at 20 euros for 50 students, and offer big discounts for the most popular stations. The company offers its clients professional technical support and a wide range of plugins.[80]

A good product is ABradio, which specializes in not only sound, but images on the Internet. At the moment, it is the leader in the Polish market. ABradio also has a network of ground-based radios working in conjunction with its network operations. On the radio side, you can find not only popular music genres, but also niches. Starting its operations in 2000, the company had only 6 stations online. Today it has over one hundred. The majority are created by professional broadcasters, ABradio employees, for business customers. It is also a significant advertising platform in Western Europe. Its services are used by various companies, such as shopping centers, at prices starting at 99zł for individuals and 700zł for companies. Radio from a very simple operation also offers its own music database. With it, the end user of the service does not have

---

[80] http://www.stream24.pl

  http://stream24.com

to worry about copyright and payment to ZAIKS. ABradio is constantly updating its database of songs. [81]

We cannot fail to mention the Live365.com Internet radio network . At this website anyone can set up a radio in a few moments, with prices starting at $6 for a private server, and $199 for commercial. The manufacturer also provides a convenient desktop application to enable a more manageable schedule. Another important element of its streaming services is full compatibility with popular players developed by RealOne, Windows Media, iTunes and Winamp, so that students can be almost one hundred percent Internet users. Currently, Live365 hosting provides more than 7,000 radio stations that carry more than 260 different music genres.[82]

An interesting addition to services running on the network is SAM Broadcaster. This application designed for PCs is used for automation of Internet radio. It has been created specifically for easy management of such servers like SHOUTcast, Windows Media, Live365, and IceCast StreamCast, and is also ideal for managing network services, which provide, inter alia, Live365, and Winamp. The application can automatically generate a programming schedule, which eliminates gaps and creates program transitions. It also has a function to play songs on request for listeners. At the moment, it supports more than 8,000 radio stations on Internet.[83]

---

[81] http://my.abradio.pl/?texty=dlaczego-my

[82] http://www.live365.com/index.live

[83] http://www.spacialaudio.com/?page=sam-broadcaster-awards

with Interdisciplinary factored system for automatic content recommendation.

Winamp[84], once a simple audio player, now has gained great popularity. Today it has become a feature-rich program for advanced tasks. As already mentioned, the creator of Winamp has also developed a standard Nullsoft SHOUTcast, linked with the application. On the Shoutcast.com site, Internet users can register their Internet radio. With this capability, almost 700,000 people have created nearly 45,000 free radio stations. Its channels include native commercial radio broadcasting, such as RMF FM and Radio Zet. SHOUTcast services are compatible with running in the cloud (Cloud Computing), while the Winamp program allows us to easily and effectively manage the outlined program. It does not have advanced features such as competitive pay programs, but it is completely free and available for virtually all system platforms, including mobile. There is also a free software called SHOUTcast Broadcaster Tools that work with the standard DNAS (Distributed Network Audio Server), allowing us to create our own server with independent services in the cloud. The server can be installed on Windows, Mac OS X or the Linux.[85]

Microsoft has very similar support for Internet radio. It uses technology developed by Windows Media. Like Nullsoft, it offers its customers a media player, network services, which via windowsmedia.com can search for available stations in our Internet area, and a server. The streaming media server is a standalone addition to the Windows Server family of operating systems . Windows Media Services (WMS)is a component of the Microsoft operating system designed for servers. Once installed, it allows easy sharing of streaming media, not only in local area networks, but also in the Internet. Currently, multimedia can be sent only in Windows Media and MP3 formats. Previously described products cannot deliver media WMS both on-demand and live. Furthermore, WMS can also act as a cache server for multimedia streaming media recordings, to authenticate users and give them access rights. An additional benefit is hassle-free access to most Internet users, because a large number of them use Windows. It is also worth to mentioning the accessibility SDK, which allows users to write their own applications using WMS. The disadvantages certainly include more work to build the radio the other products. Capabilities of the WMS server depend largely on the version of Windows. The most basic part of Windows Server 2003 Standard, and the most complex, have become part of the Windows Server 2008 Enterprise Edition or Datacenter[86].

---

[84] http://www.winamp.com

[85] http://www.Shoutcast.com/broadcast-tools

The well-known streaming media server Icecast is an open source product developed by the Xiph Foundation. The server, made available under the GNU GPL, allows you to create and manage a professional Internet radio. Icecast can be considered a sort of alternative to the WMS, and it is more powerful than SHOUTcast. Among other things, this server is enough for one process and one port to simultaneously transmit multiple streams, while SHOUTcast technology requires two different ports to function correctly. The server can also authenticate users and refer them to the appropriate channels as well as a backup if the main server does not work or is overloaded. Icecast is also compatible with applications designed to support and receive SHOUTcast technology. It supports both open file formats *.OGG and closed ones. At the moment, there are nearly 8,000 registered servers in this standard.[87]

---

[86] http://www.windowsmedia.com/RadioUI/Home.aspx?culture=pl-pl, http://msdn.microsoft.com/en-us/library/ms867201.aspx, http://www.wolk.waw.pl

[87] http://www.icecast.org/

with Interdisciplinary factored system for automatic content recommendation.

BroadWave Streaming Audio Server is an independent commercial application developed by NCH Software. The server is very simple to install and configure. It has a fairly limited relationship to other products, and allows you to broadcast up to 8 streams from a single computer. An important advantage of this application is that the radio does not require anything other than a web browser, making it much easier to reach.[88]

An alternative to the above-mentioned network services and servers that is very popular are standalone applications, which include ZaraRadio. This program is a complete set of modules needed by users to create their own radio and broadcasting it. It is most commonly used by small businesses, such as bars and restaurants, and the passionate, i.e., all those who do not have the funds needed to open a real radio station, and the desire to appear in ether.[89] As the program is made available for free, it is a good alternative to commercial applications. Just install it to begin broadcasting on the Internet, without having to worry about servers, etc.

Another application is RadioBOSS. Program elements include a fully automated, "set and forget" transmission. The application itself will generate scheduling of database records and music. The DJSoft company took care to make their software more convenient and less burdensome to use on a computer. It also has special ad units and supports the world's major transmission

---

http://pl.wikipedia.org/wiki/Icecast

http://dir.xiph.org/

[88] http://www.nch.com.au/streaming/index.html

[89] http://www.darmoweprogramy.org/903/ZaraRadio

standards. This application is only missing the capability for editing and recording interviews and programs application.[90]

The final product worth mentioning is the extensive Digas system from the German company DAVID, implemented in the Polish Japanese Institute of Information Technology and is used by RMF FM and BBC. This is an application fully designed for professional, commercial radio that combines many important features. In addition to creating a schedule, it has a DataBase manager to manage basic audio files, DigaGrabber for the transfer of music from a CD, the Multitrack Editor, a single track Editor, which is single track audio editor, and the Page Record module, which enables audio recording. For less well-off companies, they created the Digas Lite edition, which is much cheaper, but has limited functionality. [91]

# Chapter 4. Radio, listeners and advertising

Internet radio, with its characteristics and construction of commercial broadcasters provides whole new and often innovative capabilities. The form of advertising is practically independent of and no longer limited to issuing of the recordings. There is even interactive advertising available. Around the world there are several studies aimed at selecting the most relevant product promoted to the most appropriate audiences. The practical part of this paper proposes a system of issuing advertisements based on the classification of types of music buyers. The implementation was supported by a survey, which shows a number of interesting conclusions.

---

[90] http://www.djsoft.net

[91] http://www.profiaudio.com.pl/_pliki/Digas%2520Lite%2520dla%2520radia.doc
  http://www.davidsystems.com/en/

## 4.1. Targeting and selection of traditional radio advertising

On the surface, it would seem that there are many radio stations that simply focus on different genres of music, and once again, broadcast advertising. Nothing could be further from the truth. Virtually every commercial station has a clearly defined target audience and business model. It's not only about the different kinds of music broadcast on the station, which in some way defines it, but also about the very nature of the program, the nature of their subject matter or speakers, which define the radio when targeting young people or older people. Continuously conducted studies show the relationship between time of day and the age or type of audience. The data are subjected to a thorough analysis with regards to all of these factors in order to attract a specific group of radio listeners. They can then serve the appropriate times on the air, not only the best selected program, but also personalized --- as much as possible --- advertising.

### 4.1.1. Personalizing the schedule to the needs of listeners

In general, any average day of John Doe is similar in principle to the next, and this in turn to the previous --- a ringing alarm clock, morning coffee and ... favorite radio. Radio stations are trying to become leaders, to attract the best speakers and the largest number of students not only in the morning news programs, but also throughout the day. This is how leaders increase their impact on the customer as much as possible.

with Interdisciplinary factored system for automatic content recommendation.

We should be aware that the ability to choose from a variety of stations makes radio stations pay more attention to the fact that the leaders have to attract the target group. However, a typical radio can be divided into several segments of advertising throughout the day.

The first block will undoubtedly include morning, when the public wakes up and go to work. This item is the first line in most radio stations. When a student is getting ready, he or she mostly wants to know what's going on world and in their immediate vicinity, so we listen to the radio portion of condensed information. In this band there are very often current affairs programs, which are designed to encourage thinking, awaken and give rhythm to our day. This is a very important part of the whole program, because it can also include a range of information on the day's schedule, which can be an incentive to stay with the radio for a long time. You can also advertise various events or promotions for that day. Depending on the time occupied by that portion of the schedule, it is usually in the range of 6 and 8 AM.

Then comes the development of the morning program. Depending on the nature of the station, journalism, music, or the equivalent of a radio station is broadcast. This band usually extends to midday, and it usually involves a number of independent programs. Students interested in the program will also follow interesting ads with similar content.

Midday is the time when the day is slowly gaining momentum. On the radio broadcasting begins with the first major summary and analysis of what happened in the morning. From that hour, most radio stations start with affirmative approaches; there are many happy songs, light programs, often with a humorous tint, including competitions, and more. Students are encouraged to participate in afternoon shopping or events taking place in their location. The afternoon schedule usually ends around 18–19 hours.

Commercial radio stations specialize in making all kinds of charts and charts. In general, they begin to about 18-19 hours and lasts approximately 2 hours. Make the listener is not able to focus on the major issues and he does not know when he goes to the next part of the radio program, dedicated for nighttime audience.

After an hour, most of the 20 radio stations begin to stabilize and calm down a little emotion throughout the day. There are programs summarizing thematic programs that do not arouse great emotion, the music slows down, the radio becomes a lullaby.

Then all the information about daily consumption of listeners, in conjunction with the analysis of the age of the audience, their city of their origin, or their target group to which the radio data is routed, effectively allows selection of advertising content so that the time of issue is the most accurate and efficient.

### 4.1.2. Classification of the major Polish radio.

According to official data, in Poland there are currently nearly 245 radio stations, which emit signals to a wide range of listeners. In the image of modern radio, broadcasting stations consist of a range of national, local, interregional, and many smaller radio stations, which often operate on the Internet. Of course, the oldest group of radio stations includes the Polish Radio SA, and the very popular One, Two, Three and Four.

with Interdisciplinary factored system for automatic content recommendation.

There are many ways to classify radio stations operating today. You can talk about the band for signal transmission, the frequency, method of operation, and finally broadcast audiences, which characterize its broadcasts. However, the kind of music that represents specific, selected radio stations seems to be the most important factor, because it affects the type of radio listeners.

The first radio stations to examine belong to the Polish Radio group. The Polish Radio channels continue to be a kind of window into the world for many people, through which you can see it often, as many stations forget to mainstream commercials. Private stations are often subjected to so-called preventive censorship. It is based on the fact that some of the songs or deliberate messages are not forwarded to the public. All material before issue must be approved by the institution that is censoring. Although such action is contrary to the Constitution, it is unfortunately common.

Polish Radio plays above all the music created by Polish artists in different periods, from pre-war to contemporary music. Of course, music is not only Polish, because the divergence of broadcasts and tastes make the sound coming into the ether, in principle, satisfy all listeners. The types presented in broadcasting are a cross-section of music --- from jazz and classical music to the sharp sound of guitar. In the case of PR, it is important to be able to find a show that meets the tastes and expectations of the listener. It can be said that the channel is designed for students from 5 to 105 and depends only on the moment the listener decides to turn on the radio.

The second Polish Radio Program, commonly known as Deuce radio station has set the goal of fostering a high-flying culture. So PR has no shortage of important music lovers. Broadcasts can also be heard on the wider culture, which deal with art, literature, film, and theater. Broadcasts of radio plays and other copyrighted works also occupy a lot of space, with projects such as the Polish Radio Theatre. It is obvious that the young students are a rarity here.

Polish Radio channel Three is well-known and has been respected for years. Dynamics and rapid response to current events captures many listeners. On the air you can hear great music from different genres --- rock, jazz, blues, and folk are standard ones of this station. Number Three bears the marks of belonging to some commercial stations, but the level and scope of the presented works is much higher. Number Three shines because its main stream was created by the stars of popular music.

Number Four is the latest channel from the PR group. It was created by the transformation of Radio Bis, and later radio Euro. The station focuses on alternative, new voices that are not in the mainstream music business. So, if someone looking for news and trailers of new and original performances, then number Four certainly meets their needs. The radio station has no shortage of good rock, hip-hop, jazz, blues, or alternative music. It equally attracts young people, the creative, and the curious.

RMF FM is the first radio commercial from Krakow, i.e. Radio Music Facts. Its schedule, which is primarily music varies, but it is mainly pop. RMF FM broadcasts a light note, because, among other things, it is so popular among drivers. For several years, the owners of radio stations developed thematic channels, which include Classic Radio RMF, RMF MAXX. So we can say that every listener can choose for themselves a range of interesting stations in a range of music.

Radio ZET is the second-largest profit radio station, and resembles a RMF. The music it broadcasts can be classified as mainstream pop music, mostly songs from the music charts. Its audience includes 54% men. Less than 20% of student listeners come from large cities, and 40% from rural areas. The age range of the audience is mostly 15-50 years, with a median of 25-35 years. 80% of the radio audience has, at most, secondary or higher education.[92]

The radio station RADIO ESKA Music, one of the largest in the country, offers a program aimed at students aged 15 - 34 years. There is no shortage of new music genres such as pop, dance, R & B and others, which primarily arouse positive emotions. It also has a chat room, Eska Rock, directed at young people. Eski listeners are typically the residents of large cities, of which 40% are students from well-off families, and 30% are managers and white-collar workers. According to a study by SMG / KRC, 70% of Eska radio listeners have their own computer, 90% a modem, mobile phone.[93]

The above-mentioned radio stations form the mainstream, but radio does not stop there. The importance of Internet radio has significantly increased for several years. The vast majority of stations focus on very narrow audiences who listen to the selected genre. They are not, however, strong enough yet to be a threat to the giant radio market. It should be borne in mind the way radio is gaining more and more importance due to the growing need for personalized stations.

## 4.2. Communication between listener and radio

The above-mentioned need for personalization requires a constant exchange of data between the radio and the listener. Communication with students is an important element in the operation of each radio station. The development of modern technologies for interpersonal communication give the average listener more and more ways to respond to the information appearing in radio. Increasingly, it is the audience shoving station broadcasts music they would like to hear, and much more. What are the most popular forms of communication from the radio studios?

---

[92] http://www.wirtualnemedia.pl/artykul/kto-slucha-radia-zet
[93] http://www.wirtualnemedia.pl/artykul/kto-slucha-radia-eska

Surprisingly, despite the very rapid development of technology especially in the field of communication, there are still popular lists. It would seem that the institution of mail goes slowly into oblivion. It turns out, however, that many radio stations still ask the students to write letters asking for certain music. Students write happily, and the lists are frequently cited by leading programs. This applies particularly to those programs that appear every week. This form of communication in this case is extremely convenient and elegant.

The telephone is also still popular. This is because phone calls are now cheap and available and, therefore, many students simply call the radio and share information, observations, take part in competitions, and other similar actions. This is the quickest and surest form of communication that allows you to obtain feedback within a few seconds.

SMS is an extension of telephone communication. This form of contact with the audience is most commonly used during competitions or in the course of the thematic programs. It should also be noted that some radio stations have their own instant messaging, where messages are collected in a format that reaches just SMS.

Today, e-mail, of course, is essential. In fact, every editor has his or her own e-mail address to which you can send a message. This is often an informal channel of communication, because many students share very personal observations and thoughts.

All kinds of blogs are also a good form of communication run by most of the radio stations where you can add a comment, leave a trace. You could say that this is a channel of communication for the diagnosis of moods and emotions associated with specific events published on the site.

The social network Facebook has proven itself well as a communication channel. Many radio stations have their own Facebook page, where users can participate in the radio, and even individual programs. Facebook was named the 2.0 Internet[94]. It is hardly surprising, since any company that does not want to be seen as archaic needs to promote it. There is even a new form of ads running on sites like Facebook --- Social Marketing.

As we can see, communication with radio stations and indicating listener preference depends on the type of relationship we want with our favorite radio station. Some forms of contact can also use Internet stations, although their specificity can go much closer to the user through their own client applications. That's a form closer to the radio by jointly generating events on the basis of statistics and customized ads, which will be discussed later in the work.

### 4.3. Classification of the products genres

*"Without music life would be a mistake."*

---

[94] http://www.pcworld.pl/news/364997/Nadchodzi.nowy.Internet.html

with Interdisciplinary factored system for automatic content recommendation.

Friedrich Nietzsche

Modern times can, without a doubt, be called the era of extreme commercialization. Mass production of goods and services, population growth, and cheap products lead to a situation where the quality gives way to quantity production from the point of view of the plan to descend further savings (cost reduction). The most important thing is to sell. The process of obtaining the reputation of the company (the entrepreneur) by providing consumers with long-term (usually on the local market), high-quality products has given way to "build the brand", which today is defined as aggressive marketing campaigns for the new company who cannot boast of the quality of their goods sold in bulk worldwide. The key, therefore, is no longer the concept (in terms of inventive step) or quality in the classical sense, but the sale.

This state of awareness of economic players pushes them towards a greater involvement in widely understood marketing. Elements of the same marketing categories are not only economic, sociological and technical, but also psychological aspects of the consequences. No wonder no one has even broached the issue of using psychological manipulation techniques by marketers, advertising artists and dealers. In such environment, consumer attention must increase. That is, modern trade will reach for more and more subtle and sophisticated means of "cutting" a potential customer. Not only the facts (i.e. information about reality) are being manipulated, but reality itself. Most people are more or less aware of the fact that visual display, pricing, or even small things such as music seeping from the speakers in the store are designed to induce in them an unconscious (difficult or impossible to rationally justified) desire to purchase. Entire teams of professionals, researchers of human behavior, are working on them.

with Interdisciplinary factored system for automatic content recommendation.

This paper aims to inform the reader that stereotypical thinking about fans of the genre (for example, "country music listeners are people uneducated and ignorant" or "opera fans are very open people ") may or may have a basis in reality. These considerations (of course at a very high level of generality) are a contribution to reflection on the relationship between music and marketing. In the chapter on the positioning of products, some reflections and hypotheses will be made on the possibility of addressing certain products to certain types of music fans. These hypotheses are based on the premises of the characteristics and behavior of fans of certain genres. It is most likely that the market research (empirical) not only did not confirm hypotheses, but rather resent the lie. It is very likely, as it analyzes purely statistical relationships. However, the intention of the author is that this work is the pretext for the recipient to make their own reflections, or even research. The author has presented in the next section this kind of research as polemics.

### 4.3.1. The main musical genres and trends

*"Writing about music is like dancing architecture."*

Laurie Anderson

Basically, there are three main streams of music sometimes mistakenly called species (The major genres of music are about a hundred):

- Art music (art music) - This term refers mainly to classical music. This artistic music also includes some types of jazz, religious music, folk, and little-known music. Music art is often discussed in class in music theory and broadcast on public radio stations.

with Interdisciplinary factored system for automatic content recommendation.

- Popular music (popular music) - This term is used with respect to a substantial (probably most) number of musical styles that have qualified for assimilation by the general public (i.e. the recipient of mass). Economically, this music is "used" in order to make money. It is played on most commercial radio stations and television music stations.
- Traditional music (traditional music) - It is a term somewhat synonymous for folk music. It includes the music passed down from generation to generation and from a specific region.[95]

### 4.3.2. Music and Marketing

*"Today nowhere you can find silence, noticed?"*

Bryan Ferry

The fact that mostly young people (which of course does not mean that all or even most teenagers fond of the work of the team of Richard Caliber 44 or Peja) listen to hip-hop and older people choose a classic piece is probably not a surprise. However, the proposal that fans of opera and pop music are mostly women, and blues is a genre listened exclusively by men, is more interesting. It also turns out that a person with a successful personal life decides to listen to pop classics or old hits, while those defining their personal life as a kind of failure prefer rap and club music. An interesting correlation between music and sex life was also observed. Fans of opera, country, classical, and songs from musicals are mostly monogamists and sexual abstainers. It is no different from political preferences. Voters on the right tend to be fans of opera, country and jazz, which leftist voters select rock and Indian music.

---

[95] http://en.wikipedia.org/wiki/Music_genre

The largest numbers of vegetarians are f of soul music; those preferring meat are mostly the disco regulars. Fans of opera and classical music understand and agree to raising taxes; on the other hand, hip-hop fans and club members are strongly opposed to this idea. The former also read serious newspapers, while at the same time despising the tabloids. Opera lovers are people who regularly pay debts and bills. Fans of rap act contrary. Fans of jazz music, classical and musicals spend much more money on food. It is, of course, not a quantity, but rather a quality issue.

## SUMMARY OF TESTS PERFORMED ON UNIVERSITY OF LEICESTER - A MUSIC FEATURES [96]

- BLUES                Strong self-esteem, creativity, openness, gentleness, peace

- JAZZ                 Strong self-esteem, creativity, openness, peace

- CLASSICAL MUSIC      Strong self-esteem, creativity, openness, the closure, gentleness, peace

- RAP                  A strong sense of self-worth, communication

- OPERATIC MUSIC       Strong self-esteem, creativity, delicacy

---

[96] http://news.bbc.co.uk/2/hi/uk_news/scotland/7598549.stm

- COUNTRY MUSIC         Diligence, communication

- REGGAE               Strong self-esteem, creativity, laziness, communication, tenderness, anxiety

- DANCE MUSIC          Creativity, communication, rudeness

- INDEPENDENT MUSIC    Poor self-esteem, creativity, laziness, indelicacy

- ROCK/HEAVY METAL     Poor self-esteem, creativity, laziness, introvert, rudeness, peace

- POPULAR MUSIC        Strong self-esteem, lack of creativity, diligence, communication, tenderness, anxiety

- SOUL MUSIC[97]       Strong self-esteem, creativity, openness, gentleness, peace.

### 4.3.2.1. Product positioning

*"Joy of Music should never be interrupted by advertising."*

Leonard Bernstein[98]

---

[97] http://www.focus.pl/cywilizacja/zobacz/publikacje/nie-jem-miesa-wiec-lubie-soul/

with Interdisciplinary factored system for automatic content recommendation.

On the basis of the statistical characters representing the genre of music listeners, and additional information may be tempted to provide certain products and services for lovers of the genres (in brackets shows the reasons):

- BLUES     Luxurious, stylish clothes (class and composure behavior), cigars and expensive wine (taste to calm and relax)

- JAZZ     Music clubs offering services such music (in theory it applies to every listener however, the genre of music jazz complementary to the appropriate interior design can lead to above-average income from the sale of such tickets and expensive alcohol inside the premises)

- CLASSICAL MUSIC     Books and other cultural goods (tickets to museums, theater, guided tours, etc.), students classical music tend to be active buyers of cultural and other higher class goods

- RAP     Fast food (indiscrimination and laziness), contraceptives (a large number of partners sexual)

- OPERATIC MUSIC     As in the case of classical music

---

[98] http://www.cloverquotes.com/quote/by/leonard-bernstein/3868-joy-music-should-never-interrupted-commercial

with Interdisciplinary factored system for automatic content recommendation.

- COUNTRY MUSIC — Practical products (e.g., pickup instead of a car sports), clothes (love of freedom and work - for example, in the garden)

- REGGAE — Baggy clothes worn characterized (Expressing psychological traits), RTV (Laziness), alcohol (weakness of character)

- DANCE MUSIC — Mass events - games etc. (sociability, contacts), fast food (complementary with rapid, rhythmic way of life)

- INDEPENDENT MUSIC — Just as reggae music

- ROCK/HEAVY METAL — Concerts, new technologies, strange clothes, products facilitate daily life (laziness)

- POPULAR MUSIC — In fact, every product popular and promoted (Music listeners "dictated" by the companies media like to go with the mainstream)

- SOUL MUSIC — Like jazz, but without luxuries (note soul that lay at the root of the black lyrics slaves)[99]

---

[99] http://news.bbc.co.uk/2/hi/uk_news/scotland/7598549.stm

with Interdisciplinary factored system for automatic content recommendation.

*"They are only two types of artists: those who recognize Pepsi and those who do not recognize."*

Annie Lennox[100]

---

[100] http://www.brainyquote.com/quotes/quotes/a/annielenno390751.html

with Interdisciplinary factored system for automatic content recommendation.

## 4.3.4. Survey

For the purposes of this study, we prepared a special questionnaire, which was available via the website Ankietka.pl. It was taken by 1,058 Internet users, each of whom responded to 14 questions. The goal was to check whether the development of Internet radio is the future, and above all, to verify whether there is any relationship between popular music genre and the sphere products and services in which an individual may be more or less interested. This paper describes a program for radio, which has a selection of ads featured with the music broadcast.

### *4.3.4.1. The results of the survey*

In question No. 1 the vast majority (71%) of the respondents saw the future of radio optimistically. The rest of the respondents completely deleted the traditional radio in favor of the superiority of Internet radio. Traditional ways of receiving radio transmissions have many supporters. It is no wonder, as this form of delivery is convenient. You do not need to perform a series of operations or start your PC in order to connect to the radio --- just turn on the TV and enjoy your favorite radio station.

with Interdisciplinary factored system for automatic content recommendation.

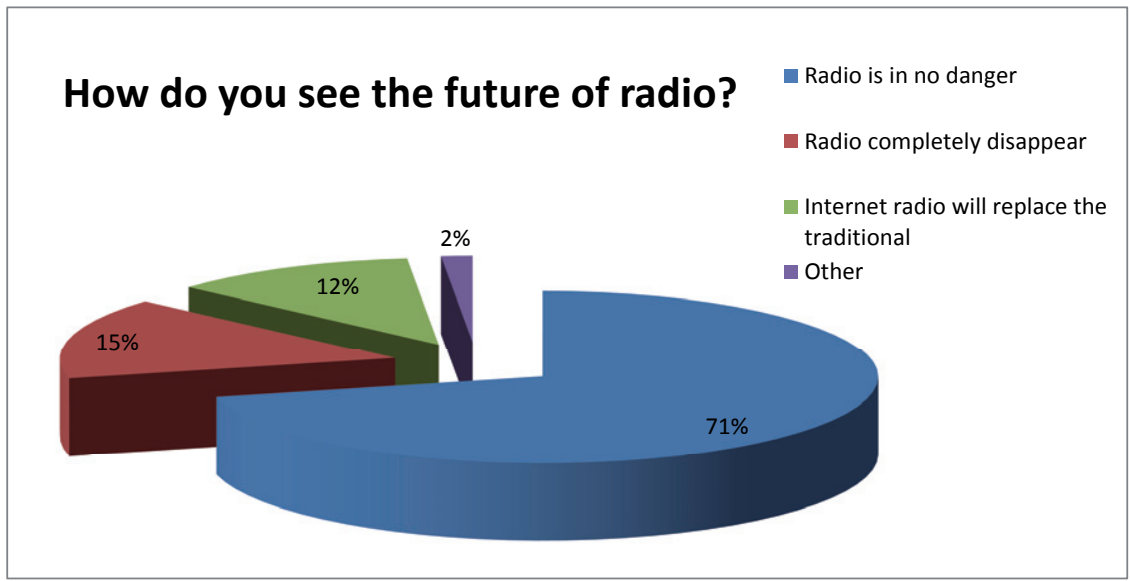

Another question concerned the preferred choice of radio: Internet or through conventional receivers. A significant portion of respondents - 70% opted for a traditional transmission, 24% - Internet-, and only 6% said that it does not like any form of radio.

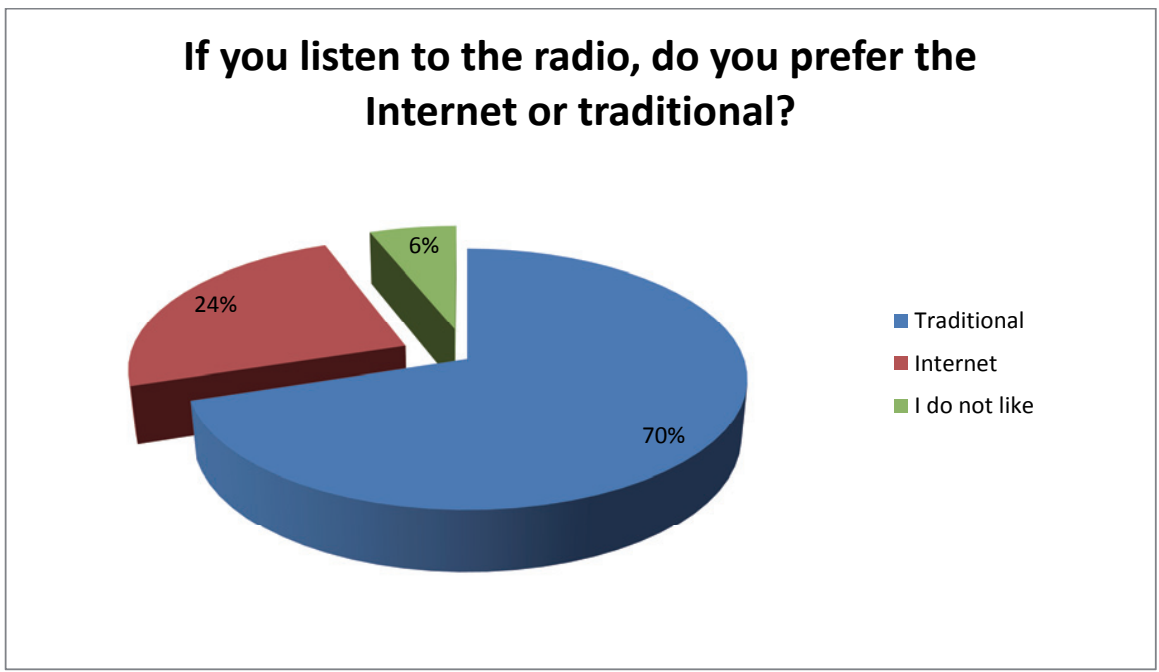

with Interdisciplinary factored system for automatic content recommendation.

The third question was very important from a marketing point of view for advertisers: "Do you think that on the basis of a person's music preference you can personalize the advertising directed to him or her?" As many as 71% of Internet users answered YES, and less than half did not agree with the thesis presented in the form of the question asked. Advertisers should choose their spots (arrangement, selection of the melodic line --- background music) also guided by the musical tastes of their potential customers. In order to know the exact group of people, which has hit the ad, you will need a series of market research studies.

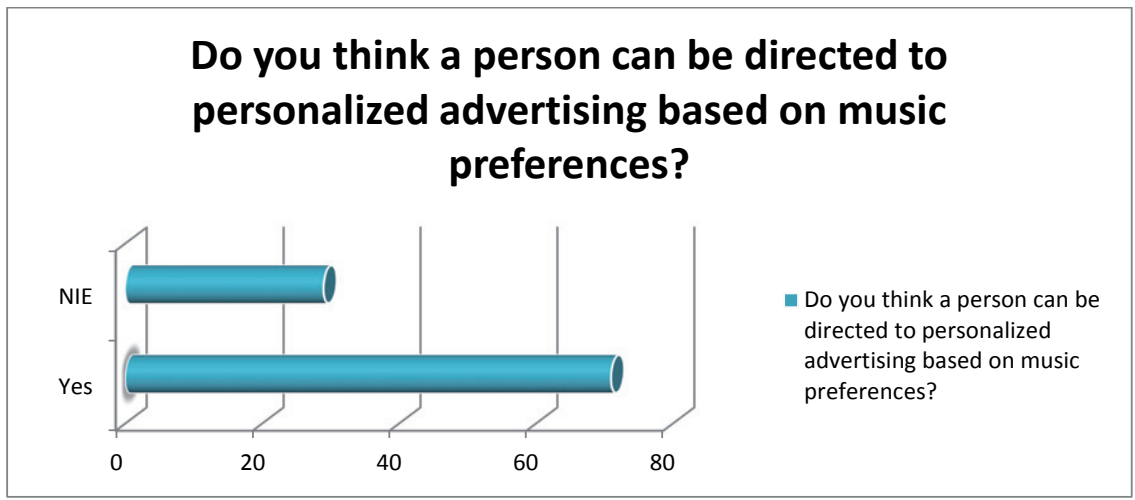

The fourth question regarding the selection of a preferred genre of music, found rock predominant (29.67%), pop insecond place (25.23%) and hip-hop in third place (8.88%). The chart below provides a detailed breakdown of these selected responses. It is surprising that the majority of respondents chose rock. Nowadays, pop music reigns. Survey results showed rock, pop and then hip-hop. This distribution is very similar in most countries, but often you find pop in place first.

with Interdisciplinary factored system for automatic content recommendation.

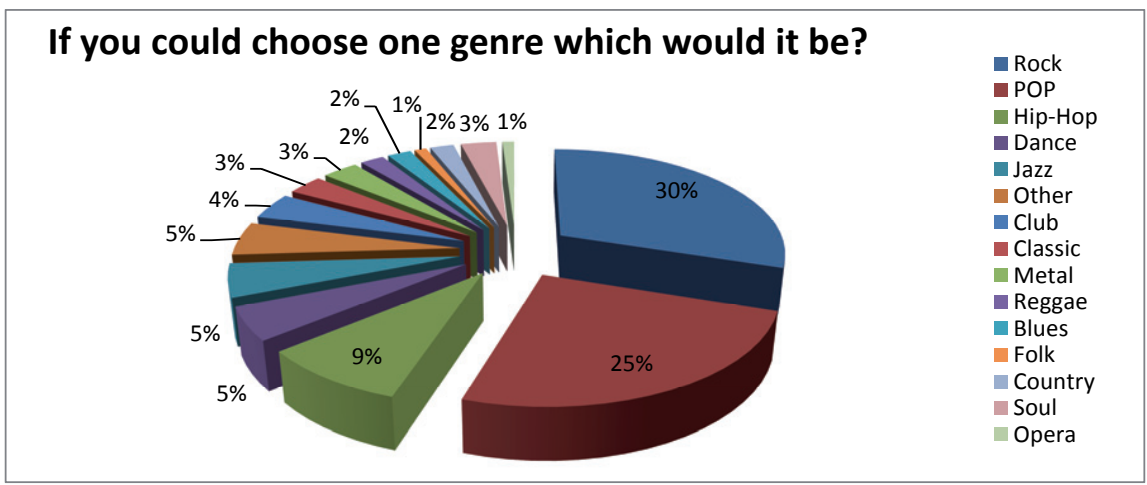

The survey also asked about their second favorite genre of music. In first place was pop (25.33%), followed by rock (24.19%) and in third place club music (6.33%).

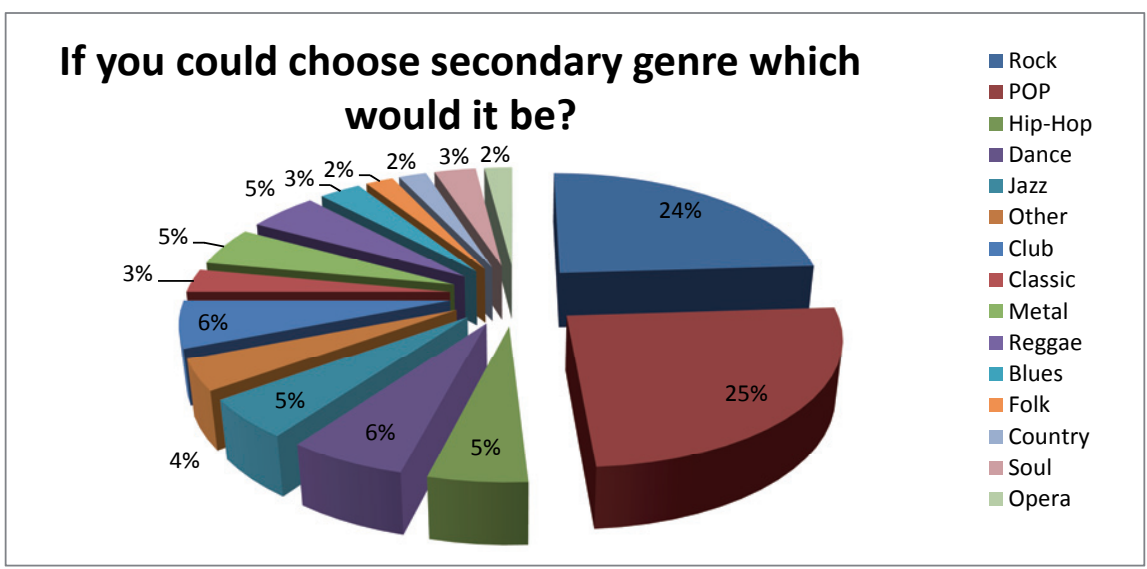

Responding to another question, respondents had to choose which of the products would be of interest. The answers varied, but the preferences were MP3 players (31.28%), MP4 (22.3%) and books (21.83%). As can be easily seen, "the most interesting" defective products are MP3 and MP4 players. Music is so ubiquitous, and the selection of appropriate equipment for this type of advertising could significantly affect the number of people interested in the offer.

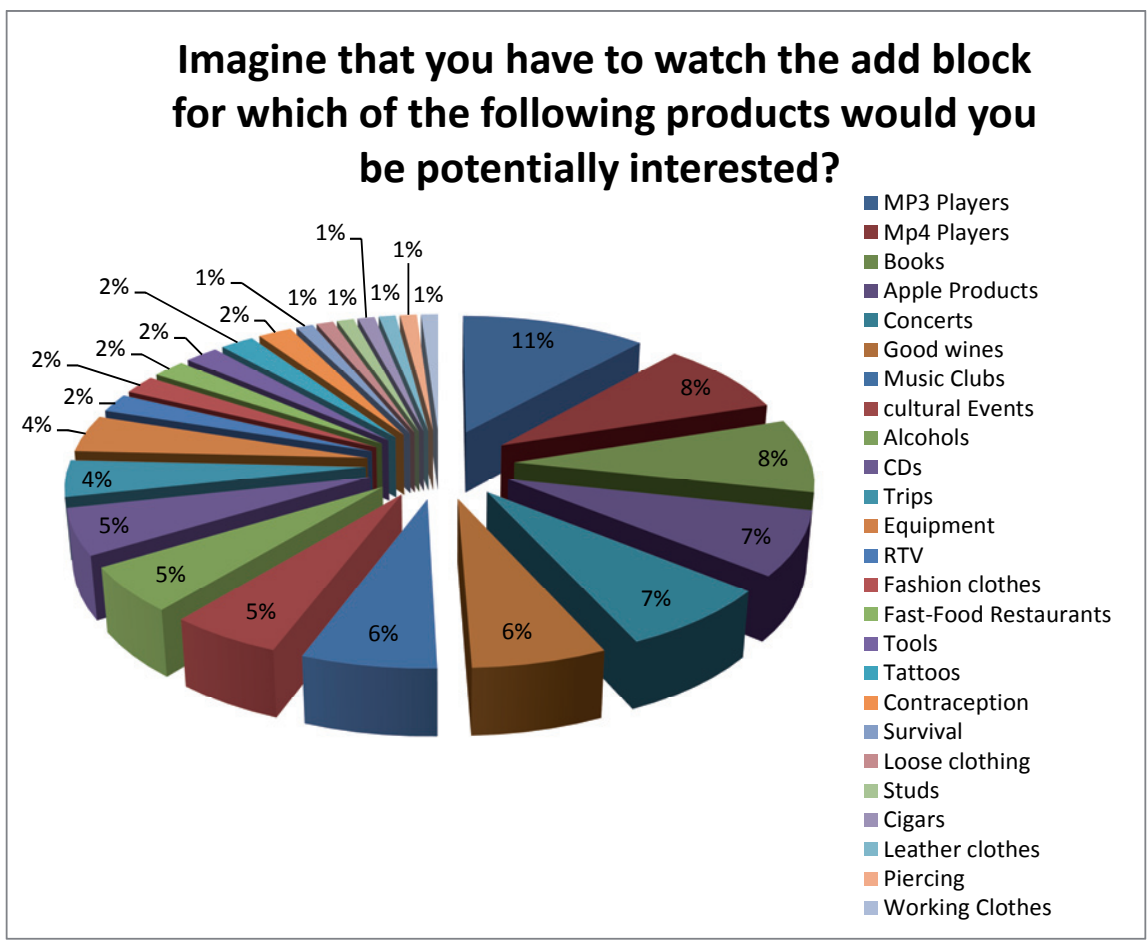

The most interesting observation, however, is that the facts clearly show, bluntly at times, that some products may be directed to the audience of the genre. By creating a cross-table between the selected genre of music and in what products the person is interested, we were able to see many interesting facts shown in Table 3 For example, it appears that the overwhelming majority of Apple fans are rock music listeners. It also turns out that the students prefer popular music and MP3 players, and hip-hoppers prefer MP4. After examining the deeper data, a conclusion can be drawn about one other interesting fact that would be an interesting research topic and seems to confirm the thesis posed by Mrs. Dr. Malgorzata Kopacz in her dissertation.

| | A | B | C | D | E | F | G | H | I | J | K | L | M | N | O | P |
|---|---|---|---|---|---|---|---|---|---|---|---|---|---|---|---|---|
| 1 | | Jazz | POP | Folk | Hip-Hop/RAP | Blues | Classic | Opera | Country | Rock | Club | Reggae | Metal | Dance | Soul | Other |
| 2 | Music Clubs | 18 | 38 | 4 | 15 | 2 | 5 | 0 | 2 | 33 | 20 | 4 | 7 | 18 | 4 | 13 |
| 3 | mp3 Players | 7 | 136 | 2 | 12 | 2 | 3 | 4 | 3 | 119 | 11 | 3 | 9 | 6 | 4 | 10 |
| 4 | mp4 Players | 4 | 23 | 3 | 47 | 2 | 3 | 0 | 1 | 122 | 8 | 2 | 3 | 9 | 2 | 7 |
| 5 | Apple products | 6 | 24 | 5 | 18 | 4 | 4 | 0 | 1 | 120 | 9 | 3 | 8 | 7 | 2 | 6 |
| 6 | Tools | 4 | 13 | 2 | 2 | 0 | 6 | 0 | 2 | 16 | 3 | 2 | 4 | 2 | 1 | 4 |
| 7 | Working Clothes | 3 | 4 | 1 | 2 | 2 | 2 | 0 | 2 | 5 | 2 | 0 | 1 | 1 | 0 | 1 |
| 8 | Fast-food Restaurants | 1 | 16 | 3 | 8 | 1 | 3 | 0 | 2 | 16 | 5 | 2 | 4 | 4 | 1 | 3 |
| 9 | Contraception | 5 | 6 | 2 | 5 | 0 | 5 | 0 | 2 | 15 | 3 | 1 | 1 | 4 | 1 | 9 |
| 10 | Skate Clothes | 5 | 7 | 1 | 4 | 2 | 1 | 0 | 1 | 7 | 2 | 6 | 1 | 4 | 1 | 0 |
| 11 | Wines | 13 | 20 | 3 | 3 | 1 | 9 | 0 | 0 | 101 | 7 | 4 | 3 | 10 | 3 | 10 |
| 12 | Good Cigar | 2 | 4 | 3 | 3 | 1 | 1 | 0 | 1 | 9 | 2 | 2 | 2 | 2 | 0 | 4 |
| 13 | Stylish Clothes | 5 | 24 | 0 | 6 | 2 | 3 | 0 | 1 | 27 | 6 | 2 | 3 | 8 | 4 | 13 |
| 14 | Books | 13 | 29 | 5 | 13 | 6 | 13 | 2 | 0 | 78 | 11 | 6 | 17 | 14 | 3 | 21 |
| 15 | Trips | 6 | 31 | 1 | 5 | 2 | 10 | 2 | 0 | 32 | 7 | 2 | 5 | 12 | 5 | 10 |
| 16 | Cultural Events | 10 | 6 | 3 | 4 | 4 | 7 | 1 | 0 | 54 | 7 | 6 | 11 | 3 | 2 | 21 |
| 17 | Survival | 3 | 4 | 1 | 6 | 3 | 1 | 0 | 0 | 17 | 2 | 4 | 5 | 2 | 0 | 6 |
| 18 | Stud | 2 | 5 | 1 | 0 | 2 | 3 | 0 | 0 | 11 | 1 | 3 | 1 | 8 | 1 | 1 |
| 19 | Tattoo Studio | 2 | 5 | 1 | 5 | 3 | 3 | 0 | 1 | 25 | 0 | 2 | 7 | 4 | 1 | 2 |
| 20 | CD | 10 | 15 | 1 | 11 | 3 | 5 | 1 | 0 | 57 | 5 | 0 | 12 | 7 | 2 | 10 |
| 21 | Piercing | 0 | 1 | 1 | 1 | 1 | 1 | 0 | 0 | 16 | 2 | 0 | 6 | 2 | 0 | 1 |
| 22 | RTV | 5 | 24 | 4 | 9 | 1 | 3 | 0 | 0 | 37 | 10 | 3 | 9 | 8 | 2 | 12 |
| 23 | Alcohols | 4 | 12 | 4 | 44 | 2 | 4 | 0 | 0 | 36 | 6 | 1 | 10 | 5 | 1 | 11 |
| 24 | Leather Clothes | 2 | 1 | 1 | 1 | 1 | 1 | 1 | 1 | 16 | 1 | 0 | 4 | 1 | 1 | 2 |
| 25 | Concerts | 12 | 26 | 4 | 11 | 6 | 6 | 0 | 0 | 72 | 10 | 4 | 20 | 7 | 4 | 20 |
| 26 | Inne | 1 | 3 | 2 | 4 | 0 | 3 | 1 | 0 | 11 | 1 | 1 | 3 | 1 | 0 | 19 |

**Table 3. Table Cross – Favorite musical genre and classification of products.**

The conclusions of the call elections for two favorite genres of music, presented in Table 4, are also noteworthy. The unit generally likes two, three, or even four streams. For example, it turns out that two quite different species are often identified by the persons participating in the survey. Hip-hoppers like the breath with rock music, whose biggest fans are often looking for a springboard to popular music.

| | A | B | C | D | E | F | G | H | I | J | K | L | M | N | O | P | Q |
|---|---|---|---|---|---|---|---|---|---|---|---|---|---|---|---|---|---|
| 1 | | \multicolumn{15}{c}{If you could choose secondary genre which would it be?} | |
| 2 | If you could choose one genre which would it be? | Jazz | POP | Folk | Hip-Hop/RAP | Blues | Classic | Opera | Country | Rock | Club | Reggae | Metal | Dance | Soul | Other | Sum |
| 3 | Jazz | 12 | 6 | 3 | 2 | 5 | 1 | 1 | 2 | 8 | 1 | 0 | 3 | 0 | 5 | 3 | 52 |
| 4 | POP | 3 | 33 | 3 | 11 | 3 | 3 | 3 | 3 | 137 | 16 | 8 | 3 | 31 | 8 | 2 | 267 |
| 5 | Folk | 1 | 1 | 2 | 4 | 0 | 2 | 1 | 3 | 2 | 0 | 0 | 1 | 0 | 0 | 0 | 17 |
| 6 | Hip-Hop/RAP | 3 | 14 | 1 | 12 | 0 | 0 | 1 | 1 | 42 | 10 | 6 | 0 | 2 | 0 | 2 | 94 |
| 7 | Blues | 6 | 0 | 0 | 1 | 5 | 1 | 2 | 1 | 4 | 0 | 1 | 0 | 0 | 0 | 0 | 23 |
| 8 | Classic | 2 | 3 | 1 | 1 | 5 | 6 | 3 | 1 | 5 | 3 | 1 | 0 | 2 | 0 | 3 | 36 |
| 9 | Opera | 0 | 0 | 1 | 1 | 0 | 3 | 1 | 1 | 0 | 0 | 0 | 1 | 0 | 0 | 0 | 8 |
| 10 | Country | 0 | 3 | 0 | 1 | 1 | 0 | 0 | 9 | 1 | 0 | 1 | 0 | 0 | 0 | 0 | 16 |
| 11 | Rock | 19 | 181 | 1 | 7 | 5 | 9 | 1 | 1 | 14 | 6 | 21 | 36 | 3 | 1 | 9 | 314 |
| 12 | Club | 1 | 6 | 1 | 2 | 0 | 2 | 2 | 2 | 5 | 9 | 3 | 0 | 12 | 1 | 0 | 46 |
| 13 | Reggae | 0 | 2 | 0 | 6 | 1 | 0 | 1 | 0 | 4 | 3 | 4 | 2 | 0 | 0 | 1 | 24 |
| 14 | Metal | 1 | 1 | 0 | 1 | 0 | 1 | 0 | 0 | 25 | 0 | 0 | 3 | 1 | 0 | 2 | 35 |
| 15 | Dance | 0 | 13 | 1 | 3 | 3 | 2 | 0 | 0 | 3 | 13 | 3 | 1 | 10 | 4 | 1 | 57 |
| 16 | Soul | 0 | 1 | 0 | 1 | 1 | 0 | 0 | 0 | 1 | 1 | 2 | 0 | 0 | 8 | 0 | 15 |
| 17 | Other | 2 | 4 | 0 | 3 | 2 | 4 | 0 | 0 | 5 | 5 | 1 | 5 | 1 | 1 | 21 | 54 |
| 18 | Sum | 50 | 268 | 16 | 56 | 31 | 34 | 16 | 24 | 256 | 67 | 51 | 55 | 62 | 28 | 44 | 1058 |

**Table 4. Table Cross – Most liked music genre.**

In answers to question 7, the determinate was a gender. Thus, more than half of the respondents were female (65.5%). Women more likely to participate in all type research market, hence the distribution of gender among the respondents.

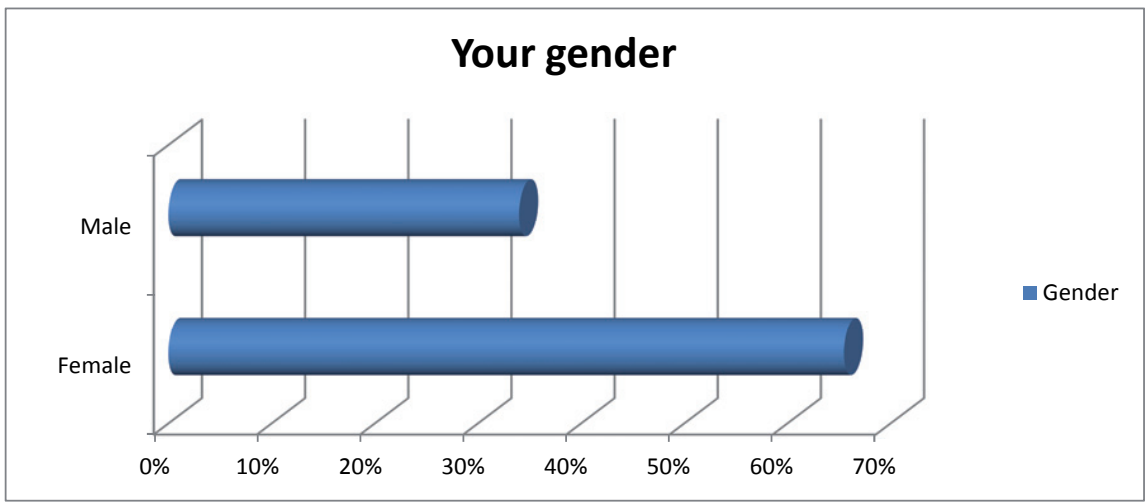

Age was also examined by including the respondent's year of birth, which was very diverse. Most respondents were born between 1980 and 1993. The average age of respondents were in the range from 18 to 25. Young, and very happy to listen to music and radio, they have their own tastes in music, and most radio programs are targeted to them.

To the question: "How many hours a week are you listening to the radio?" Up to 39.9% of all respondents answered from 1 to 5h, and 27.78% from 5 to 10h. Contrary to appearances and contemporary opinions, radio is still a very popular mean of communication. The average respondent listens to the radio a few hours a week.

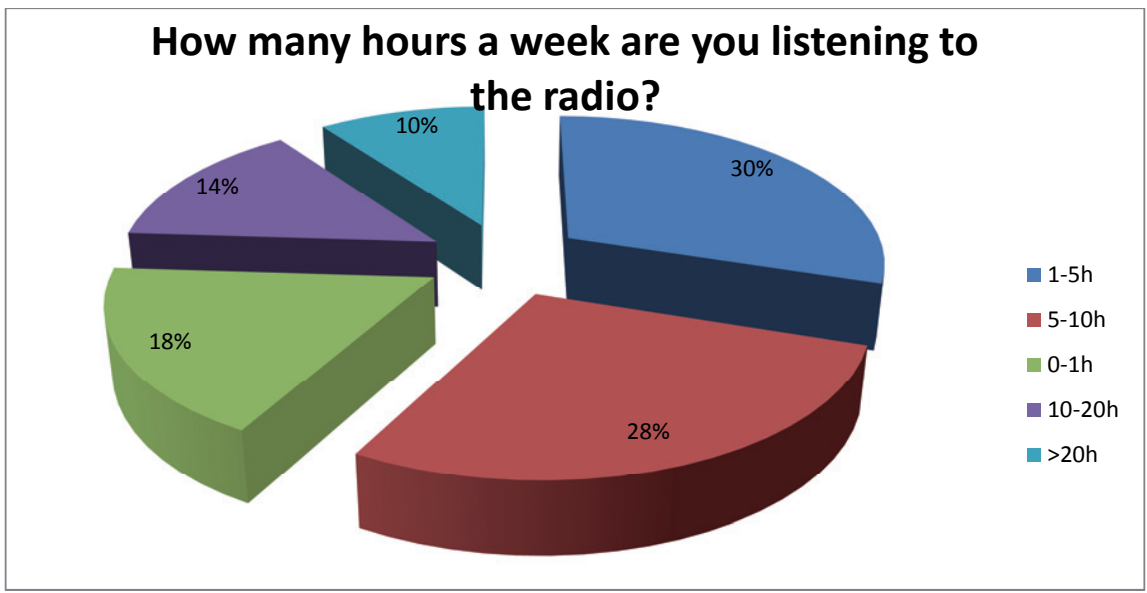

Most respondents listen to the radio when traveling by public transport (39.95%), at home (30.62%), and in the car (17.2%). Traveling is much more enjoyable when accompanied by pleasant music / radio show, hence travelers willingly listen to the radio. Radio can also help while away the daily chores and provide a break to monotony and boredom.

with Interdisciplinary factored system for automatic content recommendation.

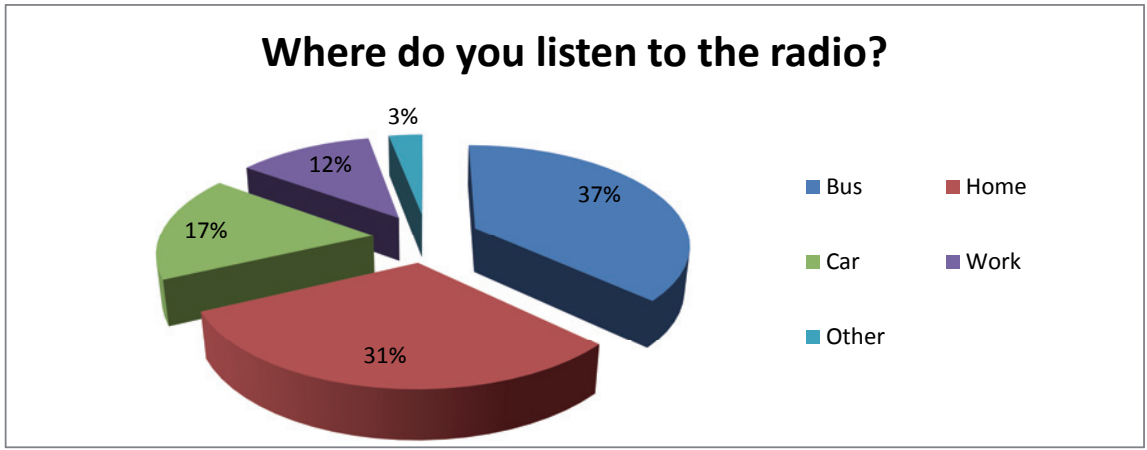

Above all, most of the respondents still use traditional radio (47.44%), but a similar number of respondents use the Internet radio (43.28%).

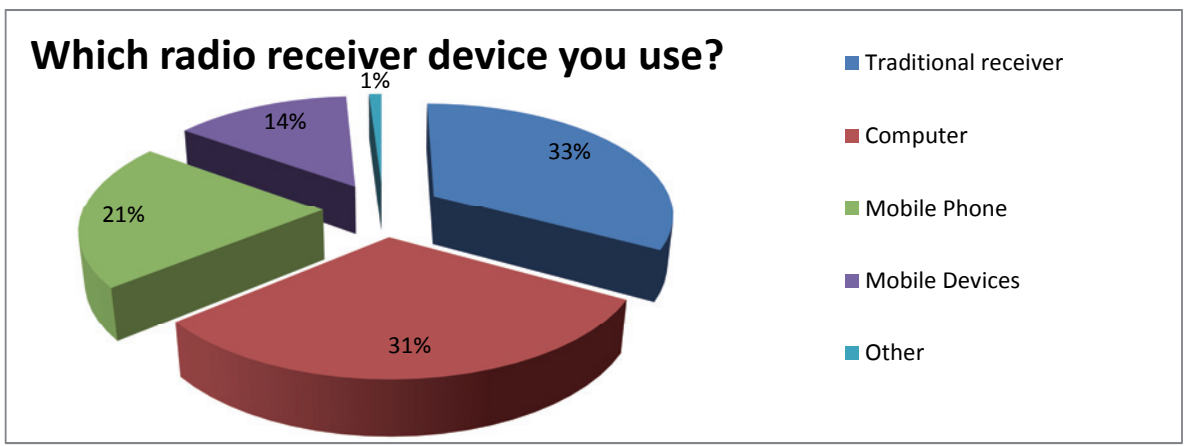

More than half of the respondents (50.61%), in the case of ubiquitous Internet access, would make use of mobile phones as a tool to connect to Internet radio.

At the same time, these same respondents did not want to give up traditional radio broadcasting. Nearly 17% of respondents use standard receivers, using only the phone. Once again, traditional radio reception has been recognized by the respondents. Despite today's technological innovations and devices that allow reception of radio via the Internet, less than 16% of respondents would choose to only receive Internet radio (over the phone).

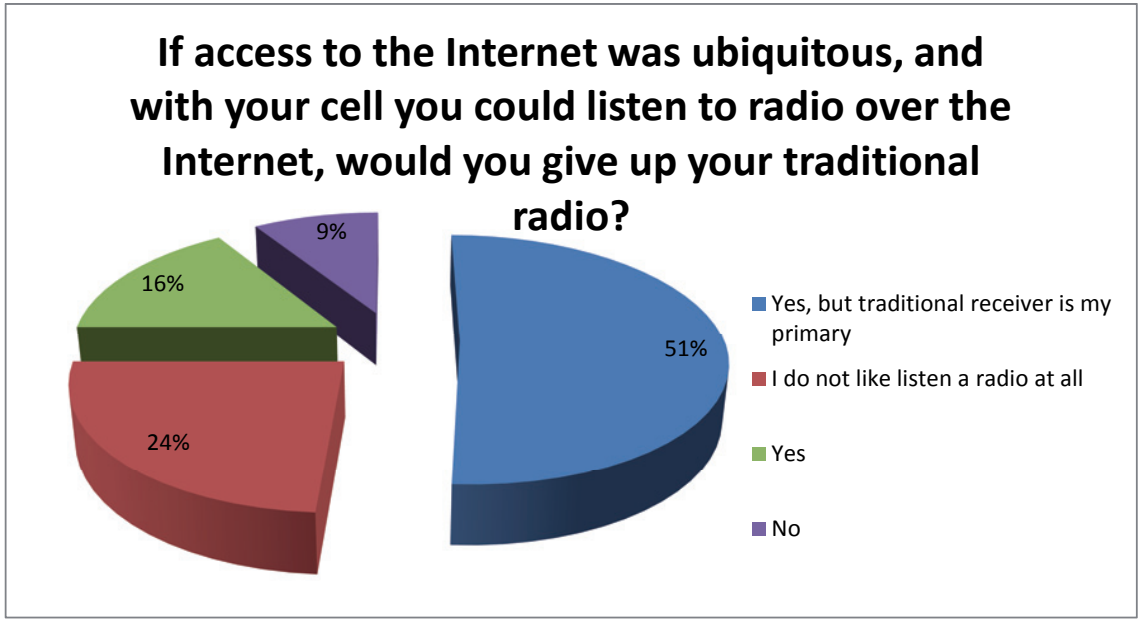

Another question referred to the well-known feature of Facebook and use of its options ("I like it", "do not like"), as a selection of favorite radio programs, music, etc. Over 70% of respondents are willing to use these options. An increasing number of people use social networking sites. They express their opinions there on various topics, comment, and talk. By clicking the "Like" we show others what we like now. Allowing users to sites such as Facebook and vote for your favorite "pieces of music" radio programs encourage them to listen to the selected radio, and it would be free advertising among others.

### *4.3.4.2. Conclusions*

The study of public opinion clearly shows that radio, especially Internet radio, has a bright future. Respondents are willing to use technological innovations, and to take an active part in the life of the community gathered around the radio. This in turn opens up completely new possibilities in the field of marketing and social psychoanalysis. It should also be noted that the majority of the respondents were young people, and it is up to them to the future. Internet Radio as a medium of mass communication can also be a great test in collaboration with popular social networking sites.

## Chapter 5. The technical aspects of radio

Fast, free and legally, you can start broadcasting your own programs and music on your Internet radio. If the radio station gets more and more popular, you will need to purchase the appropriate server. The bottom line is fast, stable, permanent connection to the Internet. It should also have a microphone, at the beginning of the activities, and even the least expensive is enough. You will also need a set of software. Applications that are necessary to run a radio station are free of charge and can be downloaded from the Internet, some of which have already been described in Chapter 3. There are many servers and many of the technologies developed specifically to run Internet radio. Many companies are trying to promote your standards. However, the most popular seems to be the SHOUTcast (developed by Nullsoft), multicast, and podcasting. It is based on the application described in the next section of this work.

The alternative to SHOUTcast would be Icecast. Both systems have their advantages and disadvantages. Software to transmit on the network, regardless of which, is free to select. Icecast Server supports streaming audio in MP3 and OGG Vorbis format, which is an open source product that can be used free of charge under the GNU GPL. SHOUTcast, in turn, only has native support for MP3 codec, which is already licensed as commercial. Furthermore, MP3 does not allow (as opposed to OGG) for audio streaming quality than higher than 48kB / s, which works well in amateur radio stations or home. Nullsoft's solution is a lot easier to set up, so that attracts a larger number of users. SHOUTcast has not been developed for a long time, as opposed to Icecast, which, thanks to continuous updates, may in the future become a very desirable product.

with Interdisciplinary factored system for automatic content recommendation.

Internet radio is not just a music playlist. To operate Internet radio it made sense to also become a website with forums and chats. Traditional streaming should also be connected to podcasting allowing the user to listen to radio away from the computer. It is worth thinking about this, because at the moment only the big commercial radio stations offer this function.

Internet radio will require a good commercial hosting for streaming. The most popular such service providers include inten.pl[101] and hostcast.net.[102]

## 5.1. Transmission technologies

### 5.1.1. UNICAST

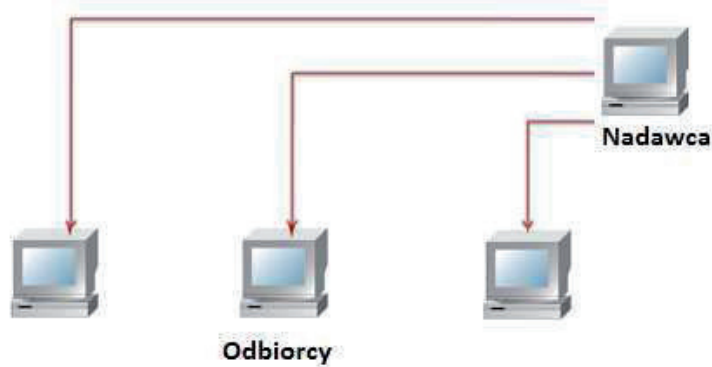

**Figure 9. Unicast**

Unicast is a type of data transmission in which single point packets are sent directly to another single point. There is exactly one sender and one receiver. Ethernet network cards are implemented precisely for this type of transmission. It is based on the protocols: TCP, HTTP, SMTP, FTP, ARP. Unicast requires a separate connection from the server to every customer. This can very quickly consume most of broadband, if the data is sent to multiple recipients at the same time. Therefore, the Ethernet is used mainly for building local area networks.

---

[101] http://www.pcworld.pl/artykuly/48807/Twoje.wlasne.radio.html

[102] http://www.pcworld.pl/artykuly/48807/Twoje.wlasne.radio.html

with Interdisciplinary factored system for automatic content recommendation.

### 5.1.2. MULTICAST

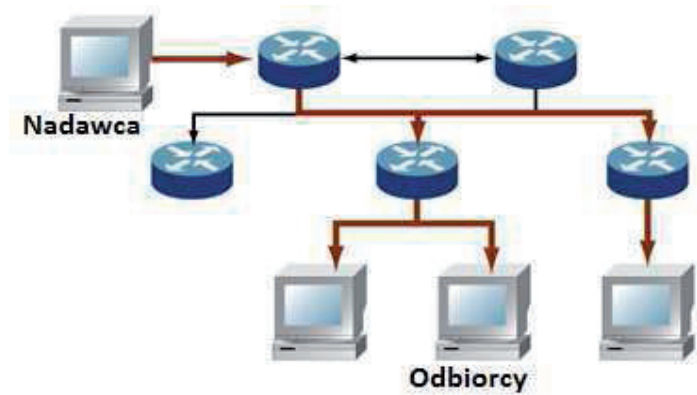

**Figure 10. Multicast**

Multicast is a way to distribute information, where the number of recipients of this information is assumed to be random. Recipients are divided into individual groups that later are available under a single IP address, a multicast group.

Unicast multicast differs from already very essence of the action and its effectiveness. For transmission to *n* listeners in the network transmits the data to a node once in unicast method *n* Times to n clients. In other words, using multiple multicast streaming, sending the same message is avoided. It offers the greatest savings on bandwidth, especially where portions of data are great.

The reference is, of course, not only teleconferences, but video transmission and broadcasting of Internet radio.

Multicast technology depends on a number of network protocols.

These can include:

with Interdisciplinary factored system for automatic content recommendation.

- IGMP **-** Internet Group Management Protocol,
- MLD - Multicast Listener Discovery
- PIM-SM - Protocol Independent Multicast - Sparse Mode
- PIM-DM - Protocol Independent Multicast - Dense Mode
- MRD - Multicast Router Discovery.

Multicast can be defined in a different way: as a way to provide data to the end-station acting as a splitter. A single host acting as a transmitter sends a signal to the other hosts that are receivers. This may be pictured as a one-to-many or as the Source Specific Multicast (SSM) model. The other transmission model, Any Source Multicast (ASM), is similar to a many-to-many where several broadcasters transmit host-to-host with multiple recipients. The SSM model is closely related to the applications broadcaster, which primarily includes radio broadcasting --- television, where it comes from individual transmitters million registered signal to receivers. ASM, in turn, refers to the best example of on-line or telephone conference or video conference. During transmission by unicast, each party must be assigned a specific IP address. In the case of multicast, such a restriction does not exist, because the transmission is directed to the whole group of addresses in the group D, which is the range 224.0.0.0 - 239.255.555.255.

Note, however, that just like multicasting, each technology has its advantages and disadvantages. This is undoubtedly the first technical problem with the network hardware that simply does not support the common IP addressing in class D. Therefore, at the moment multicasting de facto can be used in local area networks. Hope for radical change in this state of affairs is the upcoming replacement of IPv4 to IPv6, which has full support for multicast.

At the moment, globally applicable solutions combine mixed unicast and multicast. The stream is sent from the server combined 1 to 1, often tunneled to the local network hub, whose role can perform as another computer or router, where the stream is sent to the computers on a closed network using multicast. In such a way, the virtual network on the Internet, called Mbone104 (Multicast bone - Bus multicast), which is by using tunneled connections to the global transmission, allows combining both methods of transmission. [103] [104] [105]

### 5.1.3. SHOUTCAST TECHNOLOGY DESCRIPTION

SHOUTcast system is the name mentioned many times before, which was developed by Nullsoft (company that created WinAmp, among other software), and is used for the transmission of sound or images in the form of a data stream, using the HTTP protocol. The system was developed for applications like Winamp. At the moment it is the most common standard, therefore, was used in the preparation of this document.

The SHOUTcast system contains two modules:

1) SHOUTcast DSP, the source module, which is used to transmit the source data stream directly from the applications, reproducing the distribution module. The official version of the module is only available to users in the form of plug-ins for Winamp. There is also software to other players, or separate playback applications created by third parties.

---

[103] http://www.kopnet.pl/index.php?option=com_content&view=article&id=81:protokp-multicast&catid=7:przyksady-pyta&Itemid=25

[104] http://www.hotfix.pl/wideo-poprzez-wi-fi-w-trybie-unicast-n2495.htm

[105] http://www.t-k.pl/p/artykul?i26=1201&o27=1

with Interdisciplinary factored system for automatic content recommendation.

2) SHOUTcast DNAS (Distributed Network Audio called Server) is a distribution module, where the server sends a stream to customers, which can operate in two modes:

- Radio - stream received from the source module is passed to the customers who are connected to the DNAS. In this way, sound and image output from the module source applications, supported for example by the DJ, reaches the audience. It has been implemented in the application presented in this paper.
- on demand (DNAS) - after the client request is broadcast stream, the previously selected by the receiver audio file, or video that is stored on the server disk DNAS.

Stream transmission can operate in this case, through the operation of the HTTP protocol. DNAS server uses port 8000 by default. Distributed based on sound Mpeg 1 Layer 3, commonly known as MP3. It is a combination of MP3 media player, Winamp which was decided by the market success of both products. Encoding MP3 revolutionized the market and made it possible to conveniently transfer audio files. More importantly SHOUTcast is completely free software available for virtually any hardware platform, and the technology itself is very efficient and easy to use.

The sound source to the transmitter can be an MP3 file or audio signal coming from the line input, microphone, or any other source that supports Direct Sound. The latest version of the software allows you to easily switch between devices and audio mixing. Audio streams can be transmitted in bandwidth from with Interdisciplinary factored system for automatic content recommendation.

16kbps to 160kbps, which allows almost any listener to receive an adequate quality radio stream. [106]

Used in SHOUTcast as the upper-layer protocol ISO / OSI, HTTP is used to transfer web pages. The server sends the stream directly to the users (listeners) or to proxy servers that provide similar functions to a full master server with a weak link to install the server as close to the backbone network. Another interesting option is to use proxy servers (SHOUTcast Relay Servers), which are also transmitted in a single stream. Using a proxy server helps minimize load while enabling efficient Internet access for transmission to the audience. Server applications are developed for Windows, Linux based systems and Solaris.

Also interesting is that at the time when the server is running we start to send a stream, send it to a global database of Internet radios (www.shoutcast.com) information on the broadcast stream, providing the name of the radio station, the preferred style of music, the address and port of the server from which it can be taken away, flow rate, maximum and current number of listeners, and other additional information about the station. The server at certain intervals also sends messages about the current playing MP3 track name and interest on the part of Internet radio. All this information is available on the web server and automatically advertise the station.

The listener is required to have a program that plays MP3 files with support for streams. In order to connect to a server on your MP3 player, enter the address and port of the HTTP server. It takes several seconds from the moment you receive a stream until you hear the sound. This is due to the requirement to fill the buffer before playback, which protects against momentary outages, reduced flow rate, etc. At present, virtually every player, even the Web browser itself, can read the audio data stream. [107]

---

[106] http://nss.et.put.poznan.pl/study/projekty/sieci_komputerowe/Shoutcast/strona_www/index.html

### 5.1.4. PODCASTING

Podcasting is publishing audio and video content over the network in a series of episodes with a common theme. The name comes from the iPod - Apple's music player. Episodes are accompanied by a file called a "feed" that allow listeners to subscribe and automatically receive a series of new episodes. Technically speaking, that's the model "host" --- the subscriber is what differentiates podcasting from the usual peaks in web publishing. Some, however, use the term "podcasting" in relation to any distribution of audio or video appearing on the network. [108]

Creating interesting and valuable content is naturally the most labor and resource-intensive part of podcasting. It includes the planning, writing and recording the content, as well as audio and video editing and file compression.

Production requires fusing equipment and software to edit audio segments and / or video. In addition, a podcast creator must generate an RSS feed. The feed is a simple XML file that contains the name of the location in the podcast episodes. It contains the podcast file information, such as the date of publication, titles, and descriptions of each series and episode. This file can be created "by

---

[107] http://nss.et.put.poznan.pl/study/projekty/sieci_komputerowe/Shoutcast/strona_www/jak.htm

[108] http://en.wikipedia.org/wiki/Podcast

http://www.educause.edu/sites/default/files/2007/07/CMU_Podcasting_Jun07.pdf

hand" or using specialized programs.[109] Audio / Video and RSS feed are then loaded on the server. In the first podcast series, which are always open to new subscribers, the podcast creator must notify customers that the podcast is via the RSS feed. Many podcasters provide a link to the podcast on their blogs, websites, or other public places on the Internet. A podcast creator can enrich it with a lot of additional information. A listener (subscriber) to the podcast can subscribe to it using the so-called podcast aggregator (software that checks the actual podcasts and feeds them from time to time). This step is required only once --- when the listener is already added to the podcast, the end of a subscription can be made only through the command.[110]

When a listener adds a new RSS feed, the aggregator fetches all the episodes listed in the current RSS feed. Then at regular intervals, the aggregator checks for updates and downloads all episodes added later. Students can have direct access to the podcast via their computers or through their MP3/video device. Preferring the comfort of mobile devices, most aggregators will automatically download podcast files and synchronize them immediately with devices.[111]

Podcatcher (literally: "podcast catcher") calls the podcast client software program used to download a variety of multimedia files via RSS or XML. Customers' podcast are known primarily from the transfer of music files (mostly MP3) on a portable player. The first podcast client was designed in 2003 (concept began as early as 2000). The following list presents the list of the most

---

[109] http://www.educause.edu/sites/default/files/2007/07/CMU_Podcasting_Jun07.pdf

[110] http://www.educause.edu/sites/default/files/2007/07/CMU_Podcasting_Jun07.pdf

[111] http://www.educause.edu/sites/default/files/2007/07/CMU_Podcasting_Jun07.pdf

with Interdisciplinary factored system for automatic content recommendation.

popular podcast clients according to their operating environments. (There are about a hundred most important programs of this type.)

|  | **Developer** | **Operating System** | **Cost** | **Features** |
|---|---|---|---|---|
| **iTunes** | Apple Inc. | Mac OS X, Windows XP/Vista/7 | Free | Media Player, iPod updater, built-in podcatcher |
| **PodTower** | Elias Puurunen. | Mac OS X, Windows XP/Vista/7 | Free | podcatcher build with the use of Silverlight. |
| **Amarok** |  | Linux KDE | Free | Media Player, iPod updater, built-in Podcatcher |
| **PodSpider** | RapidSoftware Solution | Windows | €14.90, $17.90 | Podcatcher |
| **Odeo** | Noah Glass & Evan Williams | Mac OS X 10.4 | Free | Mac OS X, built-in Podcatcher |
| **Podcast.com** | Podcast.com | Windows/Mac OS X/Linux | Free | online |

with Interdisciplinary factored system for automatic content recommendation.

| | | | | |
|---|---|---|---|---|
| **Juice** | The Juice Team | Windows, Mac OS X 10.3+ | Free, Open Source | Podcatcher, built-in directory |
| **Wizz RSS News Reader** | Wizz Computer | f XP/2000, Mac OS X, Mandrake 10.0 | Free | Podcatcher, RSS |
| **Winamp** | Nullsoft | Windows | Free/$14.95 USD | Media Player, Music Store, can update MP3 players (including iPod) |
| **PodNova** | Active8 | Windows, Max OS X 10.3+, GNU/Linux | Free | Podcatcher, synchronizes with its web-based counterpart at PodNova.com |
| **Zune** | Microsoft | Windows XP/Vista/7 | Free | Media Player, Zune, built-in podcatcher |
| **Podceiver** | Podceiver | Windows Phone 7 | Free/$1.99 USD | Podcatcher, built-in directory, search, favorites |

**Table 5. List of most popular podcast clients**

## 5.2. Compression and the quality of program

Without a doubt, the most popular audio format is MP3, whose full name is MPEG 1/2 Audio Layer 3. Its very good audio compression results are owed to the fact that it is a lossy compression method, which uses the imperfection of the human hearing system. Depending on conditions during the compression, it is possible to lose information that a man would not be able to hear. In other words, it uses a psychoacoustic model that determines what kind of audio information are detected by the human ear and which are not. It also separates information relevant to the human ear from the irrelevant. It is assumed that a person hears the sound of the band to 20kHz. In practice, however, this value is reduced to 16 kHz. In the case of music, the sounds in excess of this amount are different depending on the audibility of the human predisposition. It is significant especially among children and young people, but fades with age. The man best hears in the band from 2 to 4 kHz. Sounds outside this range are less audible to human, so you can save the frequency extremes with much less accuracy. In the case of MP3 compression it uses the phenomenon of masking. It is based on the fact that soft sounds are suppressed by the high volume, and therefore may also be stored in a lower quality. This phenomenon occurs in the auditory system and is the increase of the detection signal masked by the presence of another signal called masker. You can distinguish simultaneous masking, which occurs when the masker occurs immediately after or before the signal. These phenomena are very closely related to the adaptation of the auditory system. This format was developed in 1991 at the Fraunhofer Institute for Integrierte Schaltungen in Germany. It is interesting that at the time of the codec used Suzanne Vega's song titled "Tom's Dinner". She served as developers adjust compression so that the sound of the human voice was the best. Files in this format have the extension *.mp3.

Another popular method of lossy compression is Vorbis. It belongs to the family of OGG codecs, and for this reason it is often used in conjunction with OGG container, bearing the name of the then combined OGG Vorbis, though, and so often it is mistakenly shortened to OGG. It is able to handle the 255-channel audio at 16 bits for frequency of 6 kHz to 48 kHz. Its license is in the public domain, so it is free and distributed under the GNU GPL. Its file extension are *.ogg or *.crew.

The lossy compression algorithm of higher quality Vorbis puts the resulting audio waves with its compatibility with the original. Compressed while sounds are not 100% reflect the high quality uncompressed behavior. This approach results in a high degree of compression at the level of 48 - 128kbps. In the pre-specified parameters, compression quality of the resulting audio file is better than MP3, but comparable to AAC. However, before you use it, you must remember that in nature, it allows only the use of VBR (as below) and that it is more stressful in the decoding than MP3.

Known and often used as WMA (Windows Media Audio)[112], it was developed by Microsoft to directly compete with MP3. The real reason for its creation was the creator of patent ensured that Microsoft could not join the MP3 format to Windows. Although the creator of the format initially claimed that WMA gives better results than MP3 compression, and CD quality is achieved by a rate equal to 64 kbps, independent tests quickly negated these views.

It turns out that below the bit rate of 96kbps, WMA sounds better than MP3, but it is far from CD quality. Above this limit of the competitive codec always was better.

When discussing the most popular codecs and audio compression methods, it is impossible not to mention the values that define compression parameters. Undoubtedly, the concept of a bit rate, which defines the number of bits transmitted per unit of time, is important. It determines how many bits of memory are used to transmit sound. In general, the better the quality of the recording, the higher the bit rate is. Modern compression algorithms use CBR (Constant BitRate) and VBR (Variable). In the case of CBR every second a stream uses exactly the same amount of memory, and in the second case, the value is changing. [113] Sound-rich fragments are encoded with a higher sampling rate. VBR has been developed by XING, not to use large amounts of memory, but to save time, such as when the audio track is less complex. A bit rate of 128Kb/s corresponds to the recording quality of a CD. In the case of MP3 compression, 2.4 MB of data is enough to save five minutes audio at a bit rate 64Kb/s.[114]

---

[112] http://pl.wikipedia.org/wiki/Windows_Media_Audio

The stereo signal is another issue worth mentioning. It can be said that there are three main varieties of signals: mono, stereo, and joint stereo. Mono and stereo are rather obvious and do not require explanation; however, joint stereo may. It is the combination of the other two varieties. The stereo signal is encoded in a separate part of the same signal for two mono channels. They are encoded in mono. However, various elements of the channels are encrypted separately. This is another way to save memory, and therefore well suited to encode the Internet radio audio stream.[115]

---

with Interdisciplinary factored system for automatic content recommendation.

## 5.3. Improving the quality of the recordings

Interference is a common problem that may occur during the recording of interviews or programs (such as the use of a microphone) in Internet radio. For professional transmission, the quality must be improved. It can serve many applications, one of which is CoolEdit PRO. If you divide a WAV file into smaller files, each which will hold a single audio track.

The most common problem is noise. There are, of course, several ways to record. They include median filtering, media based-algorithms such as prediction error (auto-regression method)processing, broadband noise reduction involving the Wiener filtering noise estimation based on signal (filtering assumes no correlation with the signal interference), Kalman filtering applied to Gaussian noise (based on minimizing the error autocovariance) spectral subtraction that can be used if we assume stationarity of the noise, and lastly noise sampling, a number of methods based on the statistical model of the signal and wavelet filtering (allowing selection of a set of wavelets based on the nature of the interference).

The hardest part is the disruption of non-stationary, additive noise when the signal and noise spectra overlap and the noise spectrum varies over time, so that we cannot acquire the sample. It is relatively easy to remove impulse noise (median filtering) --- uniform, stationary noise with a known spectrum (highly efficient spectral subtraction).

Noise, however, is not the only problem you might encounter while recording. Often it has created a gap in the recording, with missing pieces of the mask and a continuous signal. The Burga[116] method allows you to tackle this problem. It involves the use of linear prediction (both forward and backward) on the appropriate samples before and after the damage. In part of the algorithm, prediction coefficients are determined for the relevant passages, and then they are extrapolated by the continuation of the filter response with the same ingress[117]. The Burg algorithm predictor parameters are estimated without calculating the autocorrelation because they are taken from observations. The only assumptions necessary for the proper operation of the algorithm is that the time series must be stationary and the observation interval must be known.[118]

## 5.4. Network Requirements

To provide your own program on the Internet and broadcast music in good quality on the network, you need to have permanent and efficient Internet access at your disposal. A particularly important parameter is the speed of sending data to the Internet, the so-called Upload. Upload a parameter closely related to the maximum number of users who can simultaneously connect to the radio server. This is an important question, because unlimited access may lead to overloading the server, and thus its failure or malfunction. For SHOUTcast you can edit the file sc_serv.ini, in which there are three useful parameters. The first is PortBase that defines which port is sent to the audio stream. The default is port 8000, but we can change it to any other. Another parameter is a password, in which we can define the password for the user and administrators. Another

---

[116] https://ccrma.stanford.edu/~jos/sasp/Linear_Prediction_Methods.html#26412

[117] http://sirius.cs.put.poznan.pl/~inf74839/materials/pdim.pdf

[118] http://pinkaccordions.homelinux.org/staff/tp/prog/tex/examples/math/test-math.pdf

very important parameter is max Users, which controls how many people can be connected to our server. It specifies the number of available user slots.[119] [120]

Therefore, before proceeding to technical issues, a very important stage of preparation is estimating the size of the audience that may potentially be interested in our radio. Only on the basis of this analysis is it possible to be compatible with the planning and implementation of the technical infrastructure.

It should be noted that both sound quality and bandwidth are measured in kilobits per second. In the case of unicast transmissions, each connection will occupy a lot of bandwidth for a program. For example, with a 512kbps link, only two people will be able to receive broadcast-quality 192kbps. Reducing the bit rate in half will allow five people to listen to the broadcast. It is therefore necessary to find a compromise between quality and transmission bandwidth. It is worth noting that some kinds of music with less complex audio structures allow listeners to enjoy full audio quality at a substantially lower bit rate than others.[121]

---

[119] http://krapkowicefm.com/radioint.html

[120] http://www.idg.pl/news/356738/Jak.zalozyc.radio.internetowe.html

[121] http://www.idg.pl/news/356738_1/Jak.zalozyc.radio.internetowe.html.html

# Chapter 6. Implementation of radio

The implementation of Internet radio today is not a trivial task. It is not enough to program all the desired functionality. You must also take care that everyone, or at least the majority of Internet users, can make free use of our radio stations. Therefore, the creation of a fully proprietary solution, which would not be generally compatible with the accepted rules, does not make sense, because the vast majority of users would not be willing to install another application exclusively for a specific radio, one not needed for other broadcasters. So let's look at proven and widely-used technologies, their advantages and disadvantages, and then choose the best of them during implementation of your application server. Selection of technologies and available libraries will be discussed in this chapter.

## 6.1. BASS library

The Un4Seen ("unexpected") BASS audio library is highly regarded in the professional audio community. This is an extension application dealing with very limited-space operating environments running on Windows and Mac OS X. This feature turns out to be very useful if the developer wants to add a particular sound to a program or game. One conclusion emerges from our analysis: a basic library in such a situation does not prove to be sufficient, because the programmer imposes restrictions that make the end result much worse than expected. This situation occurs in particular when a sound must be added to the original (desired) program or game, which is common today.

with Interdisciplinary factored system for automatic content recommendation.

The BASS library is a tool that has been well-tested, so there is no obstacle to using it. In addition to the official technical support, user forums and other sources of information on the use of the product are available to the user. We should raise the issue of licensing. BASS library is free for non-commercial users (i.e., individuals that do not use the library for professional activities or non-profit entities). Software developers and related organizations must pay a specified fee, which, depending on the type (use in commercial activities, time use, access to technical support, etc.) and coverage (number of workstations), will cost from 100 to 2400 euro. Choosing the most options is, of course, the most expensive solution. They are used primarily by corporations for development of commercial software. The free version is a huge convenience for people who want to try to create and add sound in their comfort of home. These operations are fully acceptable and described in the product license. More importantly, the BASS license includes the rights to MP3 and other formats, so users (especially private ones) save money.

Developers who first deal with the BASS library may find that it has a lot of potential. The software comes with detailed instructions and a description of each function. The only downside may be that the included manual has been prepared in English only. There is currently no Polish version. (Of course, this should not be a problem in professional environments or for advanced computer users in these environments, because knowledge of English is common.) Developers should be interested to know that the library BASS supports four programming languages: Delphi, C++, C, and VB. These languages are sufficient for the vast majority of software developers. Before you start working with the software, you need to pay attention to the audio formats supported by the BASS library, mainly basic formats such as WAV, MP3, MP2, MP1 AIFF,

OGG, UMX, and IT. In addition, it is also possible to install add-ons, so that the program will be able to use a more widely adopted format. The biggest advantage of this library is that it is possible to use multiple audio formats simultaneously. In this situation, you receive a code that provides the capability to manipulate sound, including mute, publish, stop, and rewind. The program is good for experimenting with sound and makes it possible to determine its potential.

The library is very functional when it comes to connecting to other codecs. BASS in addition to playing WAV record files, can also add other extensions, depending on the needs and discretion of the developer. The Un4seen Company provides developers a number of extras that make working with sound even more interesting and produces effective results. Users may, however, encounter difficulties when playing Internet radio, as Bass does not support some popular formats, such as: asx, ram, pls and m3u.

The advantages and disadvantages BASS library are discussed here to enable an objective assessment of whether it is suited to your situation, its associated needs, and expectations. Note that we have provided a simple assessment of the product. Programmers working in this environment will not have problems understanding its functions. It is also important that the BASS library supports four essential languages, so virtually anyone will be able to use one of them. The BASS library, created by Un4seen, is ideal for those who want to apply a variety of effects to multiple audio tracks. BASS offers plenty of extras. Another advantage of this product is that it supports 3D sound.

The BASS library also has some disadvantages. The most prominent is a lack of support for some formats, which can cause some problems. In addition, with Interdisciplinary factored system for automatic content recommendation.

sometimes there is a problem with audio recording. According to the views of users, most are on Vista or later, since Microsoft has caused problems for several years in adapting their products to the wide variety of hardware and software. Currently, however, they are working on improving this problem, so you can expect product improvements in the near future.

On Windows BASS requires DirectX 3 or higher in order to fully exploit the opportunities and options provided by the DirectSound and DirectSound3D drivers. For MAC OS X BASS uses CoreAudio for audio calls. Therefore, version 10.3 or higher is recommended. BASS supports both PowerPC and Intel-based architectures. BASS is also available for Win64, WinCE, Linux and iOS platforms, so that applications can be written to move freely between them.

In summary, the BASS library is an ideal product for both professionals and amateurs, for commercial as well as personal applications. A big plus is that anyone interested can work with sound, thanks to the plug-in, at no extra cost. A free version can be downloaded from Un4seen website. Most users develop in Delphi, because it runs efficiently without any hassles. Anyone who has mastered the BASS library can create cross-platform applications, using plug-ins and add-ons to connect in any way, thus achieving excellent results. Despite a few flaws, in summary it can be said that the BASS library is ideal for use in most applications.

with Interdisciplinary factored system for automatic content recommendation.

## 6.2. .NET Technologies

### 6.2.1. C#

.NET technologies were created to meet the three objectives. Above all, they aim to create an environment that allows developers to create fully portable applications, as in the case of the Java virtual machine. The idea is that your application is independent of hardware and system architecture. Microsoft only supports its platforms, including the operating system for mobile devices. .NET is a powerful component library that contains common elements that occur, for example, in a graphical environment. .NET is a de facto merger of COM and ActiveX with ADO for data storage. In principle, it also facilitates the exchange of data between applications in a .NET network. ASP.NET, which is now used mainly for implementation of web applications and web services, supports this objective.

Officially, .NET has been developed exclusively for the Windows family of operating systems. It is quite popular for desktops, but much less important on mobile devices, which are dominated by Java and other similar products. Nevertheless, .NET technology is relatively young, as its development did not begin until the end of 1990s. The first version of the Microsoft .NET Framework was released in 2001, and, although it is still available for free, the fourth version is slowly gaining popularity among large developers.

A .NET environment is required on a computer that is used to run applications. In this manner, it serves as an additional layer of protection for the operating system. On the one hand, it takes over control of main memory and threads. On the other hand, it allows applications to use components called

Windows Forms, and provides a uniform mechanism to ensure access to other system resources, including databases. ADO.NET also contains a set of facilities used for communication between computers, etc. Many compilers are available for programmers. Support is also available for most popular programming languages, so .NET applications can be coded in C#, Visual Basic (VB), C++, F#, J#, Delphi, Perl, or even JavaScript.[124]

The .NET process does not directly compile to machine code, but instead uses a two-step process. In the first phase, a so-called intermediate code, which is the same for all .NET languages, is assembled. Then the JIT type compiler deals with the object code from this phase. The second stage does not necessarily take place immediately after the first, or even on the same machine.

It is noteworthy that the result of compiling C# code as well as other .NET language files are *.exe files, although their structure is nothing like a traditional executable file. In fact, the content is filled with the generated intermediate code, also known as Managed Code, whose name comes from the degree of control that the .NET platform exercises over its execution. Of course, from the point of view of the user, it does not really matter if they have the correct .NET framework version loaded on their system, if the program starts when you double-click on its icon. However, from the point of view of the system, this is a huge difference.[122]

Microsoft's .NET has become popular due to a number of useful features that greatly facilitate the development of applications. It offers a prefabricated infrastructure that allows you to overcome the most common programming problems, but also to do network programming. Many tasks that had to be programmed manually in the past are now simplified by implementing ready-made patterns using wizards. It's not just about the GUI, but also threading design, data exchange with databases, etc. Above all, it provides the ability to easily provision servers on the Internet for different functions and data through the so-called .NET Web Services. Also supports writing client applications using the new method of a Windows Forms application. This can be visualized as finished components. Windows itself is composed of components, and the developer can freely define the relationships between them. A new runtime environment has also introduced, providing very easy access to databases and other media. [123]

---

[122] http://www.besthelp.pl/x_C_I__P_3010132-3010003.html

The .NET foundation has primarily been designed to facilitate the programming of servers, services, and instant messaging, as well as many other applications. Among others, the large company Perfect Soft creates all kinds of applications using Delphi.NET, using the ASP.NET, Ajax.NET, C#, ASP.NET, XML and CSS.

Although the .NET platform is not perfect platform,.NET applications operate quickly. Created services can be debugged using tools that are specially designed for such purposes.

The primary advantages of the.NET technology are standardizing and consolidating many programming languages in one place, delivering an enormous and growing number of libraries for developers to use for production as well as prototypes, and providing great tools for collaborative programming (Visual Studio Team). It would seem that this product is a happy medium between application performance and programmer convenience.

---

[123] http://narzedziownia.softwarestudio.com.pl/Default.aspx?0001820540541

with Interdisciplinary factored system for automatic content recommendation.

IT professionals believe that .NET competes with PHP, etc., because it is primarily ASP. It seems to be the biggest competition for Java, or more precisely J2EE technology, the ONE platform developed by SUN and IBM. However, Microsoft has made a considerable contribution to the development of network services. This is primarily for Web Services and SOAP Protocol.[124]

At Microsoft, .NET is a very important development line. Even the sidetrack system developed Singularity is fully based on it. The successor to Windows XP systems were supposed to work in the.NET environment, where *.exe legacy applications would have access to resources only through the transformation of requests. Over time, it turned out that only a small portion of the system uses .NET.

At the moment we can say that there are three major development .NET platforms. The primary one is the .NET Framework available for free from Microsoft. There are also a number of programs created under the MONO platform developed by Novell with open source licenses, which in the future will be also include DotGNU Portable .NET, emerging from the GNU project.

### 6.2.2. SilverLight

Microsoft Silverlight is a web technology inherently designed to display multimedia content via a web browser. It was initially developed under the code name WPF, the Windows Presentation Foundation. Thanks to its design, it allows you to fully incorporate the JavaScript and XAML languages, and since version 2.0 it is possible to write applications for it in any language supported by the .NET platform. It also supports interpreted languages, including Ruby and Python. Currently, most applications developed for the Windows Phone 7 use it.

---

[124] http://pl.wikipedia.org/wiki/.NET_Framework

Just as in Flash, you can use a mouse or keyboard, as well as display images and sound support. Of course, it provides support for video files, including those in high-resolution FullHD. Normally, it is possible to attach music files in MP3 and WMA formats. It also supports Digital Rights Management (DRM). With the release of Silverlight 3.0 in 2009, the H.264 codec was introduced and hardware acceleration was added for 3D graphics.

Microsoft first developed this product for Windows and mobile platforms. The release of a Silverlight version for Mac OS X was an interesting move. The officially supported browsers are: Internet Explorer, Mozilla Firefox, Safari for Windows, and Safari for Mac OS X, but only on machines based on the X86 architecture. The end of January 2009 saw a Linux implementation called Moon Light 1.0 through the Mono project. Unfortunately, it is compatible only with SilverLight 1.0, so it has a limited capacity. Currently, work is in progress on MoonLight 2.0, which is an alternative to Silverlight 2.0. Unfortunately, there are currently no plans for the most recent Microsoft product.[125] It should be noted that version 5.0 was already released, which is why writing a portable application between operating systems is not as trivial a problem as it might seem due to the high fragmentation of the Silverlight version.

Today's applications are unlikely to resemble websites designed several years ago. Modern technologies and languages such as HTML 5, CSS 3, AJAX and JavaScript libraries such as ASP.NET or LQuerry allow developers to more easily develop, as well as greatly improve their usability and convenience of use.

---

[125] http://pl.wikipedia.org/wiki/Microsoft_Silverlight

with Interdisciplinary factored system for automatic content recommendation.

In addition, in 2001 Rich Internet Applications (RIA) became a popular technology. Among other things, it enables definition of advanced user interfaces, while its dynamic nature eliminates the limitations of earlier technology. Silverlight has become quite a powerful tool to build programs and web applications compatible with RIA, which works not only in web browsers, but also on mobile devices and game consoles.[126]

### 6.2.3. MONO

Mono is a project sponsored and owned by Novell. Its main objective is to develop tools for developers that are compatible with Microsoft .NET and compliant with ECMA standards. The most important part of the project includes a C# compiler and a Common Language Runtime, CL0. Although Microsoft has released its platform for other environments, this license is not compatible with the idea of open source software, which is the main reason for implementing a new solution.

The most recent version of Mono has an API that is fully compatible with .NET 2.0. While it incorporates some elements of .NET version 4.0, it is an obstacle to the development of a universal application. On the other hand, it completely opens up new possibilities for developers, because it works with multiple operating systems, including Unix, Linux, Apple Mac OSX, FreeBDS, NetBSD, Solaris, OpenBSD, PlayStation 3, Wii, and of course Windows iPhoneOS.

Currently, the Mono project provides a source code editor that has MonoDevelop templates and master pages, and predefined elements to control sites. It also provides a built-in debugger, as well as the ability to open projects from Visual Studio, strong compliance with MSBuild, and full support for the Model View Controller (MVC) pattern.

---

[126] http://msdn.microsoft.com/pl-pl/library/gg131024.aspx

### 6.2.4. MoonLight

The essence of this project is to give MoonLight Linux and Mac OS X users access to media published in the Silverlight format.[127] The cooperation established in September 2007 between Microsoft and Novell led to the development of MoonLight 1.0.[128] It is primarily a web browser plug-in for Mozilla Firefox that runs on any Linux system compatible with DX11.

Like the Moonlight project, Mono does not have all the functions of the original. Moonlight 2.0 is fully compatible with Silverlight 2.0, and probably includes functions from Silverlight 4. Among other services, these include bitmap API functions facilities, dialogue files, media queues, as well as base codecs.

In addition, some features of Mono have also been implemented in Moonlight, so developers can easily build applications specifically for Linux users. They can code in C#, Python, JavaScript, and Ruby. [129]

---

[127] http://prportal.pl/2009/02/13/moonlight-%E2%80%9Ew-pelni-blasku%E2%80%9D-na-komputerach-osobistych-z-linuksem/

[128] http://prportal.pl/2009/02/13/moonlight-%E2%80%9Ew-pelni-blasku%E2%80%9D-na-komputerach-osobistych-z-linuksem/

[129] http://di.com.pl/news/29865,0,Moonlight_2_czyli_kolejny_linuksowy_Silverlight.html

with Interdisciplinary factored system for automatic content recommendation.

## 6.3. Additional software

During the production of this document, we also used third-party software. The idea was to build radio speech into the server, which could to some extent replace the human. A successful synthesizer could announce future broadcasts or music, as well as read the contents of the presentation. This project uses the Microsoft Speech API combined with the Polish TTS developed by IVONA Software.

### 6.3.1. Microsoft Speech API (SAPI)

The Speech Application Programming Interface (SAPI) can be called a .NET development as described in Section 6.2. SAPI is an interface that allows developers to write programs that include speech recognition and synthesis. The entire system was developed by Microsoft for its embedded operating systems. The developer can use ready-made solutions from the Internet after downloading and installing the Speech SDK. It enables an easy way to write a TTS application, but unfortunately does not have a standard implementation of the Polish language. Just import the downloaded library, for example, in Visual Studio to be able to use them. They support all languages that can be used in that compiler. The most current version is SAPI 5.4. When creating an application, take care to maintain backward compatibility. SAPI version 5 is supported on Windows 2000, XP, 2003, Vista, and 7. SAPI 4 is compatible with Windows NT 4.0, Windows 98 and ME.[130]

---

[130] http://en.wikipedia.org/wiki/Microsoft_Speech_API

with Interdisciplinary factored system for automatic content recommendation.

### 6.3.2. Voices

The Polish developer of the most popular speech synthesis applications is the IVONA Software Company. This company focuses on the implementation of speech synthesis programs that combine elements of artificial intelligence and human speech, the interface between man and computer. The main product of the company is IVONA TTS, which converts written text to speech, which every year attracts a larger group of supporters. The company has developed a database containing voices in different languages, including. of course, the Polish language. It has many products used in both entertainment and business. Fromm the point of view of this project, the most important are the voices of IVONA version 2. They are designed for use with the previously described SAPI. Once a purchased voice is installed in the system, the programmer can use it. IVONA Software offers 4 Polish voices (2 female and 2 male), 3 British, 2 German, 2 Spanish, 1 Romanian, 8 U.S. (including 1 squirrel, which is styled on the popular American cartoon squirrel "Skippy the Chipmunk" [131] and 2 with a Spanish accent).[132]

## 6.4. Concept and Development

After the creation of a good application, a typical developer constantly strives to improve it, adding more and more functions. They offer not only new capabilities, but also increased personalization for the user. Development of very complex programs in this way causes problems with users. Apple has shown that users prefer a completely different solution. Users need maximum simplicity and 100% reliability, even if it results in a lack of functionality. Initial versions of iOS did not have the ability to copy and paste elements, were unable to receive MMS picture messages, and even did not support multitasking, but still dominated the market for so-called smart-phones. The software must be easy to use, with a clear and elegant GUI.

---

[131] http://www.inclusiveplanet.com/en/channelpost/4998144

[132] http://www.ivona.com/voices.php

Similar trends were noted with software on the PC. The user interface of Google Chrome, which is gaining popularity, consists of four navigation buttons and an address bar. Microsoft, another giant in the IT market, applied similar changes to their Internet Explorer 9, and also in Microsoft Office products, where complicated menu trees have been replaced by a so-called ribbon, where a friendly graphical interface presents only the most popular features.

Running algorithms takes care of everything for you. You will not even have to worry about scheduling, which will be generated automatically.

This application allows you to easily add tracks to the base software with a short and simple form. It displays, among other things, genre and language of the tracks. The same is true of ads, which can be a graphic file or SWF or GIF animation. With the information collected in the forms, the software automatically generates ads for display to users. According to a study carried out in Chapter 4, it can be assumed that some products should be addressed to specific groups of music listeners, which has become the main argument for the implementation of such functionality. The client program has the ability to click on "LIKE," and that server sends information about what a person likes recording. On this basis, the server knows the kind of music you to which you listen and draws advertising from the appropriate pool. Everything works without user intervention, acting as the sender, depending on what music the recipient of two species most "Likes".

with Interdisciplinary factored system for automatic content recommendation.

The program also allows easy recording from a microphone. Just click the appropriate button to start recording, and then again to exit and save them in MP3 format. This soundtrack is automatically added to the music library that is available when you manually create schedules.

It also implemented speech synthesis using Microsoft SAPI. By design, it allows the preparation of recordings in the Polish language using one of the IVONA TTS voices and those built into Windows. You may keep the default settings or manually select the corresponding voice from the drop-down list. Recordings are saved immediately in MP3 format and added to the music library. Before adding them, a user can listen to the recordings and, if necessary, change the text that can be entered manually or from a file tab in TTS (Text To Speech).

Creating a schedule is also automated. If the sender does not set a schedule, it chooses a randomly selected song while taking into account the statistically popular songs so that they are more often broadcast. When you create a schedule, you provide only the approximate time of its transmission. As a default, however, it is issued only when the current track is finished or in the queue of previously defined programs. This system works in a similar way to queuing commands in the CPU. The application automatically creates transitions and shifts the broadcast start time of the program. It is not until the end of the current track that it will be added next to the playlist. The current programming schedule can also preview to review, and edit. When a defined program ends, it automatically returns to shuffle.

Client applications have also been developed. It was important to be able to associate a user with his musical preferences, and to ensure ease of use. Despite the original plan to automatically generate the ID and delete it if it was inactive for a long time, it was decided to use a traditional registration and login. This might also gather personal data and contact information. Client applications are developed primarily in order to be able to transmit the contents of advertisements. Furthermore, efforts were made to choose the development platform so that programs can be transferred to other platforms, supported by web browser plug-ins. A desktop application was developed in C#, and a plug-in for your browser was implemented in Silverlight. Nevertheless, it was important to ease access to radio broadcast, so that copyright applications are not mandatory by using SHOUTcase technology, which allows you to receive broadcasts in virtually any program Media Player.

Care was also taken to transmit a signal without overloading the Internet connection through the use of functions implemented in the BASS plug-in and SHOUTcast technology. They allow convenient implementation of multicast transmission when needed. In other cases, the plug-in automatically selects unicast.

### 6.5. The use of a synthesizer on the radio

Of course, current speech synthesizers are not ready to replace man, but since most broadcasters are amateurs and speech synthesis is making more and more progress, their use is a future-oriented and tangible idea. Perhaps in the near future, synthesizers will be used in professional radio stations, for example, to read news, articles, or as speakers during the night program. The only question is what they are missing, and if they have the potential to replace people?

with Interdisciplinary factored system for automatic content recommendation.

An interesting timbre of a voice or a specific sound is one of the fundamental factors that influence our feelings about specific stations or radio programs. But can emerging technologies introducing the widespread use of speech synthesizers be as interesting?

Of course, one should not expect that synthesizers will come into common use in large stations nationwide in the near future, because the machine will not provide a character editor, and it will not react to situations in a spontaneous manner. Synthesizers can, however, be a great solution for Internet radio stations, which are usually carried out by amateurs. The benefits of using speech are measurable, by obtaining recordings of satisfactory quality. Developers continually improve the sound quality generated by synthesizers. Much has changed since the inception of the first speech synthesizer. Creating a voice-over recording using a speech synthesizer is currently very doable. So, if you want to create your own Internet radio, but do not have the appropriate technical background to provide satisfactory media quality , it may be worthwhile to consider speech synthesizers. Finally, the Web giants such as virtual news directors Poland, BBC News and Onet already use a synthesizer on almost every page that lets you read the article or information.

Of course, not every idea will appeal to listeners. For some people, eliminating the human factor in radio could prove to be unacceptable. But just check the quality of the speech generated recording and compare it with the one you can get at home. Probably a good synthesizer will surprise you. It turns out that the quality of programs created with generated speech does not differ, and often exceeds, that of homegrown editors. While nothing can replace the thrill or editorial "glitches" as an integral part of each program, for establishing loyal audiences, you might be interested in synthesizers, at least until the listeners themselves signal that we need to introduce live dialogue. On the other hand, intelligent genetic algorithms may emerge in the future that will be able to "empathize" with the situation and generate emotional intonations. It seems that the speech synthesis suits in online stations perfectly, especially in the beginning ones, as well as more advanced, probably because it is the future of audio transmissions. A very valuable paper on the subject of sound synthesis is the dissertation of K. Szklannego written in Polish at the Japanese Institute of Information Technology.[133]

## 6.6. Installation of application

Compiling the application produces several executables so you can run it on your computer. An installer is not currently available. Settings related to the technical aspects of the radio must be configured from within Visual Studio.

For the application to run properly, the target computer must be running Windows 7 or newer. It must be the Professional, Enterprise, or Ultimate versions, because only these versions have a built-in SAPI. It is also important that the system be the 32-bit edition. This is essential, because some of the libraries will not function properly in the BASS 64-bit version due to differences in memory management.

For the operating system to function properly, you will also need to install the .NET Framework version 4.0 or later, as well as Silverlight 4.0 or later. In the case of a server, you will need to open the appropriate ports in the firewall. It

---

[133] Optymalizacja funkcji kosztu w korpusowej syntezie mowy polskiej, PJWSTK 2009

with Interdisciplinary factored system for automatic content recommendation.

will be necessary to edit Visual Studio 2010 and the libraries that are included on the CD.

For proper operation of the system, you must also install the TTS SAPI for Windows. Polish voices are available to be downloaded from the Ivona.com. When all these conditions are met, the application should start and be fully functional.

## 6.7. Use and familiarization with the interface

The project was implemented in three applications: the server, a client application that supports the ad system, and a Silverlight client running in a web browser.

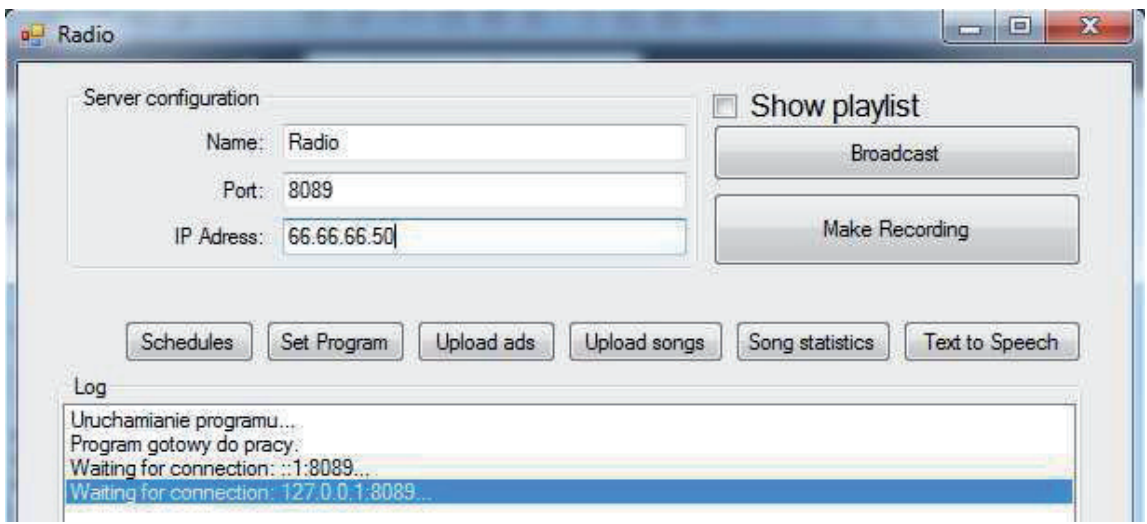

**Figure 11. The main application window.**

The above figure (Figure 11) shows the main window of the server. You have access to the main functions of the program and the basic settings, through which you can specify the name of the radio, the port on which data is being **with Interdisciplinary factored system for automatic content recommendation.**

transmitted, and the IP address of the computer where the server is installed. Here we can see the log window in which printed information as well as any errors that may occur are displayed. When you click on the "Show List" button, a window appears (Figure 12) that displays the works currently added to the playlist. By default, each new song starts playing, but you can delete any of the songs.

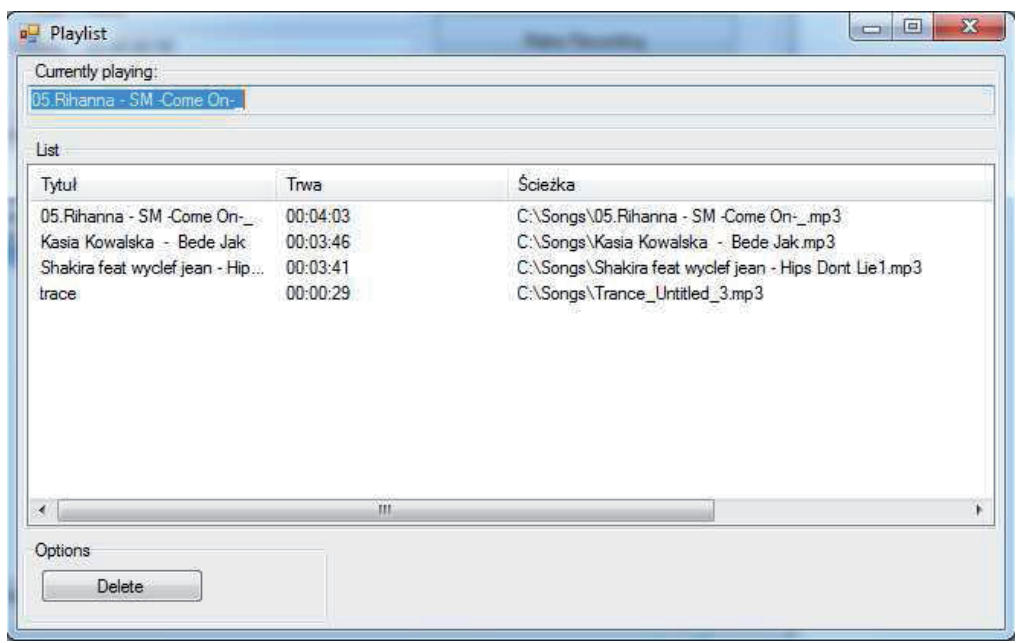

**Figure 12. Playlist.**

When the user clicks the "Broadcast" button, the radio starts transmitting. A "Record Broadcast" button is located just below this, the use of which results in the appearance of a window (Figure 13) for recording the program with a microphone. The user is allowed to select options such as file name, file format, the device from which the recorded signal comes, sample rate, and bit rate.

with Interdisciplinary factored system for automatic content recommendation.

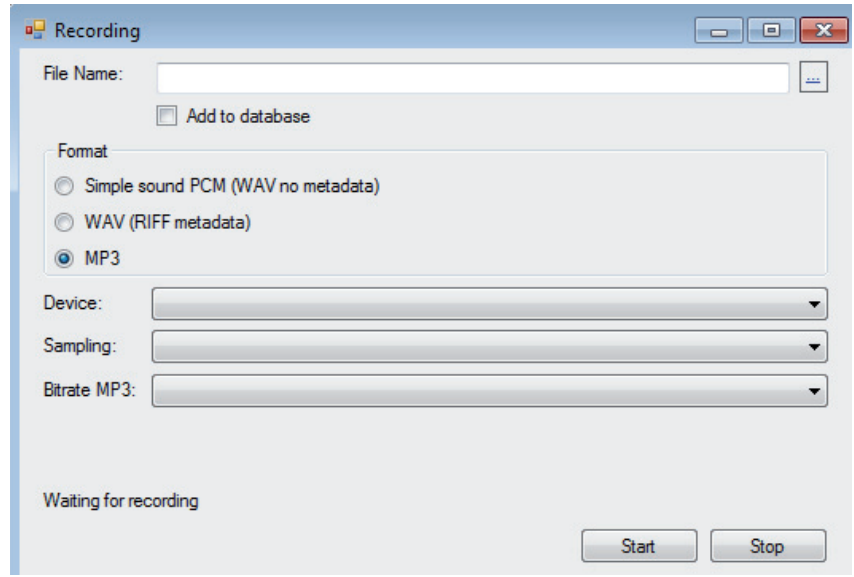

**Figure 13. The window for recording.**

Next, "Schedules" gives you an insight into the planned program (Figure 14). In this window, you can verify and change the time of transmission and fully cancel a show.

with Interdisciplinary factored system for automatic content recommendation.

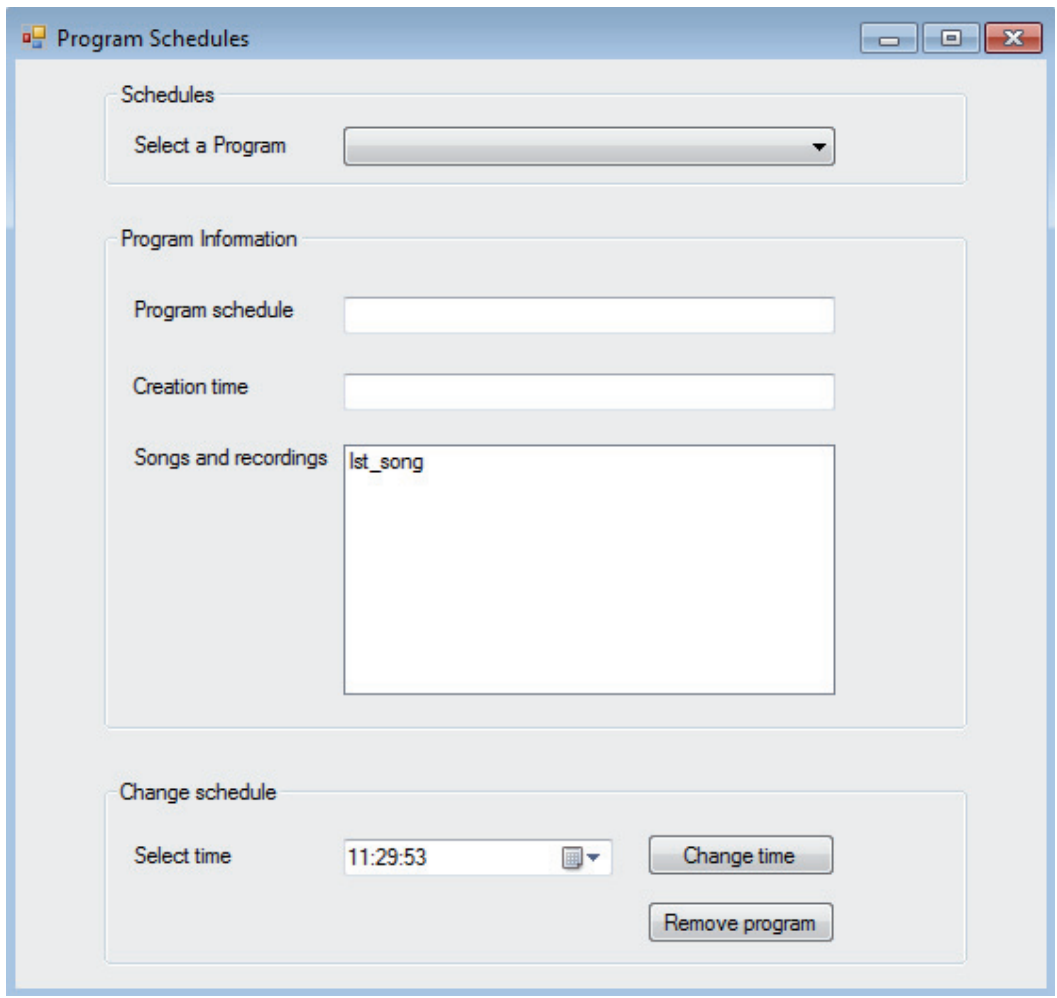

**Figure 14. Schedules Window - planning.**

Under the button "Set Program" is hidden functionality that allows you to define and plan programs (Figure 15). In the first two sections, the user can select interesting songs or recordings from a microphone or a speech synthesizer. When you click "Add to Program," it will be moved to the third section, where you can change the order of songs in a single program. After determining the hour and the date of broadcast, use the "Create Program" to save and accept the broadcast.

with Interdisciplinary factored system for automatic content recommendation.

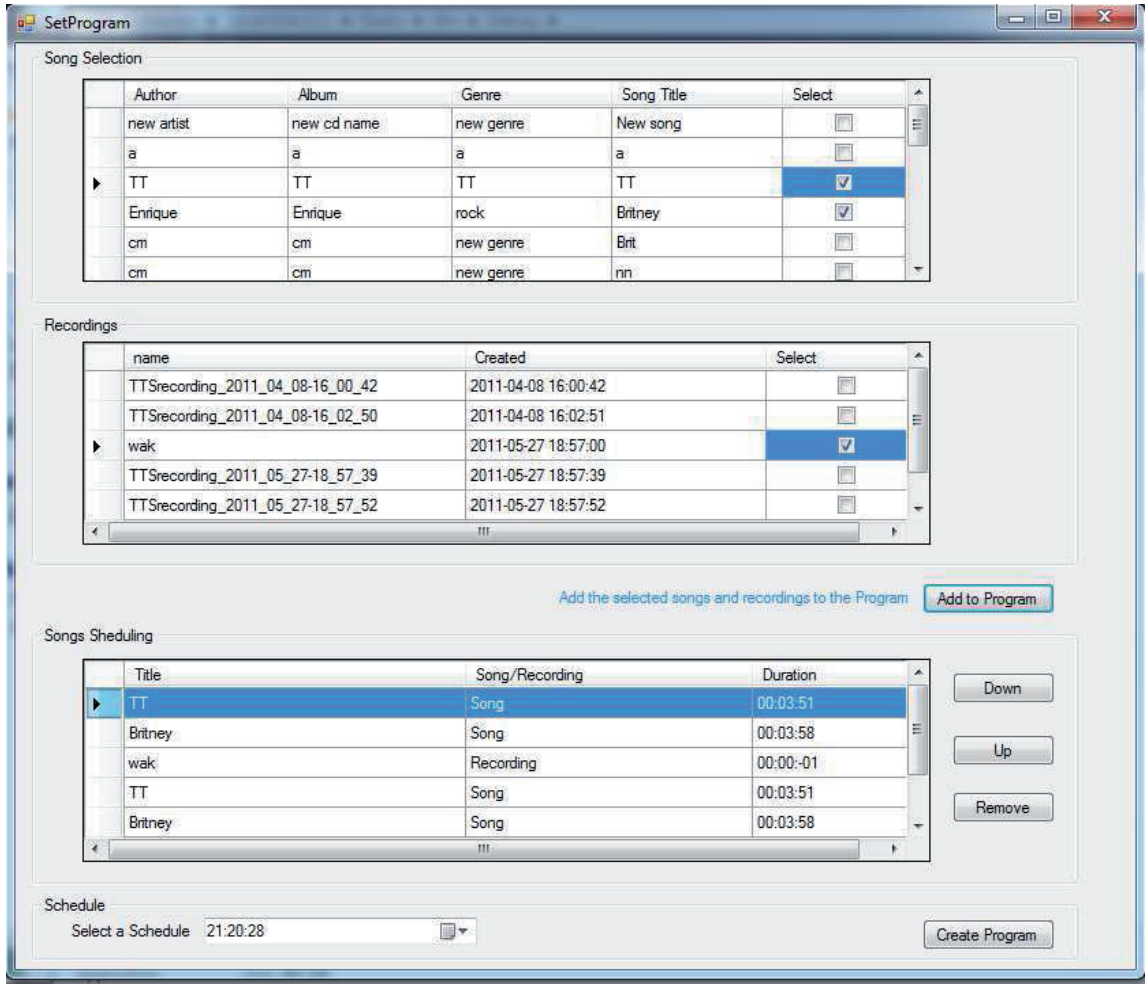

**Figure 15. Creating the program.**

The "Upload Ads" button allows you to enter new advertising content. The user must specify the path to the image file (ads), the genre for which the ad will be broadcast, and the URL of the page on which the ad will show (Figure 16). Genre is specified because of the relationship described in Chapter 4.

with Interdisciplinary factored system for automatic content recommendation.

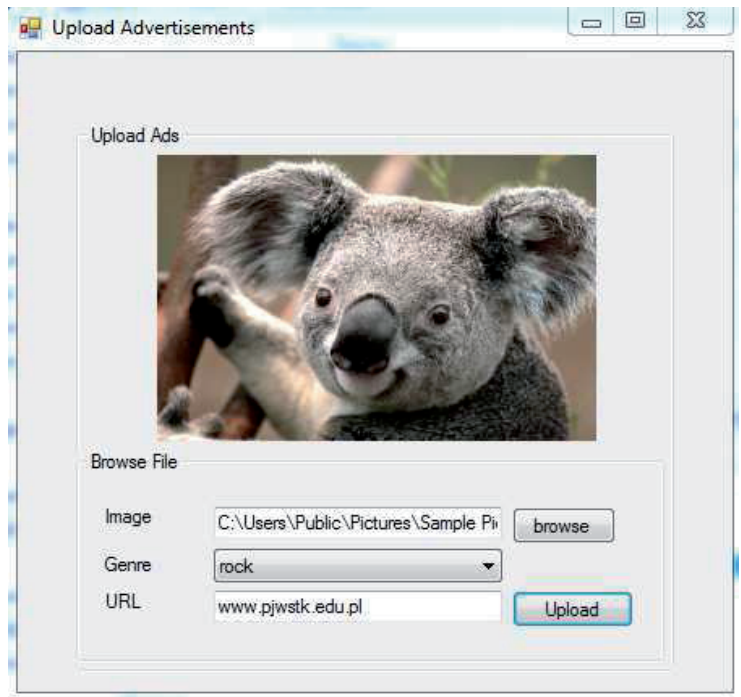

**Figure 16. Adding ads.**

The "Upload Songs" button allows you to add the next track. You can specify the language of the song, enter its title, and introduce new, or select from a list of track parameters such as artist, genre and album title (Figure 17).

with Interdisciplinary factored system for automatic content recommendation.

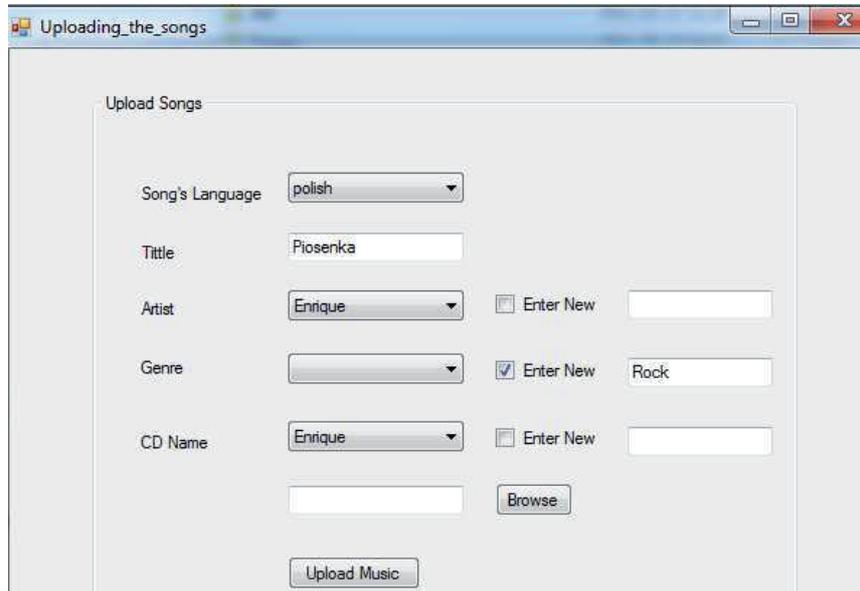

**Figure 17. Dodawanie piosenek.**

The application has also implemented very simple statistics that show a list of songs starting with the most popular (Figure 18). You can also export this data to Excel.

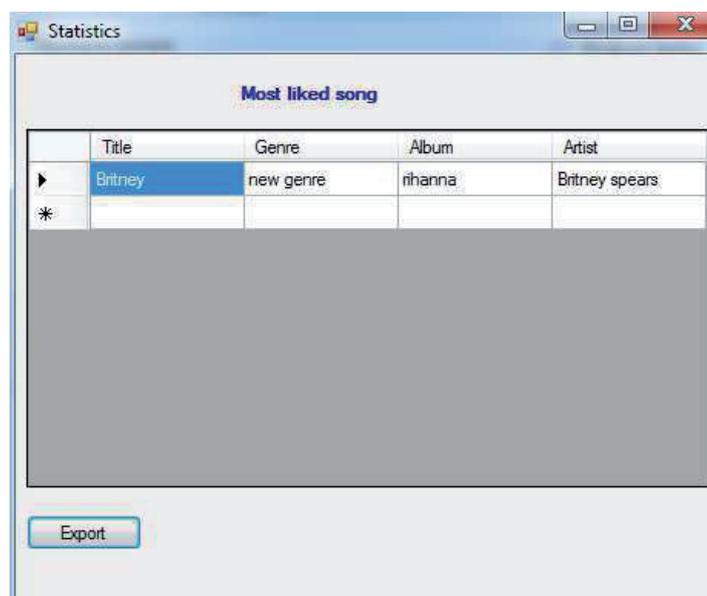

**Figure 18. Statistics.**

with Interdisciplinary factored system for automatic content recommendation.

On the server side, you can also use speech via the "Text to Speech" window (Figure 19). The synthesizer allows you to load.txt files and introduce your own content from the keyboard in both Polish and English. The user can preview to read the recording before the creation of the desired content. Depending on the language, you can select one of the available voices. You can also configure the volume of voices and the pace of reading.

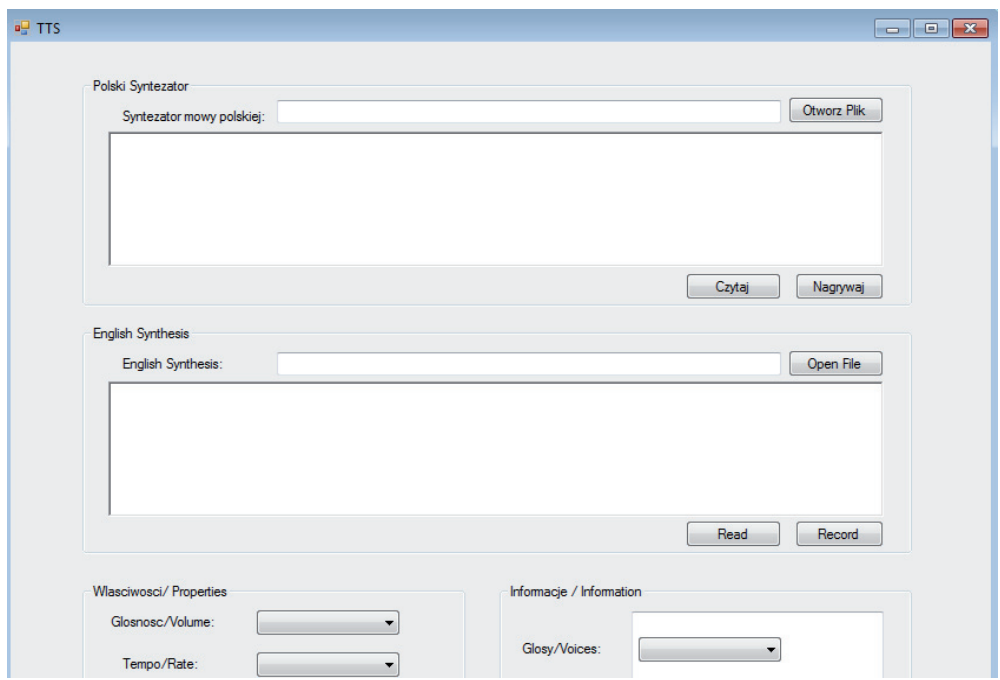

**Figure 19. Synthesis window.**

The client application is also relatively simple. It is presented in Figure 20. When it starts, you will be presented with the login screen, where you can also complete registration. It is necessary to have a user account, because otherwise personalizing content for the recipient would not be possible.

with Interdisciplinary factored system for automatic content recommendation.

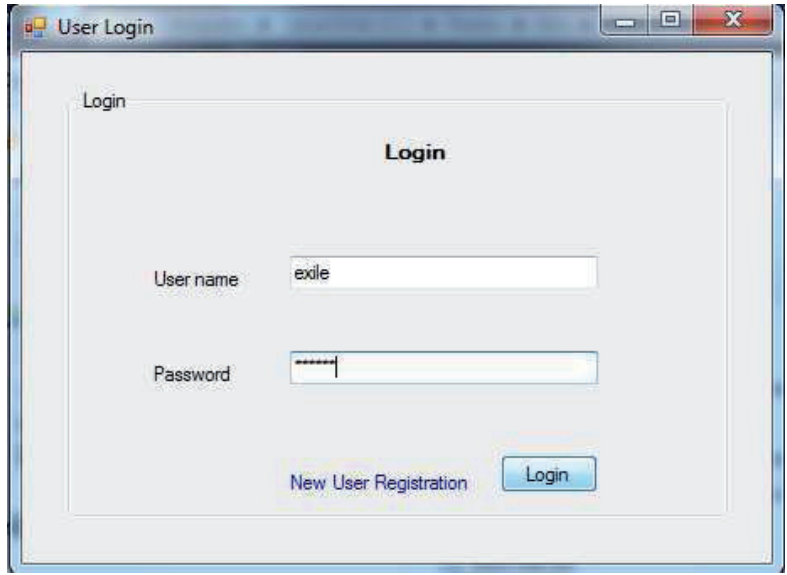

**Figure 20. Login window.**

After login you the client application will be displayed. Listeners are presented advertising and basic information about the program.

Users may use the "Connect" button to connect to the server and the "LIKE" button to vote for their favorite songs. Figure 21 shows the client application, and Figure 22 shows the Silverlight plug-in that allows receipt of radio via a web browser.

with Interdisciplinary factored system for automatic content recommendation.

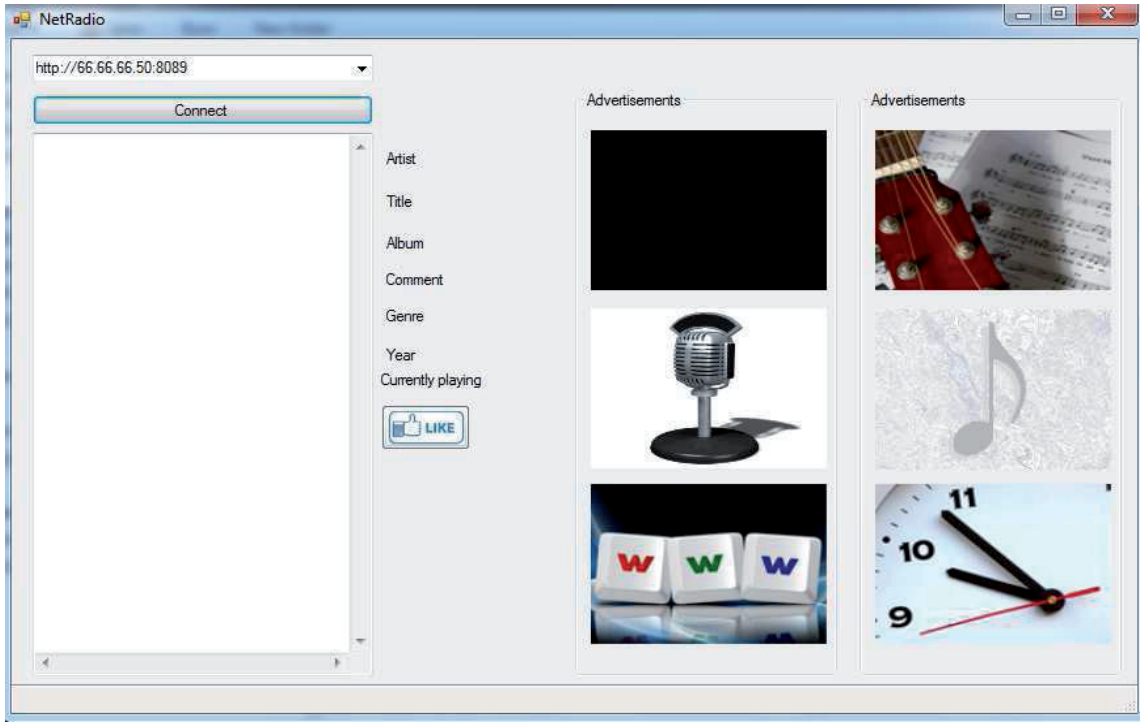

**Figure 21. The client application.**

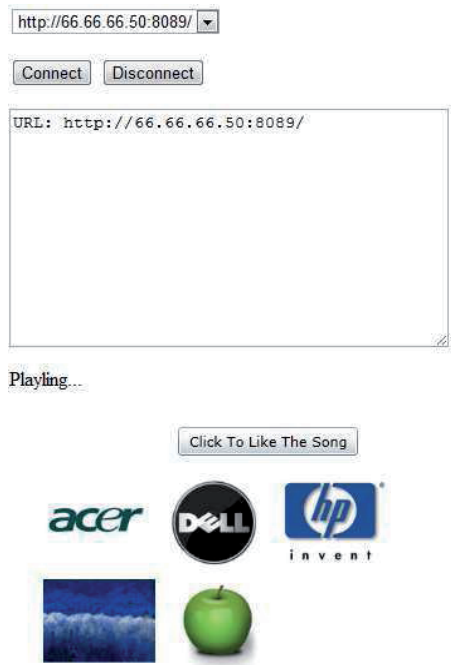

**Figure 22.Silverlight Application**

with Interdisciplinary factored system for automatic content recommendation.

## 6.8. The structure of the code and implementation

The project used Visual Studio version 2010, the .NET 4.0 language, and the BASS plug-in to implement sound transmission through the Internet. Originally, the application was to be focused only on Polish recipients and senders. However, the plan changed, and the final phase the project, including comments, was translated into English. XML tags were also used to comment on the application code in order to more easily generate documentation. A tool to generate documentation for open source applications, NDoc[134], was used to create a document as the web page found in Appendix D. Visual Studio was used to create a solution, consisting of the three projects: the Radio server, a desktop client, and a browser client Radio developed using Silverlight. It was created as a network Web Service (shown in Figure 23), which mediates the exchange of data between the client application and the server. It is responsible for authentication, user account creation, and sending advertising and information about user preferences, especially for the transmission of the audio stream.

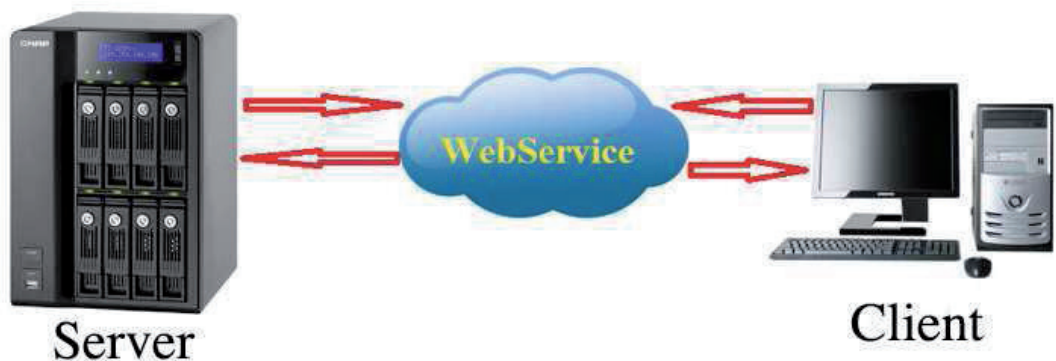

**Figure 23. Schematic of Web Service..**

---

[134] http://ndoc.sourceforge.net/

with Interdisciplinary factored system for automatic content recommendation.

The server retrieves the radio recordings from any folder on your hard drive, along with information about their location and attributes, and then stores them in XML format. The two files and playlist.xml songs.xml are filled by the Upload_music() method contained in the Uploadingthesongs.cs class. After starting, the server will be issued a playlist.xml music file. It is responsible for managing the PlaylistItem.cs class' play list. The songs.xml file, however, stores song information (title, location, etc.). Also, speech recordings will be in it. Users can vote for their favorite songs. The server stores information about that in the User_Likes.xml file. The voting procedure corresponds to the method public bool like_the_song (string title, string user, string userid), part of the class NetRadio.cs. The following methods are also involved in this process:

- check_authentication (string id string pass)
- check_Expired_like()
- check_Expired_user()

They are in the class Radioser.asmx.cs.

The following code snippet shows how to add songs to the database of songs:

```
string song_id = string.Format("song_{0}", DateTime.Now.ToFileName());
XDocument xmlDoc = XDocument.Load(initial_path + @"\Songs.xml");

xmlDoc.Element("songs").Add(new XElement("song", new XElement("id", song_id),
            new XElement("lang", cmb_langsong.SelectedItem), new XElement("title", txt_Title.Text),
       new XElement("artist", artist), new XElement("genre", genre),
        new XElement("album", album), new XElement("path", file_name),
        new XElement("added", System.DateTime.Now.ToString())));

xmlDoc.Save(initial_path + @"\Songs.xml");
```

The next section defines the method for adding songs to a playlist:

with Interdisciplinary factored system for automatic content recommendation.

```
playlist.Items.Add(new PlaylistItem()
    {
        ArePropsRetrieved = true,
        FilePath = file_name,
        //Duration = duration,
        Seconds = ddd,
        Title = txt_Title.Text
    });
playlist.Save(initial_path + @"\playlist.xml");
```

The application server, which does not have a built-in audio converter, has not imposed a bit rate. The current bit rate is taken from the MP3 file, which is why music must be prepared according to your needs. The files are transcoded into a format suitable for streaming in the classroom using the Transcode.cs ReadFileToEnd (string filePath) method. The bit rate corresponds to a field named bitrate, belonging to the BeginStreamingSongs (Letter<PlaylistItem> Playlistitems, int index) method contained in the Scheduler.cs class.

After the method of transmission is specified, the ThreadProc() in class Scheduler.cs is started, whose task is to recall the objects belonging to the playlist.

with Interdisciplinary factored system for automatic content recommendation.

```
private void ThreadProc()
    {
        while (true)
        {
            if (isAbortRequested) return;
            if (schedulestarted == true)
            {
                if (scheduler == null)
                {
                    (scheduler = new Thread(ScheduleThread)).Start();
                }
            }
            else
            {
                if (scheduler != null)
                {
                    schedulelist.Clear();
                    scheduler.Abort();
                    scheduler = null;
                }
                while (server.Playlist.Items.Count == server.CurrentIndex)
                {
                    lock (locker)
                    {
                        if (isAbortRequested) return;
                        Thread.Sleep(2000);
                        server.CurrentIndex = 0;
                    }
                }
                BeginStreamingSongs(server.Playlist.Items, server.CurrentIndex);
                server.CurrentIndex++;
            }
        }
    }
```

The algorithm checks whether the current time is not scheduled for any program, in which case it calls another thread, ScheduleThread(), which is responsible for playing the entire show. When the work is completed, it returns to the main activity thread ScheduleThread().

with Interdisciplinary factored system for automatic content recommendation.

```
ScheduleThread():

    private void ScheduleThread()
{
    int sindex = 0;
    while (true)
    {
        if (isAbortRequested) return;

        while (schedulelist.Items.Count == sindex)
        {
            schedulestarted = false;

            return;
        }
        BeginStreamingSongs(schedulelist.Items, sindex);
        sindex++;
    }
}
```

Before you start broadcasting, the algorithm waits until the song ends. The most important methods for managing these processes, found in the Scheduler.cs class, are:

- ScheduleThread()
- BeginStreamingSongs (schedulelist.Items, sindex);
- ThreadProc()
- public Scheduler (ServerWindow server)

The ad generation system uses the Ads.xml file, information such as a name, genre and URL are stored. The ads are generated based on what song is currently playing, or what genre of music the user prefers. For their generation, the network service Radioser.asmx.cs is implemented in the room class, as well as the methods Change_adds (string user) and get_userliked_genre (string user_id).

```csharp
Radioser.asmx.cs:

public ArrayList Change_adds(string user)
{
    ArrayList arlist = new ArrayList();
    ArrayList ar_user_genre = new ArrayList();

    ar_user_genre = get_userliked_genre(user);
    if (ar_user_genre.Count > 0)
    {
        foreach (XElement level1Element in XElement.Load(initial_path+@"Ads.xml").Elements("image"))
        {
            foreach (string s in ar_user_genre)
            {
                if ((level1Element.Attribute("genre").Value) == s && Convert.ToInt32(level1Element.Attribute("count").Value) < 10)
                    arlist.Add((level1Element.Attribute("path").Value));
            }
        }
    }
    else
    {
        foreach (XElement level1Element in XElement.Load(initial_path+@"Ads.xml").Elements("image"))
        {
            arlist.Add((level1Element.Attribute("path").Value));
        }
    }
    if (arlist.Count == 0)
    {
        foreach (XElement levelElement in XElement.Load(initial_path + @"Ads.xml").Elements("image"))
        {
            arlist.Add((levelElement.Attribute("path").Value));
        }
    }

    return arlist;
}
```

They examine what genre of music is currently playing, and present relevant ads in intervals specified in the method timer1_Tick (object sender, EventArgs e), part of the NetRadio.cs class.

with Interdisciplinary factored system for automatic content recommendation.

```csharp
private void timer1_Tick(object sender, EventArgs e)
{
    ArrayList list_add = new ArrayList();
    object[] list_ad = radio_service.Change_adds(user_id);
    foreach (var i in list_ad)
    {
        list_add.Add(i);
    }
    //list_add =radio_service.Change_adds(user_id);
    try
    {
        pictureBox1.Load(@list_add[counter1].ToString());
        pictureBox2.Load(@list_add[counter2].ToString());
        pictureBox3.Load(@list_add[counter3].ToString());
        pictureBox4.Load(@list_add[counter4].ToString());
        pictureBox5.Load(@list_add[counter5].ToString());
    }
    catch { }

    counter1 = counter1 + 5;
    counter2 = counter2 + 5;
    counter3 = counter3 + 5;
    counter4 = counter4 + 5;
    counter5 = counter5 + 5;

    if (counter1 >= list_add.Count)
        counter1 = 0;

    if (counter2 >= list_add.Count)
        counter2 = 1;
    if (counter3 >= list_add.Count)
        counter3 = 2;
    if (counter4 >= list_add.Count)
        counter4 = 3;
    if (counter5 >= list_add.Count)
        counter5 = 4;
}
```

An administrator can check the most popular genres in the users file User_Likes.xml. If you do not have a favorite genre, advertising will be issued only based on the song currently being broadcast. The test method that corresponds to the user's preferences is get_userliked_genre (string user_id), part of Radioser.asmx.cs. To connect to the server using the basic port 8089, which like the IP address can be changed in the class ServerConfiguration.cs. Establishing streaming to each customer requires a separate port for each listener. These parameters are negotiated using the Start()methods in the Client.cs Listener.cs class.

SHOUTcase technology has been implemented on the client side in the classroom Client.cs. It starts with Client.Start method().

```csharp
    public void Start()
{
  try
  {
    var recBuff = new byte[100];
    var recCount = tcpClient.Client.Receive(recBuff);
    var recText = ASCIIEncoding.ASCII.GetString(recBuff, 0, recCount);

    var startPos = recText.IndexOf("HTTP", 1);
    if (startPos < 0)
    {
      ClientError("Nie odnaleziono nagłówka HTTP w żądaniu klienta.");
      return;
    }

    var httpVersion = recText.Substring(startPos, 8).ToLowerInvariant();

    if (!recText.ToLowerInvariant().Contains("get"))
    {
      ClientError("Żądanie klienta jest nieprawidłowe, obsługiwane jest tylko żądanie GET.");
      return;
    }

    var shoutCastHeader = ShoutcastHelpers.GenerateShoutCastHeader(
      ServerConfiguration.UserAddress, httpVersion, ServerConfiguration.PortNumber.ToString());
```

with Interdisciplinary factored system for automatic content recommendation.

```csharp
if (!SendToClient(tcpClient, shoutCastHeader, "Nie można przesłać nagłówka shoutcast.")) return;

stream = tcpClient.GetStream();
if (!stream.CanWrite)
{
    ClientError("Nie można pisać do strumienia danych klienta.");
    return;
}

var packets = new List<BufferedPacket>(AudioBuffer.Packets);
AudioBuffer.PacketReady += OnNewPacketReady;
foreach (var p in packets)
{
    if (!SendToClient(tcpClient, p.GetRawData(), "Nie można wysłać pakietu danych do klienta.")) return;
}

clientendpointaddress = tcpClient.Client.RemoteEndPoint.ToString();
}
catch (Exception ex)
{
    //AudioBuffer.PacketReady -= OnNewPacketReady;
    Close();
    clientendpointaddress = "";
}
}
```

Then the method SendToClient(), which is responsible for establishing calls to the appropriate port and caching, runs.

with Interdisciplinary factored system for automatic content recommendation.

```csharp
    private bool SendToClient(TcpClient client, byte[] buffer, string errorMessage)
{
    var isOk = false;
    try
    {
        if (client != null && client.Connected == true) //BT
        {
            if (client.Client.Send(buffer) != buffer.Length)
            {
                Log.OneLine(errorMessage);
            }
            else
            {
                isOk = true;

            }
        }

    }
        catch (System.Net.Sockets.SocketException se)
    {
    }
    }

    void OnNewPacketReady(object sender, EventArgs e)
{
    if (tcpClient != null && tcpClient.Connected == true)
    {
        if (!SendToClient(tcpClient, AudioBuffer.Last.GetRawData(), "")) return;
    }
}
```

In addition to sound, metadata are transmitted, using the synchronization methods in the NetRadio.cs Class:

- MetaSync (int handle, int channel, int date IntPtr user)
- UpdateTagDelegate();
- UpdateTagDisplay()

The most important feature for transmitting sound is the BASS library. First of all, there are two imported files, basswma.dll and Bass.Net.dll. Plug-ins are called by the method Bass.BASS_StreamCreateFile NetRadio.cs(), which prepares the file for transmission and Bass.BASS_ChannelPlay (stream1, false), which starts the transmission.

with Interdisciplinary factored system for automatic content recommendation.

```csharp
        private void NetRadio_Load(object sender, System.EventArgs e)
{
    comboBoxURL.SelectedIndex = 0;
    if (Utils.HighWord(Bass.BASS_GetVersion()) != Bass.BASSVERSION)
    {
        MessageBox.Show(this, "Wrong Bass Version!");
    }

    Bass.BASS_SetConfigPtr(BASSConfig.BASS_CONFIG_NET_AGENT, _myUserAgentPtr);

    Bass.BASS_SetConfig(BASSConfig.BASS_CONFIG_NET_PREBUF, 0); // so that we can display the buffering%
    Bass.BASS_SetConfig(BASSConfig.BASS_CONFIG_NET_PLAYLIST, 1);

    if (Bass.BASS_Init(-1, 44100, BASSInit.BASS_DEVICE_DEFAULT, this.Handle))
    {

        if (Bass.BASS_SetConfig(BASSConfig.BASS_CONFIG_WMA_PREBUF, 0) == false)
        {
            Console.WriteLine("ERROR: " + Enum.GetName(typeof(BASSError), Bass.BASS_ErrorGetCode()));
        }
        // we alraedy create the user callback methods...
        myStreamCreateURL = new DOWNLOADPROC(MyDownloadProc);
    }
    else
        MessageBox.Show(this, "Bass_Init error!");
}

        private void button1_Click(object sender, System.EventArgs e)
{
    Bass.BASS_StreamFree(_Stream);
    this.textBox1.Text = "";
    _url = this.comboBoxURL.Text;
    // test BASS_StreamCreateURL
```

```csharp
bool isWMA = false;
if (_url != String.Empty)
{
    this.textBox1.Text += "URL: " + _url + Environment.NewLine;
    // create the stream
    _Stream = Bass.BASS_StreamCreateURL(_url, 0, BASSFlag.BASS_STREAM_STATUS, myStreamCreateURL, IntPtr.Zero);
    if (_Stream == 0)
    {
        // try WMA streams...
        _Stream = BassWma.BASS_WMA_StreamCreateFile(_url, 0, 0, BASSFlag.BASS_DEFAULT);
        if (_Stream != 0)
            isWMA = true;

        else
        {
            // error
            this.statusBar1.Text = "ERROR...";
            return;
        }
    }
    _tagInfo = new TAG_INFO(_url);
    BASS_CHANNELINFO info = Bass.BASS_ChannelGetInfo(_Stream);
    if (info.ctype == BASSChannelType.BASS_CTYPE_STREAM_WMA)
        isWMA = true;
    // ok, do some pre-buffering...
    this.statusBar1.Text = "Buffering...";
    if (!isWMA)
    {
        // display buffering for MP3, OGG...
        while (true)
        {
            long len = Bass.BASS_StreamGetFilePosition(_Stream, BASSStreamFilePosition.BASS_FILEPOS_END);
            if (len == -1)
                break; // typical for WMA streams
            // percentage of buffer filled
            float progress = ((
                Bass.BASS_StreamGetFilePosition(_Stream, BASSStreamFilePosition.BASS_FILEPOS_DOWNLOAD) -
                Bass.BASS_StreamGetFilePosition(_Stream, BASSStreamFilePosition.BASS_FILEPOS_CURRENT)
                ) * 100f) / len;

            if (progress > 75f)
            {
                break; // over 75% full, enough
            }

            this.statusBar1.Text = String.Format("Buffering... {0}%", progress);
        }
    }
    else
    {
        // display buffering for WMA...
        while (true)
        {
            long len = Bass.BASS_StreamGetFilePosition(_Stream, BASSStreamFilePosition.BASS_FILEPOS_WMA_BUFFER);
            if (len == -1L)
                break;
            // percentage of buffer filled
            if (len > 75L)
            {
                break; // over 75% full, enough
            }

            this.statusBar1.Text = String.Format("Buffering... {0}%", len);
        }
    }
```

```csharp
// get the meta tags (manually - will not work for WMA streams here)
string[] icy = Bass.BASS_ChannelGetTagsICY(_Stream);
if (icy == null)
{
    // try http...
    icy = Bass.BASS_ChannelGetTagsHTTP(_Stream);
}
if (icy != null)
{
    foreach (string tag in icy)
    {
        this.textBox1.Text += "ICY: " + tag + Environment.NewLine;
    }
}
// get the initial meta data (streamed title...)
icy = Bass.BASS_ChannelGetTagsMETA(_Stream);
if (icy != null)
{
    foreach (string tag in icy)
    {
        this.textBox1.Text += "Meta: " + tag + Environment.NewLine;
    }
}
else
{
    // an ogg stream meta can be obtained here
    icy = Bass.BASS_ChannelGetTagsOGG(_Stream);
    if (icy != null)
    {
        foreach (string tag in icy)
        {
            this.textBox1.Text += "Meta: " + tag + Environment.NewLine;
        }
    }
}

// alternatively to the above, you might use the TAG_INFO (see BassTags add-on)
// This will also work for WMA streams here ;-)
if (BassTags.BASS_TAG_GetFromURL(_Stream, _tagInfo))
{
    // and display what we get
    this.textBoxAlbum.Text = _tagInfo.album;
    this.textBoxArtist.Text = _tagInfo.artist;
    this.textBoxTitle.Text = _tagInfo.title;
    this.textBoxComment.Text = _tagInfo.comment;
    this.textBoxGenre.Text = _tagInfo.genre;
    this.textBoxYear.Text = _tagInfo.year;
}

// set a sync to get the title updates out of the meta data...
mySync = new SYNCPROC(MetaSync);
Bass.BASS_ChannelSetSync(_Stream, BASSSync.BASS_SYNC_META, 0, mySync, IntPtr.Zero);
Bass.BASS_ChannelSetSync(_Stream, BASSSync.BASS_SYNC_WMA_CHANGE, 0, mySync, IntPtr.Zero);
```

with Interdisciplinary factored system for automatic content recommendation.

```csharp
// start recording...
int rechandle = 0;
if (Bass.BASS_RecordInit(-1))
{
    _byteswritten = 0;
    myRecProc = new RECORDPROC(MyRecoring);
    rechandle = Bass.BASS_RecordStart(44100, 2, BASSFlag.BASS_RECORD_PAUSE, myRecProc, IntPtr.Zero);
}
this.statusBar1.Text = "Playling...";
//// play the stream
Bass.BASS_ChannelPlay(_Stream, false);
//// record the stream
Bass.BASS_ChannelPlay(rechandle, false);
timer1.Enabled = true;
timer2.Enabled = true;
timer3.Enabled = true;
}
}
```

A detailed description of all the classes, methods, and even variables is included on the DVD as Appendix C to this paper.

### 6.9. Tests

Each application requires thorough testing of both reliability and performance. The first is the time-consuming process, often impossible to achieve one hundred percent before you deploy the application to be used, of catching all faults. Their removal is difficult logistical task, requiring passage through all the possible behavior of the user, which is desired before releasing that application. One limitation was the size of the test environment. In the case of Internet radio, where there are likely to be hundreds or even thousands of customers, the best solution would be to test the application on such a group, but this was not possible due to limited resources. Instead, testing virtualization using the Microsoft Hyper-V and RemoteFX technologies was used. We created 11 virtual machines, one server running under Windows 7 (Service Pack 1) and with Interdisciplinary factored system for automatic content recommendation.

10 workstations also under the control of the same system. This was the basis of an attempt to estimate hardware requirements and to test reliability.

### 6.9.1. Performance tests

There are three tools available to check the performance of your application. They include the Task Manager built into Windows, as well as the Performance Monitor, also available on this system. Another beneficial tool is the application NetLimiter 3 Pro, which accurately estimates the amount of incoming and outgoing data for a specific application. Our virtual testing machine had a single 2.4 GHz processor and 1GB of RAM.

A radio server developed through the use of plug-ins and a powerful development platform is characterized by a relatively low, and importantly very stable, demand for resources. During the transfer of data to 10 clients, overall CPU usage for the entire computer as shown by Task Manager does not exceed 5% (Figure 24).

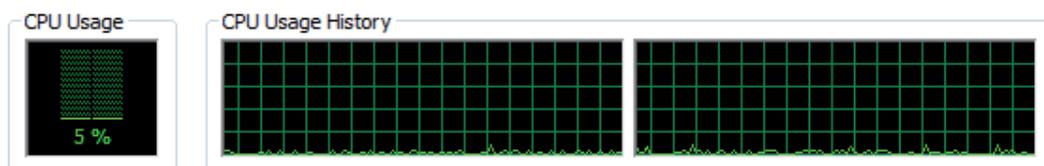

**Figure 24. CPU**

At the same time, the Event Monitor Processes tab in CPU usage indicates that the server is 0, and the value does not go over 1%. The memory requirement is 17MB RAM (Figure 25).

with Interdisciplinary factored system for automatic content recommendation.

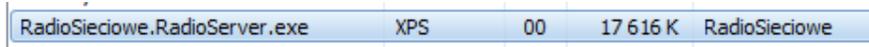

**Figure 25. RAM**

In the case of a single audio stream transmission, the NetLimiter application also shows that the data flow is stable. The demand for transfer can be described by the equation:

$T = n*(A+B)$, where:

T - the resulting transfer

n - the number of concurrent connections to the server

A – the bit rate of the data

B - the amount of data needed to synchronize the clients with the server

The chart below (Figure 27) and Figure 26 shows that the data needed to synchronize did not exceed 1 kbits, and the transmission is stable. The only exceptions were the need for the link, represented as individual bars in the graphs, at the moment when the client is connecting to it and are sent advertising.

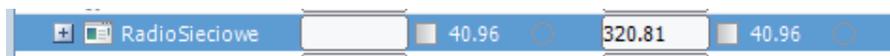

**Figure 26. Demand for outgoing transmission link to one customer.**

with Interdisciplinary factored system for automatic content recommendation.

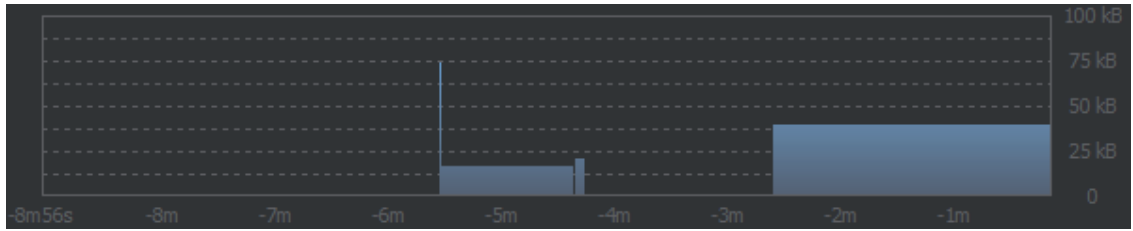

Figure 27. Demand for outgoing transmission link to one customer.

Previous measurements confirm the correctness of the Windows Performance Monitor (Figure 28). It presents a chart with similar CPU usage and a constant need for RAM memory for the transfer.

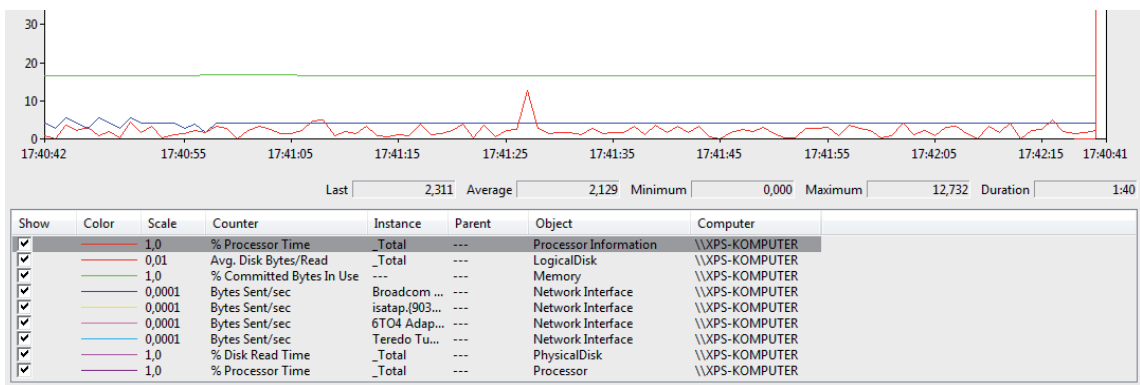

Figure 28. Indication of the performance of the Performance Monitor **application**

Since the client application processor resource requirements amount to about 1-2% in the case, this does not include extensive Flash ads. A greater demand for memory RAM, which is connected with the need to buffer the broadcast advertising content in RAM requires levels around 40 Mb.

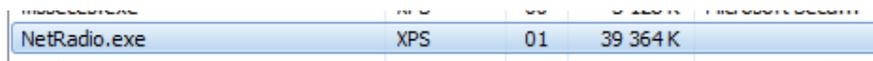

Figure 29. Measuring client application requirements.

with Interdisciplinary factored system for automatic content recommendation.

### 6.9.2. Reliability tests

During stability testing of the application, using the 11 virtual machines, 5 of them benefited from the existence of the client application, while Winamp was used to monitor the signal . Attempts were made to check the functionality of each individual, and then transmit the broadcast for a period of 30 minutes, calculated from the change in the server configuration. The other item of interest is whether there are any correlations between the settings that could cause the server to malfunction. We made every effort to handle exceptions that may occur during unusual circumstances, such as loss of Internet connectivity, sudden disconnection of the customer, for example, due to computer failure or communications interruption between the client and the server. All the above-mentioned problems have been addressed.

### 6.9.3. Approximation of hardware requirements

Approximation of hardware requirements is not a trivial task due to the limited testing. However, you can assume that each computer meets the minimum requirements to run Windows 7, and to be a machine that fully handles the server and the client program. Minimum requirements are shown in Table 5.

|  | Server | Client |
|---|---|---|
| Minimum | - Windows 7 32-bit<br>- support for SAPI<br>- .NET 4.0<br>- IIS6<br>- 1 voice of IVONA for synthesis<br>- 1GHz processor<br>- 1GB of RAM<br>- 100MB of free space of Disk (not counting songs and advertisements) | - Windows 7 32-bit<br>- .NET 4.0<br>- 1GHz processor<br>- 512 MB of RAM<br>- 50MB of free space of Disk |
| Recommended | - Windows 7 32-bit<br>- SAPI<br>- .NET 4.0<br>- IIS7<br>- 1 IVONA voice for synthesis<br>- dual-core processor (up to 500 customers), quad (above 500 clients).<br>- 1GB RAM + 2MB for each listener<br>- 100MB of free disk space (not counting | - Windows 7 32-bit<br>- .NET 4.0<br>- processor 1GHz<br>- 1GB RAM<br>- 50MB of free space on Disk |

| | songs and advertisements) | |
|---|---|---|

**Table 6. Approximate hardware requirements.**

# SUMMARY

This study has discussed in detail Internet radio and traditional radio, which has been known for more than 100 years in communications.

Issues discussed include the construction of Internet radio and techniques used in the design of such solutions, and the demonstrated characteristics of publishing and receiving music,. A key element of this effort was a unique Internet radio application design, in addition to viewing this issue from a practical side.

The need for radio innovation in the context of the changing reality, economic development, and individual needs of its customers has been shown. This paper has depicted a detail picture of the human psyche, a market that wants to manipulate, as well as marketing techniques that can be used effectively to influence the emotional human.

Also, the legal and financial aspects are very important influences on the development of Internet radio, with the result that the reader can see the problems of the twenty-first century through the prism of the competition amount emerging technologies.

with Interdisciplinary factored system for automatic content recommendation.

This paper shows that there are both opportunities for and threats because of the development of Internet radio and other electronic media. Our understanding of the technical and legal aspects are immature, because though Internet radio is gaining recognition around the world, it is not yet very popular, and is most commonly used alongside traditional or amateur radio.

The history of radio and the birth of its Internet counterpart show that innovation in communications that is not impossible to achieve. However, you have to create the right legal and social conditions to fully technically developed it. It could be the next tradition established in the evolutionary model.

# TABLE OF FIGURES

# LIST OF TABLES

# AKNOWLEDGEMENTS

*Krzysztof Wołk was supported by the European Community from the European Social Fund within the Interkadra project UDA-POKL-04.01.01-00-014/10-00.*

with Interdisciplinary factored system for automatic content recommendation.

with Interdisciplinary factored system for automatic content recommendation.